\documentclass[useAMS,usegraphicx,usenatbib]{mn2e}
\usepackage{times}

\newcommand{\Msun}{\ensuremath{\,{\rm M}_\odot}}                  
\newcommand{\Rsun}{\ensuremath{\,{\rm R}_\odot}}                  
\newcommand{\Lsun}{\ensuremath{\,{\rm L}_\odot}}                  
\newcommand{\psun}{\ensuremath{\,\rho_\odot}}                     
\newcommand{\Mjup}{\ensuremath{\,{\rm M}_{\rm Jup}}}              
\newcommand{\Rjup}{\ensuremath{\,{\rm R}_{\rm Jup}}}              
\newcommand{\pjup}{\ensuremath{\,\rho_{\rm Jup}}}                 
\newcommand{\Mearth}{\ensuremath{\,{\rm M}_\oplus}}               
\newcommand{\Teff}{\ensuremath{T_{\rm eff}}}                      
\newcommand{\Teffsun}{\ensuremath{T_{\rm eff,\odot}}}             
\newcommand{\Teq}{\ensuremath{T_{\rm eq}^{\,\prime}}}             
\newcommand{\safronov}{\ensuremath{\Theta}}                       
\newcommand{\logg}{\ensuremath{\log g}}                           
\newcommand{\FeH}{\ensuremath{\left[\frac{\rm Fe}{\rm H}\right]}} 
\newcommand{\MoH}{\ensuremath{\left[\frac{\rm M}{\rm H}\right]}}  
\newcommand{\Porb}{\ensuremath{P_{\rm orb}}}                      
\newcommand{\kms}{\,km\,s$^{-1}$}                                 
\newcommand{\ms}{\,m\,s$^{-1}$}                                   
\newcommand{\mss}{\,m\,s$^{-2}$}                                  
\newcommand{\as}{\ensuremath{^{\prime\prime}}}                    
\newcommand{\chir}{\ensuremath{\chi_\nu^{\,2}}}                   
\newcommand{\lten}{\ensuremath{\log_{10}}}                        

\newcommand{\mc}[1]{\multicolumn{2}{c}{#1}}
\newcommand{\mcc}[1]{\multicolumn{3}{c}{#1}}
\newcommand{\er}[3]{\ensuremath{#1^{+#2}_{-#3}}}
\newcommand{\err}[5]{\ensuremath{#1\,^{+#2}_{-#3}\,^{+#4}_{-#5}}}
\newcommand{\erc}[3]{\mc{\ensuremath{#1^{+#2}_{-#3}}}}

\newcommand{\ercc}[3]{\mcc{\ensuremath{#1^{+#2}_{-#3}}}}
\newcommand{\ermcc}[5]{\mcc{\ensuremath{{#1\,^{+#2}_{-#3}}\,^{+#4}_{-#5}}}}
\newcommand{\kepler}{{\it Kepler}}
\newcommand{\spitzer}{{\it Spitzer}}
\newcommand{\corot}{CoRoT}
\newcommand{\reff}[1]{{#1}}

\title[Homogeneous studies of transiting exoplanets. IV.]
      {Homogeneous studies of transiting extrasolar planets. \\ IV. Thirty systems with space-based light curves}

\author[John Southworth]
       {John Southworth\thanks{E-mail: jkt@astro.keele.ac.uk} \\
        Astrophysics Group, Keele University, Staffordshire, ST5 5BG, UK}

\begin{document} \maketitle 

\begin{abstract}
I calculate the physical properties of 32 transiting extrasolar planet and brown-dwarf systems from existing photometric observations and measured spectroscopic parameters. The systems studied include fifteen observed by the \corot\ satellite, ten by \kepler\ and five by the {\it Deep Impact} spacecraft. Inclusion of the objects studied in previous papers leads to a sample of \reff{58} transiting systems with homogeneously measured properties. The \kepler\ data include observations from Quarter 2, and my analyses of several of the systems are the first to be based on short-cadence data from this satellite.

The light curves are modelled using the {\sc jktebop} code, with attention paid to the treatment of limb darkening, contaminating light, orbital eccentricity, correlated noise, and numerical integration over long exposure times. The physical properties are derived from the light curve parameters, spectroscopic characteristics of the host star, and constraints from five sets of theoretical stellar model predictions. An alternative approach using a calibration from eclipsing binary star systems is explored and found to give comparable results \reff{whilst imposing a much smaller computational burden}.

My results are in good agreement with published properties for most of the transiting systems, but discrepancies are identified for \corot-5, \corot-8, \corot-13, Kepler-5 and Kepler-7. Many of the errorbars quoted in the literature are underestimated. Refined orbital ephemerides are given for \corot-8 and for the \kepler\ planets. Asteroseismic constraints on the density of the host stars are in good agreement with the photometric equivalents for HD\,17156 and TrES-2, but not for HAT-P-7 and HAT-P-11.

Complete error budgets are generated for each transiting system, allowing identification of the observations best-suited to improve measurements of their physical properties. Whilst most systems would benefit from further photometry and spectroscopy, HD\,17156, HD\,80606, HAT-P-7 and TrES-2 are now extremely well characterised. HAT-P-11 is an exceptional candidate for studying starspots. The orbital ephemerides of some transiting systems are becoming uncertain and they should be re-observed in the near future.

\reff{The primary results from the current work and from previous papers in the series have been placed in an online catalogue, from where they can be obtained in a range of formats for reference and further study. TEPCat is available at {\tt http://www.astro.keele.ac.uk/$\sim$jkt/tepcat/}}
\end{abstract}

\begin{keywords}
stars: planetary systems --- stars: binaries: eclipsing --- stars: binaries: spectroscopic --- stars: fundamental parameters
\end{keywords}


\section{Introduction}                                                                                                              \label{sec:intro}

Of the 550 planets known to orbit stars other than our Sun\footnote{\tt www.exoplanet.eu}, the transiting systems are the most interesting ones. Transiting extrasolar planets (TEPs) are the only planets outside our Solar system whose masses and radii, and thus surface gravities and mean densities, can be measured to reasonable precision. Observational selection effects mean that most known TEPs orbit very close to their parent star and thus have highly irradiated atmospheres whose physical properties can be scrutinised using high-quality astronomical observations.

One hindrance to the study of TEPs is that measurement of their physical properties requires not only transit light curves and radial velocity measurements of the parent stars, but also some sort of additional constraint. This is ordinarily obtained by forcing the properties of the star to match predictions from theoretical stellar models, guided by an effective temperature (\Teff) and metal abundance (\FeH) obtained from spectral analysis. The dependence on stellar theory leads to systematic errors which can be sizeable for some of the measured quantities, and also allows inhomogeneities to occur between studies which use different theoretical predictions or apply the constraint in a different way. This in turn compromises statistical studies of transiting planets.

For these reasons I am measuring the properties of the known TEPs using strictly homogeneous methods. Paper\,I \citep{Me08mn} discussed the methodology used to model the transit light curves, paying particular attention to error analysis and the treatment of limb darkening, and applied them to the 14 systems with good observational data at the time. Paper\,II \citep{Me09mn} explored the application of constraints from seven different sets of theoretical model predictions and alternatively an empirical mass--radius relation obtained from eclipsing binaries. Three of the sets of theoretical predictions were \reff{selected to aid in the determination} of the physical properties of the same 14 systems, with detailed error budgets including random and systematic contributions. In Paper\,III \citep{Me10mn} I extended the number of systems to 30 and the number of theoretical predictions used to five sets, improving both the statistical weight of the ensemble and the precision of the systematic errorbars.

In the current work I enlarge the number of TEPs with homogeneous properties to \reff{58}, concentrating on those which have been observed by the space missions \kepler\ \citep{Borucki+10sci}, \corot\ \citep{Baglin+06conf} and EPOCH \citep{Christiansen+09iaus}. In Paper\,II I outlined the concept of applying an observational mass--radius relation instead of using constraints from stellar theory, resulting in totally empirical measurements of the properties of TEP systems. This approach was not very successful (see Paper\,III) because the mass-radius relation did not allow for stellar evolution and had to be calibrated on low-mass eclipsing binary systems whose properties (primarily radius) are affected by magnetic \reff{activity} arising from comparatively fast rotation. In the current work I follow the alternative approach of \citet{Enoch+10aa} which relates the radius of the star to its density, \Teff\ and \FeH. This technique is not purely empirical, as it incorporates parameters derived using spectral synthesis techniques for both the TEP hosts and the calibrating sample, but returns results in much better agreement with the default method using theoretical predictions. Finally, I explore the opportunities for further study of the systems studied in this work.

For a small number of TEPs it is possible to either partially or totally avoid systematic errors: the study of transit timing variations (TTVs) allow the masses of some TEPs to be constrained directly \citep{HolmanMurray05sci,Lissauer+11natur}; and radial velocities of the {\em planet} HD\,209458\,b have been measured by \citet{Snellen+10natur} from \reff{infrared absorption} lines, allowing the system properties to be calculated in an identical way to double-lined eclipsing binary star systems. These methods are, however, not applicable to the predominant population of TEPs which show no detectable TTVs and are not amenable to direct velocity measurements.


\section{Analysis of the light curves}                                                                                                 \label{sec:lc}

I have modelled the light curves of each TEP using the methods espoused in Paper\,I. In short, the {\sc jktebop}\footnote{{\sc jktebop} is written in {\sc fortran77} and the source code is available at {\tt http://www.astro.keele.ac.uk/$\sim$jkt/codes/jktebop.html}} code \citep{Me++04mn,Me++04mn2} is used to model the available transit light curves. The components are approximated as biaxial spheroids whose shapes are governed by the mass ratio, $q$. The results in this work are all extremely insensitive to the values adopted for $q$.

The main parameters of a {\sc jktebop} fit are the orbital inclination, $i$, and the fractional radii of the star and planet, $r_{\rm A}$ and $r_{\rm b}$. These are defined as
\begin{equation}
r_{\rm A} = \frac{R_{\rm A}}{a} \qquad \qquad r_{\rm b} = \frac{R_{\rm b}}{a}
\end{equation}
where $R_{\rm A}$ and $R_{\rm b}$ are the volume-equivalent stellar and planetary radii and $a$ is the orbital semimajor axis. In {\sc jktebop} the fractional radii are re-parameterised as their sum and ratio:
\begin{equation}
r_{\rm A} + r_{\rm b} \qquad \qquad k = \frac{r_{\rm b}}{r_{\rm A}} = \frac{R_{\rm b}}{R_{\rm A}}
\end{equation}
as these are less strongly correlated. In general the orbital period, $\Porb$, is taken from the literature and the time of transit midpoint, $T_0$, is included as a fitted parameter.

Each light curve is fitted with a number of different approaches to limb darkening (LD). 1$\sigma$ errorbars are obtained using 1000 Monte Carlo simulations \citep{Me+04mn3,Me+05mn}. Errorbars are also calculated using a residual permutation (or ``prayer bead'') algorithm \citep{Jenkins++02apj} which accounts for correlated observational noise, and the largest of the two alternatives is adopted for each parameter.

The LD of the star has an important influence on transit light curves. For each light curve, solutions are obtained using five different LD laws, each with three different approaches to the limb darkening coefficients (LDCs): (1) both fixed (hereafter `LD-fixed'); (2) the linear one ($u_{\rm A}$) fitted and the nonlinear one ($v_{\rm A}$) fixed but perturbed by $\pm$0.05 in the error analysis simulations (`LD-fit/fix'); and (3) both coefficients fitted (`LD-fitted'). Initial or fixed values for the LDCs are bilinearly interpolated in \Teff\ and \logg\ within the tabulated theoretical predictions included in the {\sc jktld}\footnote{{\sc jktld} is written in {\sc fortran77} and is the source code is available at {\tt http://www.astro.keele.ac.uk/$\sim$jkt/codes/jktld.html}} code.

Once the best of the three alternatives (LD-fixed, LD-fit/fix, LD-fitted) is identified, the combined solution for that option is calculated by taking the mean of the solutions for the four two-coefficient LD laws and by taking the largest errorbar from these solutions plus a contribution to account for the scatter in the parameter values. In most cases the LD-fit/fix solutions turn out to be the best compromise between severing the dependence on theoretical calculations and trying to fit too many parameters to the data.

Some TEPs have a non-circular orbit which changes the duration of the transit but has a negligible effect on the light curve shape \citep{Kipping08mn}. I account for this by adding published constraints on orbital eccentricity ($e$) and periastron longitude ($\omega$), with the parameter combinations $e\cos\omega$ and $e\sin\omega$ when possible, using the approach outlined in Paper\,III and \citet{Me+09apj}.

An additional annoyance for some systems is light from a nearby star contaminating the photometry \citep[e.g.][]{Daemgen+09aa}. This `third light' makes the transit shallower but cannot be fitted for in the light curve due to strong degeneracy with $r_{\rm A}$, $r_{\rm b}$ and $i$ (see Paper\,III). When third light is known to exist it is accounted for using the method put forward by \citet{Me+10mn}.

\subsection{Numerical integration of light curves}                                                                              \label{sec:lc:numint}

\begin{figure} \includegraphics[width=\columnwidth,angle=0]{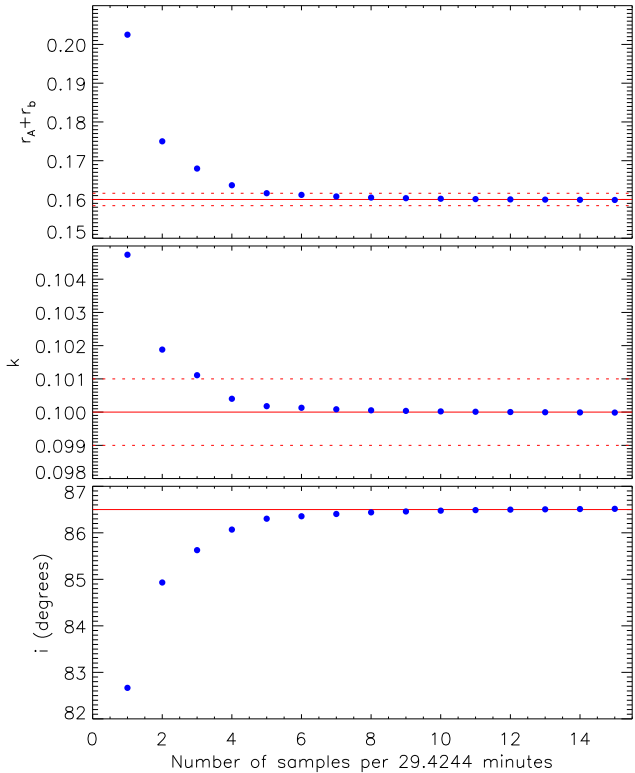}
\caption{\label{fig:lc:numint:diff1} Parameters of light curve solutions of an
undersampled light curve, closely resembling that of Kepler-6, with different
numbers of numerical integration points used in the solution. Unbroken lines
show the correct parameter values and dotted lines show the $\pm$1\% intervals
for these parameters.} \end{figure}

\begin{figure} \includegraphics[width=\columnwidth,angle=0]{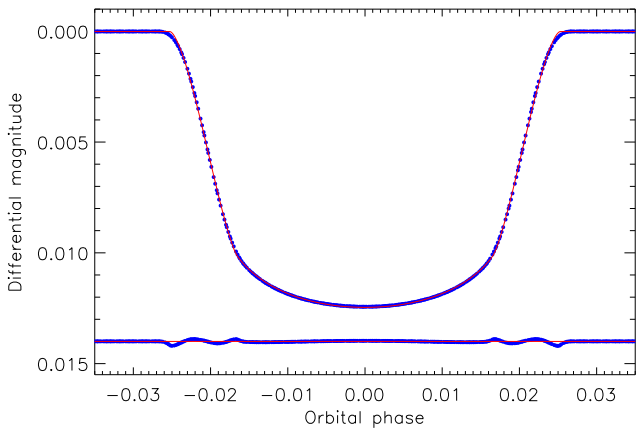}
\caption{\label{fig:lc:numint:diff2} Plot of a model light curve closely resembling
that of Kepler-6, but covering 32 instead of 12 orbital periods (blue points). This is
compared to the best fit obtained without using numerical integration (red line). The
residuals of the fit are plotted at the base of the figure offset from zero.} \end{figure}

In some light curves the sampling rate is a significant fraction of the transit duration, leading to a smearing of the transit shape. If left uncorrected this could cause errors of up to 30\% (worst-case scenario) in the physical properties of TEPs. The prime candidates for this problem are the \kepler\ planets, whose long-cadence photometric points consist of summations of 270 consecutive datapoints leading to an overall sampling rate of one datum every 29.4244\,min \citep{Jenkins+10apj}. Some of the \corot\ satellite data are also affected, as the standard cadence for this instrument is 512\,s. Most \corot\ TEPs also have short-cadence sampling with a rate of 32\,s; these data do not suffer from temporal undersampling.

For the current work I have modified {\sc jktebop} to optionally perform numerical integration to cope with the \kepler\ and \corot\ data. The approach is to calculate a given number of model datapoints ($N_{\rm int}$) evenly spread over a given time interval, and sum them to create an integrated datapoint which can be directly compared with observations. The next question is: how fine a time sampling is necessary? I created a dataset very similar to that of the TEP Kepler-6 by generating a model light curve ($r_{\rm A}+r_{\rm b} = 0.16$, $k = 0.10$, $i = 86.5^\circ$, quadratic LD), extending it over 12 orbital cycles, and summing it into 29.4\,min bins. This was then fitted with $N_{\rm int}$ varied from 1 (i.e.\ no numerical integration) to 15 (equivalent to a 1.96\,min sampling rate). The resulting values of $r_{\rm A}+r_{\rm b}$, $k$ and $i$ are plotted in Fig.\,\ref{fig:lc:numint:diff1} and show that $r_{\rm A}+r_{\rm b}$ is more affected than $k$ and $i$. Using $N_{\rm int} = 10$ means we incur an error of only 0.1\% in $r_{\rm A}+r_{\rm b}$; using $N_{\rm int} = 5$ would engender a 1\% error. TEPs with shorter orbital periods (i.e.\ quicker transits) or higher $i$ (sharper partial phases) will be more strongly affected.

Fig.\,\ref{fig:lc:numint:diff2} shows the Kepler-6-like model light curve extended to cover 32 orbital cycles and summed into 29.4\,min bins. This has been fitted by {\sc jktebop} but without performing numerical integration, in order to demonstrate the effect of neglecting the undersampling. The fitted model is unable to correctly reproduce the synthetic data during the partial phases of the transit, the discrepancy being worst at the first and last contact points where the light curve derivative is of greatest magnitude. This suggests that it would be possible to {\em fit} for the amount of numerical integration needed for a high-quality dataset, although it is very unlikely that such an option will ever be useful. The amount of numerical integration needed will also be quite sensitive to LD, particularly around the limb of the star.


\section{Calculation of physical properties}                                                                                       \label{sec:absdim}

\subsection{Via stellar models}                                                                                              \label{sec:absdim:model}

Analysis of a transit light curve gives the quantities \Porb, $T_0$, $r_{\rm A}$, $r_{\rm b}$ and $i$. Radial velocity (RV) measurements of the parent star yield its orbital velocity amplitude, $K_{\rm A}$. Measuring the physical properties of the system requires an additional constraint, which is normally taken from theoretical stellar evolutionary models. The observed stellar effective temperature, \Teff\ and metal abundance, \FeH, are useful to guide this process, which is discussed in detail in Paper\,II and Paper\,III.

I use the velocity amplitude of the {\em planet} ($K_{\rm b}$) to control the solution process. A starting value is guessed, and is combined with the measured \Porb, $r_{\rm A}$, $r_{\rm b}$, $i$ and $K_{\rm A}$ to obtain the physical properties of the system using standard formulae \citep[e.g.][]{Hilditch01book}. The resulting calculated stellar mass ($M_{\rm A}$) and measured \FeH\ are used to obtain the expected stellar radius ($R_{\rm A}$) and \Teff\ by interpolation in a set of tabulated theoretical predictions. $K_{\rm b}$ is then iteratively refined to obtain the best agreement with the calculated $R_{\rm A}$ and observed \Teff\ by minimising the figure of merit:
\begin{equation}
{\rm fom} = \left[\frac{r_{\rm A}^{\rm (obs)}-(R_{\rm A}^{\rm (calc}/a)}{\sigma{\rm (r_{\rm A}^{\rm (obs)})}}\right]^2 +
            \left[\frac{\Teff^{\rm (obs)}-\Teff^{\rm (pred)}}{\sigma(\Teff^{\rm (obs)})}\right]^2
\end{equation}
This process is performed for a range of ages from the zero-age to the terminal-age main sequence (curtailed at a maximum of 20\,Gyr) to find the overall best fit. The code which performs this step ({\sc jktabsdim}) has been profiled in order to improve its speed, which has allowed the step size in age to be decreased from 0.1\,Gyr to 0.01\,Gyr \reff{for the current work.}

The uncertainties on the input parameters to {\sc jktabsdim} are propagated using a perturbation analysis \citep{Me++05aa}, resulting in a complete error budget for every output parameter. This allows identification of which type of observations would be best to improve our understanding of each TEP (see Sect.\,\ref{sec:followup}).

Apart from the random errors, which are calculated using the propagation analysis, systematic errors arise from the use of theoretical stellar models. These can be estimated by running solutions with a range of different model predictions. As in Paper\,III I use five sets of model predictions: {\it Claret} \citep{Claret04aa,Claret05aa,Claret06aa2,Claret07aa2}, {\it Y$^2$} \citep{Demarque+04apjs}, {\it Teramo} \citep{Pietrinferni+04apj}, {\it VRSS} \citep{Vandenberg++06apjs} and {\it DSEP} \citep{Dotter+08apjs}. {\sc jktabsdim} is run with each of these five sets, and the final results are taken to be the unweighted mean of the individual values for each output quantity. The statistical error is taken to be the largest of the individual uncertainties from the perturbation analysis, and the systematic error to be the standard deviation of the values from each of the model sets. The final results therefore rest evenly on the predictions of all five model sets. Since Paper\,III I have obtained additional tabulations for the {\it Claret} models, for fractional metal abundances of $Z=0.005$ and $Z=0.015$, to allow for \FeH\ values down to $-0.60$. I have also identified and fixed a mistake in my reformatting of the {\it VRSS} models which caused the wrong $M_{\rm A}$ values to be used for a small number of tabulations.

The final results from five runs of {\sc jktabsdim} with different theoretical model sets is the following parameters with statistical and systematic errorbars: the mass, radius, surface gravity and density of the star ($M_{\rm A}$, $R_{\rm A}$, $\log g_{\rm A}$, $\rho_{\rm A}$) and of the planet ($M_{\rm b}$, $R_{\rm b}$, $g_{\rm b}$, $\rho_{\rm b}$). In additional to this I calculate a surrogate for the equilibrium temperature for the planet:
\begin{equation} \label{eq:teq}
\Teq = \Teff \left(\frac{R_{\rm A}}{2a}\right)^{1/2} = \Teff \left(\frac{r_{\rm A}}{2}\right)^{1/2}
\end{equation}
and also its \citet{Safronov72} number:
\begin{equation}
\Theta = \frac{1}{2} \left(\frac{V_{\rm esc}}{V_{\rm orb}}\right)^2
       = \left(\frac{a}{R_{\rm b}}\right) \left(\frac{M_{\rm b}}{M_{\rm A}}\right)
       = \frac{1}{r_{\rm b}} \frac{M_{\rm b}}{M_{\rm A}}
\end{equation}
Three quantities are independent of stellar theory: $g_{\rm b}$ \citep{Me++07mn}, $\rho_{\rm A}$ \citep{SeagerMallen03apj} and \Teq\ (Paper\,III).

\subsection{Via eclipsing binary relations}                                                                                     \label{sec:absdim:eb}

\begin{table*} \caption{\label{tab:absdim:eb:coeff} Coefficients of the equations
for $\lten M$ and $\lten R$ derived using eclipsing binary star systems.
Extra significant figures are provided to guard against round-off errors.}
\begin{tabular}{l r@{\,$\pm$\,}l r@{\,$\pm$\,}l r@{\,$\pm$\,}l r@{\,$\pm$\,}l}
\hline \hline
Calibration parameter    & \mc{$\lten M$}   & \mc{$\lten M$}  & \mc{$\lten R$}   & \mc{$\lten R$}  \\
Valid mass range (\Msun) & \mc{0.2 to 27.0} & \mc{0.2 to 3.0} & \mc{0.2 to 27.0} & \mc{0.2 to 3.0} \\
\hline
$c_1$ &    0.01092 & 0.00176 &    0.01384 & 0.00263 &   0.003759 & 0.000933&    0.007055 & 0.000930\\
$c_2$ &    1.0826  & 0.0178  &    1.0569  & 0.0268  &   0.36325  & 0.00711 &    0.37796  & 0.00729 \\
$c_3$ &    0.4028  & 0.0259  &    0.360   & 0.107   &   0.1317   & 0.0102  &         \mc{ }        \\
$c_4$ & $-$0.15139 & 0.00536 & $-$0.16236 & 0.00936 &$-$0.38296  & 0.00161 & $-$0.37880  & 0.00175 \\
$c_5$ & $-$0.0008  & 0.00159 & $-$0.01405 & 0.00589 &        \mc{ }        &         \mc{ }        \\
$c_6$ &    0.1803  & 0.0105  &    0.1755  & 0.0102  &   0.06020  & 0.00417 &    0.06004  & 0.00422 \\
\hline
Scatter (dex)            &   \mc{0.0286}    &    \mc{0.0268}   &   \mc{0.00953}  &   \mc{0.00907}  \\
\hline \hline \end{tabular} \end{table*}

In Paper\,II I found a way to bypass the use of stellar models entirely, by defining an empirical mass--radius relation based on well-studied and unevolved detached eclipsing binary star systems (dEBs). The chief advantages of this approach were tractability and the avoidence of a dependence on stellar theory. The primary disadvantage was that the results were much inferior to those calculated using stellar models. This arose because the known well-studied dEBs tend to have substantially larger radii than predicted by the models (see e.g.\ \citealt{Lopez07apj,Ribas+08conf}), so a more massive star was needed to reproduce the density obtained from the light curve solutions. Stellar evolution was also not allowed for, so the systems with more evolved stars were found to be rather more massive than they actually are. These problems make a simple mass--radius relation untenable.

An alternative approach would be to use the empirical relations for dEBs defined by \citet{Torres++10aarv}, which give $\log M$ and $\log R$ in terms of $\log\Teff$, \logg\ and \FeH. These account for evolution and also for metal abundance, and have a modest scatter of $\sigma = 0.027$ in $\log M$ and $\sigma = 0.014$ in $\log R$. As noted by \citet{Enoch+10aa}, a better approach for TEPs would be to replace \logg\ with $\log\rho$ as the former quantity is rather tricky to derive from spectra of solar-like stars, whereas the latter quantity is almost directly measurable from transit light curves \citep{SeagerMallen03apj}. \citet{Enoch+10aa} found that this modified approach yields lower scatters of $\sigma = 0.023$ in $\log M$ and $\sigma = 0.009$ in $\log R$.

\begin{figure} \includegraphics[width=\columnwidth,angle=0]{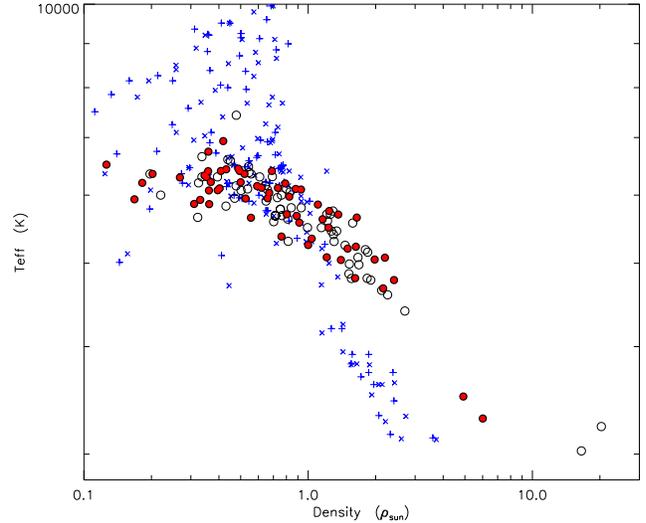}
\caption{\label{fig:absdim:eb:teffrho} \reff{Plot of density versus \Teff\ for TEP host
stars compared to well-studied dEBs. Filled (red) circles represent data from Paper\,III
and the current work and open circles show data from the literature. Plusses and crosses
(blue) denote the primary and secondary components of dEBs, respectively.} } \end{figure}

There are several possible criticisms of the implementation of this method by \citet{Enoch+10aa}. Firstly, they included only those dEBs with metallicity measurements (19 out of the 95 systems in the compilation by \citealt{Torres++10aarv}) so could suffer from small-number statistics. Also, many of the TEP host stars are of mass $\la$1.3\Msun\ where metal abundance has a much smaller effect on the stellar properties\footnote{\reff{The value of 1.3\Msun\ below which metal abundance is comparatively unimportant was obtained by plotting {\it VRSS} zero-age main sequence isochrones for $Z$ values ranging from 0.01 to 0.05}}. Secondly, they included component stars of dEBs with masses up to 15\Msun, which are of \reff{uncertain} value for studying the currently know TEP hosts. Finally, a plot of density versus \Teff\ does not allow one to conclude that the dEBs are good calibrators of TEP host stars: the former might be systematically less dense than the latter (see Fig.\,\ref{fig:absdim:eb:teffrho}). This leads to concerns that the relations have to be {\em extrapolated} to TEP hosts rather than {\em interpolated}. However, the \citet{Enoch+10aa} approach yields results which are quick to calculate and are in good agreement with stellar theory, so is worthy of further investigation.

I have therefore obtained my own calibration of $\lten M$ and $\lten R$ in terms of $\lten\Teff$, $\lten\rho$ and \MoH. The calibration sample is taken from DEBCat\footnote{Catalogue of well-studied detached eclipsing binary star systems: {\tt http://www.astro.keele.ac.uk/$\sim$jkt/debcat/}} and includes all stars but one\footnote{ASAS\,J0528+0338 was dropped from the sample as it is a pre-main-sequence system.} with masses up to 3\Msun. Objects without a metallicity measurement were assigned $\MoH = 0 \pm 100$. The final sample contains 90 dEBs (180 stars), and benefits in particular from the six new low-mass dEBs studied by \citet{Kraus+11apj}.

The adopted equations are:
\begin{eqnarray}
\lten M  &  =  &  c_1  +  c_2 \lten X  +  c_3 (\lten X)^2  \nonumber\\
         &     &  +  c_4 \lten \rho  +  c_5 (\lten \rho)^2  +  c_6 \MoH
\label{eq:logm} \end{eqnarray}
\begin{eqnarray}
\lten R  &  =  &  c_1  +  c_2 \lten X  +  c_3 (\lten X)^2  \nonumber\\
         &     &  +  c_4 \lten \rho  +  c_5 (\lten \rho)^2  +  c_6 \MoH
\label{eq:logr} \end{eqnarray}
where $X = \lten(\Teff/\Teffsun)$, $\Teffsun = 5781$\,K, and mass, radius and density are given in solar units. The coefficients of the fit, $c_i$, were found using a downhill simplex algorithm and are given in Table\,\ref{tab:absdim:eb:coeff}. The equation adopted for the current paper is for $\lten R$ with a range of validity of 0.2--3.0\Msun. The scatter around the calibration is only 0.009\,dex, in good agreement with \citet{Enoch+10aa}. The coefficients are much more precise because of the larger calibration sample, but are not directly comparable because of a different choice of normalisation parameter for \Teff. The calibration has been implemented into the {\sc jktabsdim} code, and its scatter is propagated through into the final results using the perturbation method.

\subsection{Constants and units}                                                                                            \label{sec:absdim:units}

In previous papers in this series\footnote{Paper\,II and Paper\,III, and also \citet{Me+09mn,Me+09mn2,Me+09apj,Me+10mn,Me+11aa}.} the densities of TEPs were given relative to that of Jupiter. The density of Jupiter was calculated incorrectly, using the equatorial radius rather than the volume-equivalent radius. The former is larger by 2.26\% due to the oblateness of the planet, leading to a scale in which the density of the planet Jupiter was counterintuitively 1.0694\pjup. Starting with the present work I account for this effect, leading to planetary densities which are lower by 6.94\%. No other quantities are affected by this oversight. In Table\,\ref{tab:absdim:densities} I give the corrected densities of those planets which were studied in previous papers (see Paper\,III). It is likely that this error has occurred several times before in the literature, and in these cases $\rho_{\rm b}$ should be divided by 1.0694. In the Appendix I make no attempt to perform this correction to published values as it is not clear which of them are affected.

\begin{table} \centering \caption{ \label{tab:absdim:densities}
Corrected planetary densities of the TEPs studied in Paper\,III.}
\begin{tabular}{ll}
\hline \hline
System      & $\rho_{\rm b}$ (\pjup) \\
\hline
GJ\,436     & \er{1.41 }{0.18 }{0.02 } \\
HAT-P-1     & \er{0.271}{0.025}{0.002} \\
HAT-P-2     & \er{4.8  }{1.4  }{0.0  } \\
HD\,149026  & \err{1.57 }{0.67 }{0.53 }{0.01 }{0.01 } \\
HD\,189733  & \err{0.706}{0.062}{0.062}{0.008}{0.008} \\
HD\,209458  & \er{0.254}{0.004}{0.002} \\
OGLE-TR-10  & \er{0.125}{0.036}{0.000} \\
OGLE-TR-56  & \er{0.70 }{0.32 }{0.32 } \\
OGLE-TR-111 & \er{0.40 }{0.10 }{0.00 } \\
OGLE-TR-113 & \er{0.85 }{0.15 }{0.01 } \\
OGLE-TR-132 & \er{0.55 }{0.22 }{0.00 } \\
OGLE-TR-182 & \er{0.311}{0.104}{0.002} \\
OGLE-TR-211 & \err{0.348}{0.109}{0.124}{0.001}{0.001} \\
OGLE-TR-L9  & \er{0.96 }{0.35 }{0.01 } \\
TrES-1      & \er{0.536}{0.052}{0.009} \\
TrES-2      & \er{0.584}{0.048}{0.006} \\
TrES-3      & \err{0.804}{0.047}{0.053}{0.006}{0.004} \\
TrES-4      & \er{0.138}{0.037}{0.001} \\
WASP-1      & \err{0.246}{0.055}{0.033}{0.001}{0.001} \\
WASP-2      & \er{0.70 }{0.13 }{0.01 } \\
WASP-3      & \er{0.63 }{0.10 }{0.00 } \\
WASP-4      & \err{0.463}{0.014}{0.017}{0.005}{0.006} \\
WASP-5      & \er{0.93 }{0.13 }{0.01 } \\
WASP-10     & \er{2.43 }{0.36 }{0.05 } \\
WASP-18     & \er{6.21 }{0.84 }{0.05 } \\
XO-1        & \er{0.492}{0.059}{0.005} \\
XO-2        & \err{0.532}{0.111}{0.067}{0.012}{0.010} \\
XO-3        & \er{5.69 }{0.63 }{0.06 } \\
XO-4        & \err{0.66 }{0.12 }{0.37 }{0.01 }{0.00 } \\
XO-5        & \er{0.79 }{0.17 }{0.00 } \\
\hline \hline \end{tabular} \end{table}

This seems an appropriate point to specify the values of all constants and adopted quantities used in the present series of papers. The measured physical properties of the star are given in solar units for mass, radius and mean density, and in c.g.s.\ for \logg. For the planet, mass, radius and density are given in Jupiter units, and surface gravity in \mss. Table\,\ref{tab:absdim:units} gives all constants used, plus the ensuing values of some quantities of interest. \reff{Since this paper was submitted, \citet{HarmanecPrsa11xxx} have proposed a standardisation of the physical constants used in astronomy. They propose that a nominal solar mass, radius and luminosity are defined as exact quantities for use by all researchers. The values they propose are in full agreement with those used in the current work.}

\begin{table} \caption{\label{tab:absdim:units} Physical constants and adopted
quantities used in the current work. The lower part of the Table gives the
value of several quantities resulting from the adopted physical constants.
\newline {\bf References:}
(1) US National Institute of Standards and Technology (2006 constants);
(2) \reff{IERS Conventions 2010 \citep{PetitLuzum10};}
(3) 2010 Astronomical Almanac;
(4) \citet{BrownChristensen98apj};
(5) \citet{Bahcall++95rvmp};
(6) JPL DE405 Ephemerides;
(7) NASA NSSDC Jupiter fact sheet.}
\setlength{\tabcolsep}{4pt}
\begin{tabular}{l l l l}
\hline \hline
Parameter                  & Value                       & Unit                     & Ref. \\
\hline
Stefan-Boltzmann constant  & $5.67040 \times 10^{-8}$    & W m$^{-2}$ K$^{-4}$      & 1    \\
$G$                        & $6.67428 \times 10^{-11}$   & m$^3$ kg$^{-1}$ s$^{-1}$ & 1    \\
$GM_\odot$                 & $1.32712421 \times 10^{20}$ & m$^3$ s$^{-2}$           & 2    \\
AU                         & $1.49597871 \times 10^{11}$ & m                        & 3    \\
$R_\odot$                  & $6.95508 \times 10^{9}$     & m                        & 4    \\
$L_\odot$                  & $3.844 \times 10^{26}$      & W                        & 5    \\
$R_{\rm Jup}$ (equatorial) & $7.1492 \times 10^{8}$      & m                        & 3    \\
$M_\odot$/$M_{\rm Jup}$    & $1.0473486 \times 10^{3}$   &                          & 6    \\
Volume of Jupiter          & $1.43128 \times 10^{24}$    & m$^3$                    & 7    \\
\hline
$M_\odot$                  & $1.98842 \times 10^{30}$    & kg                       &      \\
$\log g_\odot$             & $4.43831$                   & (c.g.s.)                 &      \\
$\rho_\odot$               & $1410.95$                   & kg m$^{-3}$              &      \\
$M_{\rm Jup}$              & $1.89852 \times 10^{27}$    & kg                       &      \\
$g_{\rm Jup}$ (equatorial) & $24.7916$                   & m s$^{-2}$               &      \\
$\rho_{\rm Jup}$           & $1326.45$                   & kg m$^{-3}$              &      \\
\hline \hline \end{tabular}\end{table}


\section{Light curve morphology}                                                                                              \label{sec:terminology}

\begin{figure} \includegraphics[width=\columnwidth,angle=0]{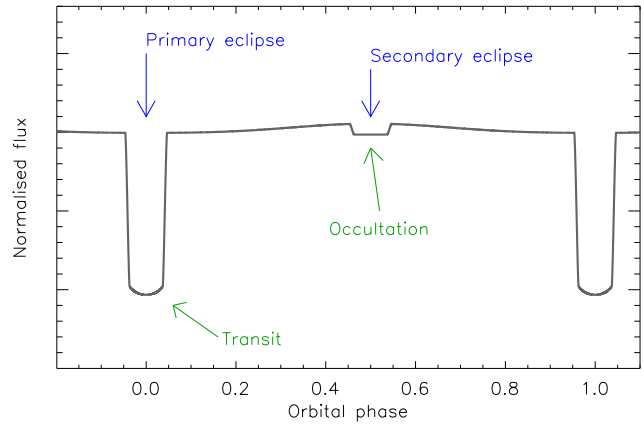}
\caption{\label{fig:schematic} Representation of a transit light curve with
the main features indicated and labelled.} \end{figure}

The literature contains a proliferation of terms for the different types of eclipses in TEP systems. The correct terminology has been established for many years for eclipsing binary systems \citep[e.g.][]{Hilditch01book} -- and a TEP is simply a special case of an eclipsing binary. A `transit' is when a smaller object passes in front of a larger object, for example a planet in front of a star. An `occultation' is when a smaller object passes behind a larger object. Transits and occultations are two types of eclipses, and the third type is a `partial eclipse'. Fig.\,\ref{fig:schematic} is a schematic representation of the situation.

The definition of `primary eclipse' is when the hotter object is behind the cooler object, and `secondary eclipse' denotes to the reverse situation. This ordinarily means that the primary eclipse is deeper than the secondary, although exceptions to the rule are possible in the case of eccentric orbits (where different surface areas are covered at the two types of eclipse). \reff{For almost all TEPs\footnote{\reff{A few TEPs undergo partial eclipses and therefore are technically not ``transiting'' planets (e.g.\ WASP-34; \citealt{Smalley+11aa}).}}, primary eclipses are transits and secondary eclipses are occultations.}

Eclipsing systems can have only one transit per orbit\footnote{It is technically possible to have two transits per orbit due to gravitational lensing in binary systems containing degenerate objects \citep{Marsh01mn} but no such system is known.}, so references to a `primary transit' or `anti-transit' are misleading. The phrase `secondary transit' is incorrect and should not be used.

\begin{table*} \caption{\label{tab:teps:lcpar} Parameters from the light curve analyses presented
here and in previous works, and used here to determine the physical properties of the TEPs. The
orbital periods are either from this work or from the literature, and the bracketed numbers represent
the uncertainty in the preceding digits. Systems for which orbital eccentricity or third light was
accounted for are indicated with a $\star$ in the column marked ``$e$?'' or ``$L_3$?'', respectively.}
\begin{tabular}{l l c c r@{\,$\pm$\,}l r@{\,$\pm$\,}l r@{\,$\pm$\,}l l}
\hline \hline
System & Orbital period & $e$? & $L_3$? & \mc{Orbital inclination,} & \mc{Fractional stellar}  & \mc{Fractional planetary} & Reference \\
       &     (days)     &      &        &    \mc{$i$ (degrees)}     & \mc{radius, $r_{\rm A}$} & \mc{radius, $r_{\rm b}$}  &           \\
\hline
\corot-1      & 1.5089686 (6)   &         & $\star$ & 84.42 & 0.31            & 0.2073 & 0.0020              & 0.02924 & 0.00041               & This work        \\
\corot-2      & 1.7429935 (10)  & $\star$ & $\star$ & 87.45 & 0.34            & 0.1478 & 0.0023              & 0.02462 & 0.00035               & This work        \\
\corot-3      & 4.2567994 (35)  &         & $\star$ & 85.80 & 0.77            & 0.1266 & 0.0067              & 0.00857 & 0.00056               & This work        \\
\corot-4      & 9.20205 (37)    &         & $\star$ & \erc{89.96}{0.04}{0.79} & \erc{0.0585}{0.0046}{0.0015} & \erc{0.00608}{0.00060}{0.00020} & This work        \\
\corot-5      & 4.0378962 (19)  & $\star$ & $\star$ & 86.24 & 0.53            & 0.0977 & 0.0067              & 0.01129 & 0.00091               & This work        \\
\corot-6      & 8.886593 (4)    &         & $\star$ & 88.88 & 0.25            & 0.0567 & 0.0013              & 0.00662 & 0.00019               & This work        \\
\corot-7      & 0.853585 (24)   &         &         & 79.6 & 3.2              & 0.264 & 0.039                & 0.0047 & 0.0012                 & This work        \\
\corot-8      & 6.21229 (3)     &         & $\star$ & 87.44 & 0.56            & 0.0659 & 0.0056              & 0.00537 & 0.00060               & This work        \\
\corot-9      & 95.2738 (14)    & $\star$ &         & 89.97 & 0.13            & 0.01083 & 0.00072            & 0.001233 & 0.000089             & This work        \\
\corot-10     & 13.2406 (2)     & $\star$ & $\star$ & 88.57 & 0.18            & 0.0326 & 0.0023              & 0.00424 & 0.00038               & This work        \\
\corot-11     & 2.994330 (11)   &         & $\star$ & 83.13 & 0.19            & 0.1452 & 0.0022              & 0.01549 & 0.00023               & This work        \\
\corot-12     & 2.828042 (13)   &         & $\star$ & 85.79 & 0.43            & 0.1235 & 0.0035              & 0.01638 & 0.00074               & This work        \\
\corot-13     & 4.035190 (30)   &         & $\star$ & 85.27 & 0.47            & 0.1161 & 0.0053              & 0.01173 & 0.00063               & This work        \\
\corot-14     & 1.51214 (13)    &         & $\star$ & 79.7 & 1.4              & 0.206 & 0.019                & 0.0181 & 0.0013                 & This work        \\
\corot-15     & 3.06036 (3)     &         & $\star$ & \erc{89.9}{0.1}{5.0}    & \erc{0.138}{0.034}{0.009}    & \erc{0.0109}{0.0034}{0.0008}    & This work        \\
HAT-P-4       & 3.0565195 (25)  &         &         & \erc{89.2}{0.8}{1.5}    & \erc{0.1666}{0.0080}{0.0027} & \erc{0.01431}{0.00072}{0.00028} & This work        \\
HAT-P-7       & 2.2047304 (24)  &         &         & 83.40 & 0.12            & 0.23880 & 0.00095            & 0.018401 & 0.000090             & This work        \\
HAT-P-11      & 4.88781501 (68) & $\star$ &         & 89.36 & 0.36            & 0.06148 & 0.00082            & 0.003604 & 0.000071             & This work        \\
HD\,17156     & 21.216398 (16)  & $\star$ &         & 86.94 & 0.34            & 0.04222 & 0.00079            & 0.003107 & 0.000081             & This work        \\
HD\,80606     & 111.4367 (4)    & $\star$ &         & 89.232 & 0.029          & 0.01056 & 0.00024            & 0.001045 & 0.000019             & This work        \\
Kepler-4      & 3.213658 (38)   &         & $\star$ & \erc{89.2}{0.8}{2.6}    & \erc{0.153}{0.026}{0.010}    & \erc{0.00391}{0.00076}{0.00034} & This work        \\
Kepler-5      & 3.548469 (15)   &         & $\star$ & \erc{87.1}{1.0}{0.6}    & \erc{0.1445}{0.0042}{0.0033} & \erc{0.01163}{0.00032}{0.00027} & This work        \\
Kepler-6      & 3.2347020 (33)  &         & $\star$ & \erc{89.9}{0.1}{0.7}    & \erc{0.1321}{0.0017}{0.0006} & \erc{0.01259}{0.00022}{0.00007} & This work        \\
Kepler-7      & 4.8854948 (82)  &         & $\star$ & 85.31 & 0.43            & 0.1494 & 0.0035              & 0.01250 & 0.00041               & This work        \\
Kepler-8      & 3.5225047 (76)  &         & $\star$ & 84.23 & 0.16            & 0.1432 & 0.0018              & 0.01360 & 0.00021               & This work        \\
KOI-428       & 6.87349 (64)    &         &         & \erc{86.5}{3.5}{3.6}    & \erc{0.139}{0.029}{0.019}    & \erc{0.0079}{0.0021}{0.0013}    & This work        \\
LHS\,6343     & 12.71382 (4)    & $\star$ & $\star$ & 89.247 & 0.088          & 0.02388 & 0.00090            & 0.00489 & 0.00018               & This work        \\
TrES-2        & 2.47061323 (7)  &         & $\star$ & 83.925 & 0.030          & 0.12568 & 0.00041            & 0.015979 & 0.000027             & This work        \\
TrES-3        & 1.30618700 (72) &         &         & 81.93 & 0.13            & 0.1682 & 0.0014              & 0.02750 & 0.00035               & This work        \\
WASP-3        & 1.8468373 (14)  &         &         & 83.72 & 0.39            & 0.1994 & 0.0032              & 0.02125 & 0.00041               & This work        \\
WASP-7        & 4.9546416 (35)  &         &         & 87.03 & 0.93            & 0.1102 & 0.0061              & 0.01053 & 0.00070               & \citet{Me+11aa}  \\
XO-4          & 4.1250828 (40)  &         &         & \erc{89.9}{0.1}{3.9}    & \erc{0.1300}{0.0283}{0.0051} & \erc{0.01124}{0.00334}{0.00054} & Paper\,III       \\
\hline \hline \end{tabular} \end{table*}

\begin{table*} \caption{\label{tab:teps:spec} Measured quantities for the
parent stars which were adopted in the analysis presented in this work.}
\begin{tabular}{l r@{\,$\pm$\,}l l r@{\,$\pm$\,}l l r@{\,$\pm$\,}l l}
\hline \hline
System & \multicolumn{3}{l}{Velocity amplitude (\ms)} & \mc{\Teff\ (K)} & Reference & \mc{\FeH} & Reference \\
\hline
\corot-1    &   188    &  11        & \citet{Barge+08aa}    & 5950 & 150 & \citet{Barge+08aa}         & $-$0.30 & 0.25 & \citet{Barge+08aa}     \\
\corot-2    &   603    &  18        & \citet{Gillon+10aa}   & 5696 &  70 & \citet{Chavero+10aa}       &    0.03 & 0.06 & \citet{Chavero+10aa}   \\
\corot-3    &  2170    &  27        & \citet{Triaud+09aa}   & 6740 & 140 & \citet{Deleuil+08aa}       & $-$0.02 & 0.06 & \citet{Deleuil+08aa}   \\
\corot-4    &    63    &   6        & \citet{Aigrain+08aa}  & 6190 &  60 & \citet{Moutou+08aa}        &    0.05 & 0.07 & \citet{Moutou+08aa}    \\
\corot-5    & \erc{59.1}{6.2}{3.1}  & \citet{Rauer+10aa}    & 6100 &  95 & \citet{Rauer+10aa}         & $-$0.25 & 0.06 & \citet{Rauer+10aa}     \\
\corot-6    &   280    &  30        & \citet{Fridlund+10aa} & 6090 &  70 & \citet{Fridlund+10aa}      & $-$0.20 & 0.10 & \citet{Fridlund+10aa}  \\
\corot-7    &     5.04 &   1.09     & \citet{Hatzes+10aa}   & 5250 &  60 & \citet{Bruntt+10aa}        &    0.12 & 0.06 & \citet{Bruntt+10aa}    \\
\corot-8    &    26    &   4        & \citet{Borde+10aa}    & 5080 &  80 & \citet{Borde+10aa}         &    0.31 & 0.05 & \citet{Borde+10aa}     \\
\corot-9    &    38    &   3        & \citet{Deeg+10natur}  & 5625 &  80 & \citet{Deeg+10natur}       & $-$0.01 & 0.06 & \citet{Deeg+10natur}   \\
\corot-10   &   301    &  10        & \citet{Bonomo+10aa}   & 5075 &  75 & \citet{Bonomo+10aa}        &    0.26 & 0.07 & \citet{Bonomo+10aa}    \\
\corot-11   &   280    &  40        & \citet{Gandolfi+10aa} & 6440 & 120 & \citet{Gandolfi+10aa}      & $-$0.03 & 0.08 & \citet{Gandolfi+10aa}  \\
\corot-12   & \erc{125.5}{8.0}{7.5} & \citet{Gillon+10aa2}  & 5675 &  80 & \citet{Gillon+10aa2}       &    0.16 & 0.10 & \citet{Gillon+10aa2}   \\
\corot-13   &   157.8  &   7.7      & \citet{Cabrera+10aa}  & 5945 &  90 & \citet{Cabrera+10aa}       &    0.01 & 0.07 & \citet{Cabrera+10aa}   \\
\corot-14   &  1230    &  34        & \citet{Tingley+11aa}  & 6035 & 100 & \citet{Tingley+11aa}       &    0.05 & 0.15 & \citet{Tingley+11aa}   \\
\corot-15   &  7360    & 110        & \citet{Bouchy+11aa}   & 6350 & 200 & \citet{Bouchy+11aa}        &    0.1  & 0.2  & \citet{Bouchy+11aa}    \\
HAT-P-4     &    81.1  &   1.9      & \citet{Kovacs+07apj}  & 5860 &  80 & \citet{Kovacs+07apj}       &    0.24 & 0.08 & \citet{Kovacs+07apj}   \\
HAT-P-7     &   211.8  &   2.6      & \citet{Winn+09apj3}   & 6350 &  80 & \citet{Pal+08apj}          &    0.26 & 0.08 & \citet{Pal+08apj}      \\
HAT-P-11    &    11.8  &   0.9      & \citet{Hirano+11pasj} & 4780 &  50 & \citet{Bakos+10apj}        &    0.31 & 0.05 & \citet{Bakos+10apj}    \\
HD\,17156   &   272.7  &   2.1      & \citet{Winn+09apj4}   & 6079 &  56 & \citet{Fischer+07apj}      &    0.24 & 0.05 & \citet{Fischer+07apj}  \\
HD\,80606   &   476.1  &   2.2      & \citet{Winn+09apj2}   & 5574 &  72 & \citet{Santos++04aa}       &    0.34 & 0.05 & \citet{Gonzalez++10mn} \\
Kepler-4    & \erc{9.3}{1.1}{1.3}   & \citet{Borucki+10apj} & 5857 & 120 & \citet{Borucki+10apj}      &    0.17 & 0.06 & \citet{Borucki+10apj}  \\
Kepler-5    &   227.5  &   2.8      & \citet{Koch+10apj2}   & 6297 &  60 & \citet{Koch+10apj2}        &    0.04 & 0.06 & \citet{Koch+10apj2}    \\
Kepler-6    &    80.9  &   2.6      & \citet{Dunham+10apj}  & 5647 &  50 & \citet{Dunham+10apj}       &    0.34 & 0.05 & \citet{Dunham+10apj}   \\
Kepler-7    &    42.9  &   3.5      & \citet{Latham+10apj}  & 5933 &  50 & \citet{Latham+10apj}       &    0.11 & 0.05 & \citet{Latham+10apj}   \\
Kepler-8    &    68.4  &  12.0      & \citet{Jenkins+10apj3}& 6213 & 150 & \citet{Jenkins+10apj3}     & $-$0.055& 0.05 & \citet{Jenkins+10apj3} \\
KOI-428     &   179    &  27        & \citet{Santerne+11aa} & 6510 & 100 & \citet{Santerne+11aa}  &\erc{0.10}{0.15}{0.10}&\citet{Santerne+11aa} \\
LHS\,6343   &  9600    & 300        & \citet{Johnson+11apj} & 3300 & 200 & This work                  &    0.04 & 0.08 & \citet{Johnson+11apj}  \\
TrES-2      &  181.3   &   2.6      & \citet{Odonovan+06apj}& 5850 &  50 & \citet{Sozzetti+07apj}     & $-$0.15 & 0.10 & \citet{Sozzetti+07apj} \\
TrES-3      &  369     &  11        & \citet{Sozzetti+09apj}& 5650 &  75 & \citet{Sozzetti+09apj}     & $-$0.19 & 0.08 & \citet{Sozzetti+09apj} \\
WASP-3      &  286.5   &   7.8      & This work             & 6400 & 100 & \citet{Pollacco+08mn}      &    0.00 & 0.20 & \citet{Pollacco+08mn}  \\
WASP-7      &   97     &  13        & \citet{Hellier+09apj} & 6400 & 100 & \citet{Hellier+09apj}      &    0.00 & 0.10 & \citet{Hellier+09apj}  \\
XO-4        &  165.8   &   6.2      & \citet{Narita+10pasj} & 6397 &  70 & \citet{Mccullough+08xxx}   & $-$0.04 & 0.05 &\citet{Mccullough+08xxx}\\
\hline \hline \end{tabular} \end{table*}


\section{Data acquisition}                                                                                                           \label{sec:data}

The \corot\ data used here are the N2 public version obtained from the public archive\footnote{\tt http://idoc-corot.ias.u-psud.fr/} on 2010/12/03 (except for \corot-14 which was obtained on 2011/01/11) and interpreted using the N2 data description document\footnote{\tt http://idoc-corotn2-public.ias.u-psud.fr/jsp/doc/\\DescriptionN2v1.3.pdf}. The \corot\ data are of two types: short-cadence and long-cadence. The total integration times for the two cadences are 32\,s and 512\,s, respectively. Many of the \corot\ TEPs have both types of data, and these are treated separately in each case.

The \kepler\ data used are the public data obtained from the Multimission Archive at STScI (MAST\footnote{\tt http://archive.stsci.edu/kepler/data\_search/search.php}). The \kepler\ quarter 0 (Q0) and quarter 1 (Q1) data were used in the original version of this work, but the analysis was revised to include quarter 2 (Q2) data when these became available. These data were downloaded from MAST on 2011/04/11. As with \corot, the \kepler\ data come in short-cadence and long-cadence flavours. The total integration times are 58.84876\,s \citep{Gilliland+10apj} and 29.4244\,min \citep{Jenkins+10apj}, respectively.

In order to account for slow variations in the mean flux level of the systems in the \corot\ and \kepler\ data (both astrophysical and instrumental), the data surrounding each transit were \reff{extracted from} the full time-series and normalised to unit flux using a straight-line fit to the out-of-transit datapoints. In general the number of points used was roughly double the number in the transit, split equally before and after the transit. A straight line was used because it was adequate for the job and it cannot distort transit profiles. In a few cases entire transits were rejected because of a jump in observed flux during the transit or because the observations did not cover all of the transit event.

In cases where the number of datapoint was too large for Monte Carlo simulations to be completed in a reasonable length of time, the datasets were reduced down to a manageable size. This was done by converting the timestamps to orbital phase, sorting them, and then binning $N$ consecutive datapoints where $N$ ranged from 20 to 100. The value of $N$ was carefully chosen to ensure the sampling rate was still sufficiently high that numerical integration was not needed. The averaging process tends to convert correlated noise into white noise, but does not cause a problem for the error estimates of the resulting parameters. The phased and binned datapoints will be referred to as `normal points' below, a term which was once used extensively in studies of eclipsing binaries. \reff{At the request of the referee I have analysed the \corot\ 32s-sampled data for \corot-1 both with and without phase-binning, in order to check that the binning process does not affect the solutions. The resulting light curve parameters differ by negligible amounts, supporting the use of phase-binning below.}


\section{Results for individual systems}                                                                                             \label{sec:teps}

In this section I present the photometric ({\sc jktebop}) and absolute-dimensions ({\sc jktabsdim}) analyses of \reff{32} TEPs based on published high-quality data obtained from space missions, complemented with ground-based data where possible. The results are obtained using the same methods as those in Paper\,III, leading to homogeneous measurements for a sample of \reff{58} TEPs. The final {\sc jktebop} results of \reff{the 32} TEPs are collected in Table\,\ref{tab:teps:lcpar}, which also includes the orbital periods and indicates for which systems a non-circular orbit was adopted. The adopted spectroscopic parameters (\Teff, \FeH\ and $K_{\rm A}$) are given in Table\,\ref{tab:teps:spec}, with a lower limit on the errorbars of $\pm$50\,K for \Teff\ and $\pm$0.05\,dex for \FeH\ (see Paper\,II). Extensive tables of results, plus a comparison with literature values, can be found in the online-only Appendix.


\subsection{\corot-1}                                                                                                         \label{sec:teps:corot1}

\begin{figure} \includegraphics[width=\columnwidth,angle=0]{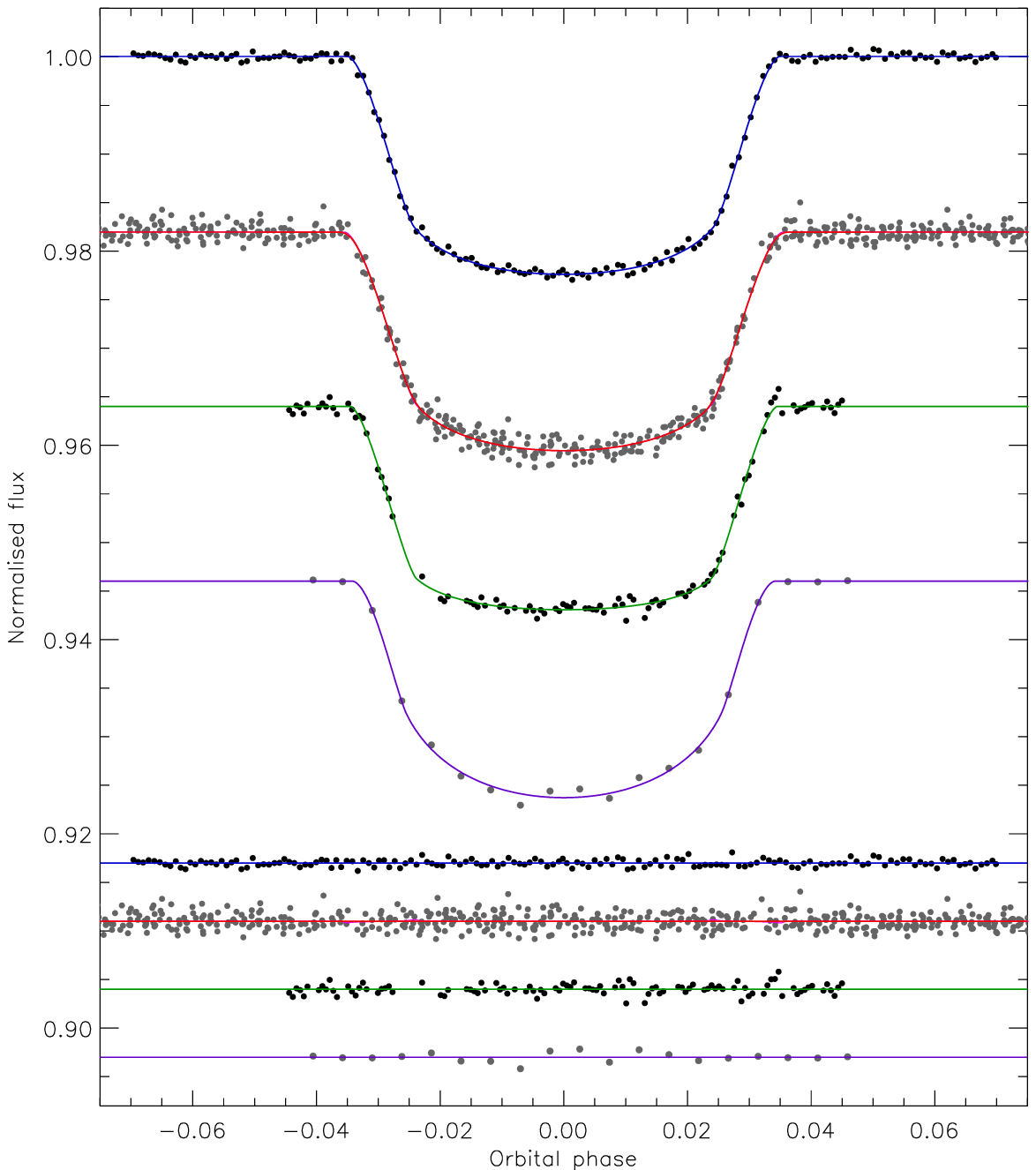}
\caption{\label{fig:corot1:lc} Phased light curves of \corot-1 compared to the
best fit found using {\sc jktebop} and the quadratic LD law. The residuals are
plotted at the base of the figure, offset from unity. The purple line through
some data show the best-fitting model without numerical integration -- in these
cases the difference between this model with and without numerical integration
is shown by another purple line through the residuals. From top to bottom the
light curves are the binned \corot\ 32\,s data, the \corot\ 512\,s data, the
FORS2 $R$-band data and the FORS2 $B$-band data.} \end{figure}

\newcounter{appref}
\setcounter{appref}{1}

\corot-1 was the first TEP discovered by the \corot\ satellite \citep{Barge+08aa}, and was originally called \corot-Exo-1. The \corot\ data have been subjected to TTV analyses by \citet{Bean09aa} and \citet{Csizmadia+10aa} with null results. \citet{Gillon+09aa2} observed a transit in the $R$-band using VLT/FORS2 and an occultation at 2.09\,$\mu$m using VLT/HAWKI. The transit data are of very high precision (0.52\,mmag scatter) but do not fully sample the transit. \citet{Pont+10mn} presented the same $R$-band data supplemented with a smaller number of $B$-band observations of the same transit taken with the same instrument. \citet{Pont+10mn} also obtained RV observations during a transit and detected a Rossiter-McLaughlin (RM) effect which show that the orbital axis of \corot-1\,b is not aligned with the stellar spin axis. Occultation studies by \citet{Alonso+09aa} and \citet{Rogers+09apj} and \citet{Deming+11apj} have found no evidence for orbital eccentricity, so I assumed that the orbit is circular.

The \corot\ data show 20 transits observed at long cadence (512\,s) and 17 observed at short cadence (32\,s). Each transit was normalised (Sect.\,\ref{sec:data}) and the two cadences were treated separately in the {\sc jktebop} analysis. After a preliminary fit the 32\,s data were phase-binned with each phased datapoint representing 50 original ones. Seven out of 546 of the 512\,s datapoints were rejected as 4$\sigma$ outliers. The 512\,s data were modelled using numerical integration with $N_{\rm int} = 3$. The $R$-band and $B$-band data were also studied (Fig.\,\ref{fig:corot1:lc}).

The results for the \corot\ 32\,s data are given in Table\,A\arabic{appref}\addtocounter{appref}{1} and are of sufficiently high quality for the LD-fitted solutions to be adopted. LD-fit/fix solutions were adopted for the \corot\ 512\,s (Table\,A\arabic{appref}\addtocounter{appref}{1}) and the VLT $R$-band (Table\,A\arabic{appref}\addtocounter{appref}{1}) data, and in both cases correlated noise was found to be unimportant. Correlated noise is moderately important for the $B$-band data (the residual-permutation errorbars are larger than the Monte-Carlo errorbars) but the data were good enough to allow the LD-fit/fix solutions to be adopted (Table\,A\arabic{appref}\addtocounter{appref}{1}).

The photometric results are given in Table\,A\arabic{appref}\addtocounter{appref}{1} and show good agreement except for $k$ (which has a reduced $\chi^2$ of $\chir=3.2$). This phenomenon has been noted several times before (Paper\,I, Paper\,II, \citealt{Me+09mn2}) and is attributable to correlated residuals (both instrumental and astrophysical) in the data. The $B$-band data cannot match their counterparts so I combine the other three sets of results together to get the final photometric parameters. These are in good agreement with those of the discovery paper \citep{Barge+08aa} and of \citet{Bean09aa}, but not with \citet{Gillon+09aa2}.

The physical properties of \corot-1 have been calculated using {\sc jktabsdim}. The individual solutions are given in Table\,A\arabic{appref}\addtocounter{appref}{1} and compared with literature values, where a good agreement is found (except for $\rho_{\rm A}$ from \citealt{Gillon+09aa2} due to their somewhat different photometric results). My measurements of $g_{\rm b}$, which is quite similar to the Earth's surface gravity, and \safronov\ appear to be the first ones published. Our understanding of the system would benefit from further spectroscopy to give improved values of \Teff, $K_{\rm A}$ and \FeH\ in particular.


\subsection{\corot-2}                                                                                                         \label{sec:teps:corot2}

\begin{figure} \includegraphics[width=\columnwidth,angle=0]{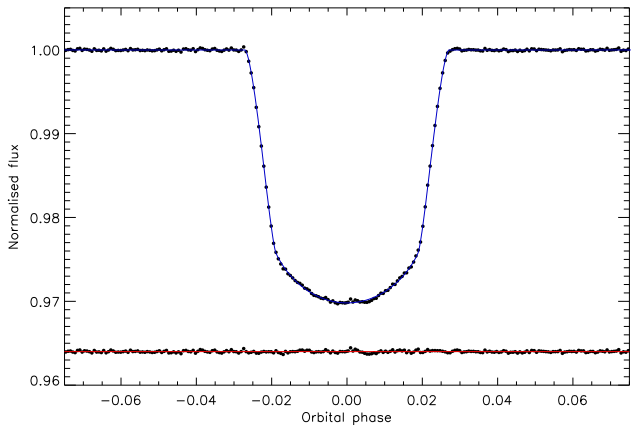}
\caption{\label{fig:corot2:lc} The 32\,s light curve of \corot-2 compared to the
{\sc jktebop} best fit. Other comments are the same as Fig.\,\ref{fig:corot1:lc}.}
\end{figure}

The discovery of \corot-2 was announced by \citet{Alonso+08aa2} based on a 142\,d light curve from the \corot\ satellite which exhibited transit events every 1.74\,d and significant starspot activity on longer timescales. \corot-2\,A is a young star hosting a relatively massive planet (3.6\Mjup). \citet{Bouchy+08aa} obtained RV measurements through transit and modelled the RM effect to find that the planetary orbital and projected stellar spin axes are aligned to within 7.2$^\circ$ (1.6$\sigma$). The occultation of the planet has been found in the \corot\ data by \citet{Alonso+09aa2} and \citet{Snellen++10aa}, and its time of occurrence is consistent with a circular orbit. \citet{Gillon+10aa} presented {\it Spitzer} observations of the secondary eclipse at 4.5\,$\mu$m and 8\,$\mu$m, finding a small but significant eccentricity characterised as $e\cos\omega = \er{-0.00291}{0.00063}{0.00061}$ and $e\sin\omega = \er{0.0139}{0.0079}{0.0084}$. Similar results have been obtained by \citet{Deming+11apj} from Warm-\spitzer\ observations at 3.6\,$\mu$m: $e\cos\omega = -0.0030 \pm 0.0004$.

The spot activity of \corot-2\,A deserves mention. It is an active star with a rotation period of only 4.5\,d and probable differential rotation \citep{Lanza+09aa,Frohlich+09aa,Huber+10aa,SilvaLanza11aa}. One of the standard assumptions of modelling transit light curves of TEPs is that the surface brightness of those parts of the star eclipsed by the planet is the same as that of the rest of the star, modulo effects such as LD and gravity darkening. In the case of starspots this is certainly not a reliable assumption. However, the effect will average out over a large number of transits if the starspots do not show a preference for particular \reff{latitudes}. But we know from the Sun that this is not the case: sunspots appear mostly within 30$^\circ$ of the equator, and their preferred latitudes vary throughout the 11\,yr solar activity cycle\footnote{\reff{The spot characteristics of the Sun are nicely captured in the Maunder butterfly diagram. An up-to-date version can be found at: {\tt http://solarscience.msfc.nasa.gov/images/bfly.gif}}}. \reff{A different effect is noticed in the study of active stars \citep[e.g.][]{Olah+97aa,Barnes05mn} and eclipsing binaries, where light curve solutions often favour large polar spots \citep{Hilditch01book}}. It is not usually possible to account for this problem in TEP studies because of the difficulty in detecting starspots outside the area eclipsed by the planet. The effect of starspots on the analysis of the \corot-2 system has been studied by \citet{Czesla+09aa} and \citet{Huber+10aa}, who agree that the spot coverage on the chord of the planet transit is greater than the average for the stellar disc, and that the planet is therefore a few percent larger than standard analyses would suggest.

In the current work I have modelled the \corot\ 32\,s light curve, which covers 79 transits. The 512\,s data are not used as they only spread over three transits with partial coverage. 148 points of the 32\,s data were rejected by a 4$\sigma$ clip and the remaining 50\,666 points were phase-binned by a factor of 200 into 254 normal points. In the {\sc jktebop} analysis I did not account for starspots, as this is beyond the scope of the current work, so have produced a baseline solution which is more easily comparable to the results for other TEPs. This is equivalent to assuming that the starspots affect all parts of the star equally on average. A third light of $L_3 = 0.053 \pm 0.003$ \citep{Alonso+08aa2} and the small eccentricity \citep{Gillon+10aa} is accounted for in the ways described in Paper\,III. The VLT data from \citet{Gillon+10aa} were not modelled because they only cover half a transit.

For the phase-binned 32\,s data I found that correlated noise is not important and that the LD-fitted solutions are reliable enough to be adopted as the final photometric results (Table\,A\arabic{appref}\addtocounter{appref}{1}). A comparison with literature measurements (Table\,A\arabic{appref}\addtocounter{appref}{1}) shows an acceptable agreement except compared to the more sophisticated analysis of \citet{Czesla+09aa}. My best fit to the 32\,s data is plotted in Fig.\,\ref{fig:corot2:lc}.

The young age of \corot-2 (it is expected to be of Pleiades age) manifests itself in the {\sc jktabsdim} analysis by the best solutions in some cases being on the zero-age edge of the grids of theoretical predictions. A modestly lower \Teff\ would alleviate this problem whilst not having a significant effect on the resulting physical properties. The edge effects can lead to underestimated errorbars for the affected solutions (see Paper\,II) but this is not a problem because the final results then rest on the errorbars from the unaffected solutions. The edge effect do, though, cause an increase in the systematic errorbars (Table\,A\arabic{appref}\addtocounter{appref}{1}).

\corot-2\,A would benefit from an improved \Teff\ measurement: the discovery value of $5625 \pm 120$\,K \citep{Alonso+08aa2} is comparatively uncertain, the alternative measurement of $5608 \pm 37$\,K by \citet{Ammler+09aa} accompanies a \logg\ which is too high (4.71 versus 4.53), and the value of $5696 \pm 70$\,K found by \citet{Chavero+10aa} is higher so causes stronger edge effects. A more precise measurement of $K_{\rm A}$ would be useful.


\subsection{\corot-3}                                                                                                         \label{sec:teps:corot3}

\begin{figure} \includegraphics[width=\columnwidth,angle=0]{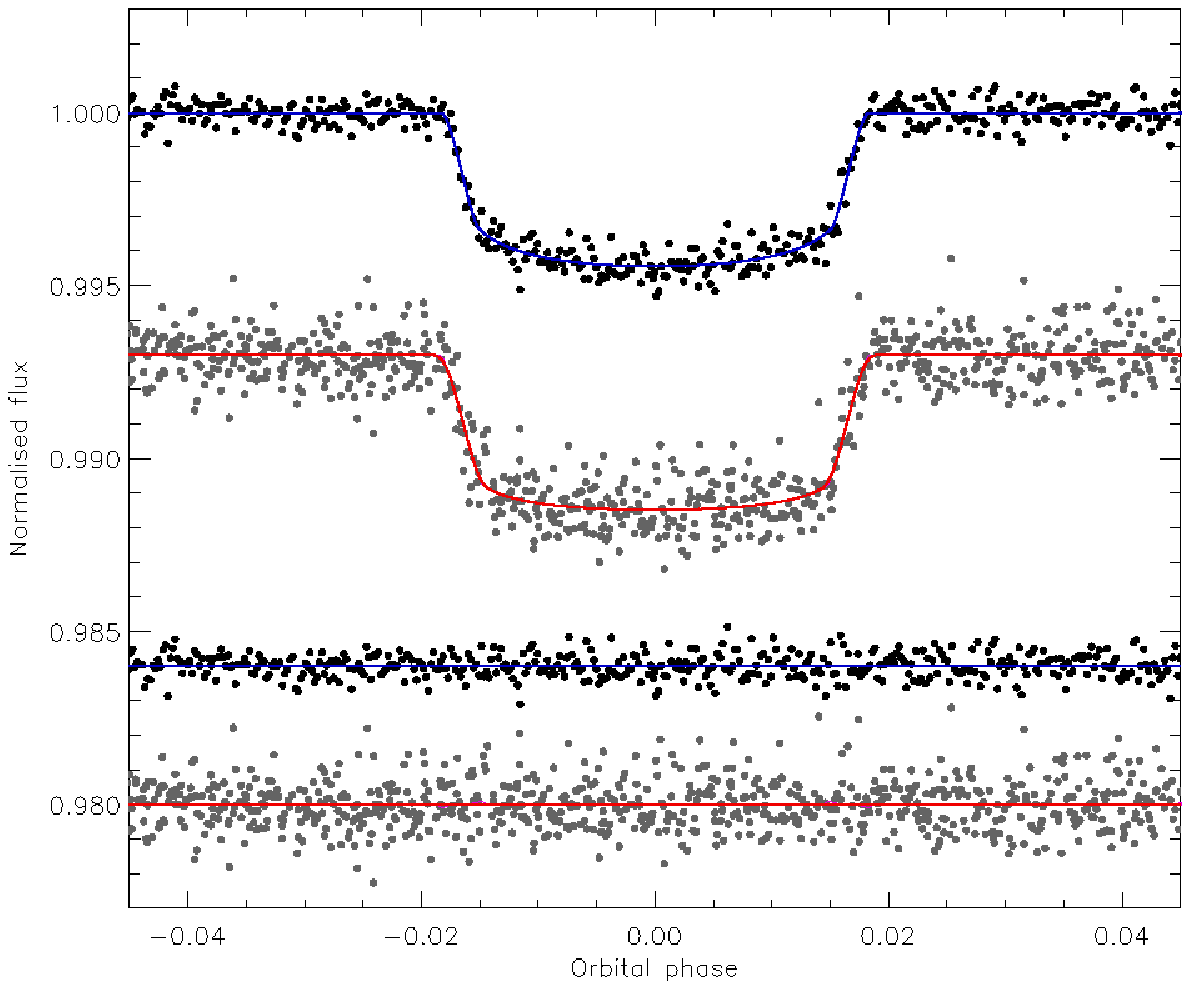}
\caption{\label{fig:corot3:lc} The 32\,s (upper) and 512\,s (lower)
\corot\ light curves of \corot-3. Other comments are the same as
Fig.\,\ref{fig:corot1:lc}.} \end{figure}

\corot-3 was announced by \citet{Deleuil+08aa} and was the first brown-dwarf-mass object with a precisely measured mass and radius. Its F3\,V host star is rather hot (6740\,K), making photometry and RV measurements difficult due to a shallow transit and high stellar $v \sin i$. An RM study has been presented by \citet{Triaud+09aa}, whose measurements are consistent with a circular and axially aligned orbit.

The 32\,s data cover 19 transits with a total of 196\,691 datapoints. Of the 19\,046 points in the regions of the transits, 20 were rejected by a 4$\sigma$ clip and the remainder were phase-binned by a factor of 40 to get 477 normal points. The 512\,s data cover 16 transits and were fitted using $N_{\rm int} = 3$. A circular orbit was assumed and a third light of $L_3 = 0.082 \pm 0.007$ was adopted \citep{Deleuil+08aa}.

The results for the 32\,s data are given in Table\,A\arabic{appref}\addtocounter{appref}{1}: correlated noise is not important (as usual for phase-binned datapoints) and the LD-fitted solutions yield the lowest scatter and reasonable LDCs. For the 512\,s data (Table\,A\arabic{appref}\addtocounter{appref}{1}) correlated noise is again unimportant and the LD-fit/fix solutions are best. The best fits are plotted in Fig.\,\ref{fig:corot3:lc}. The final photometric parameters are the weighted means of the 32\,s and 512\,s values. They are compared to published values in Table\,A\arabic{appref}\addtocounter{appref}{1}, where a very good agreement is found.

The measured physical properties of \corot-3 are given in Table\,A\arabic{appref}\addtocounter{appref}{1} and agree well with literature values. I provide the first measurement of \Teq\ and \safronov. The star could do with a better \Teff\ measurement, and an improved light curve would also be useful.


\subsection{\corot-4}                                                                                                         \label{sec:teps:corot4}

\begin{figure} \includegraphics[width=\columnwidth,angle=0]{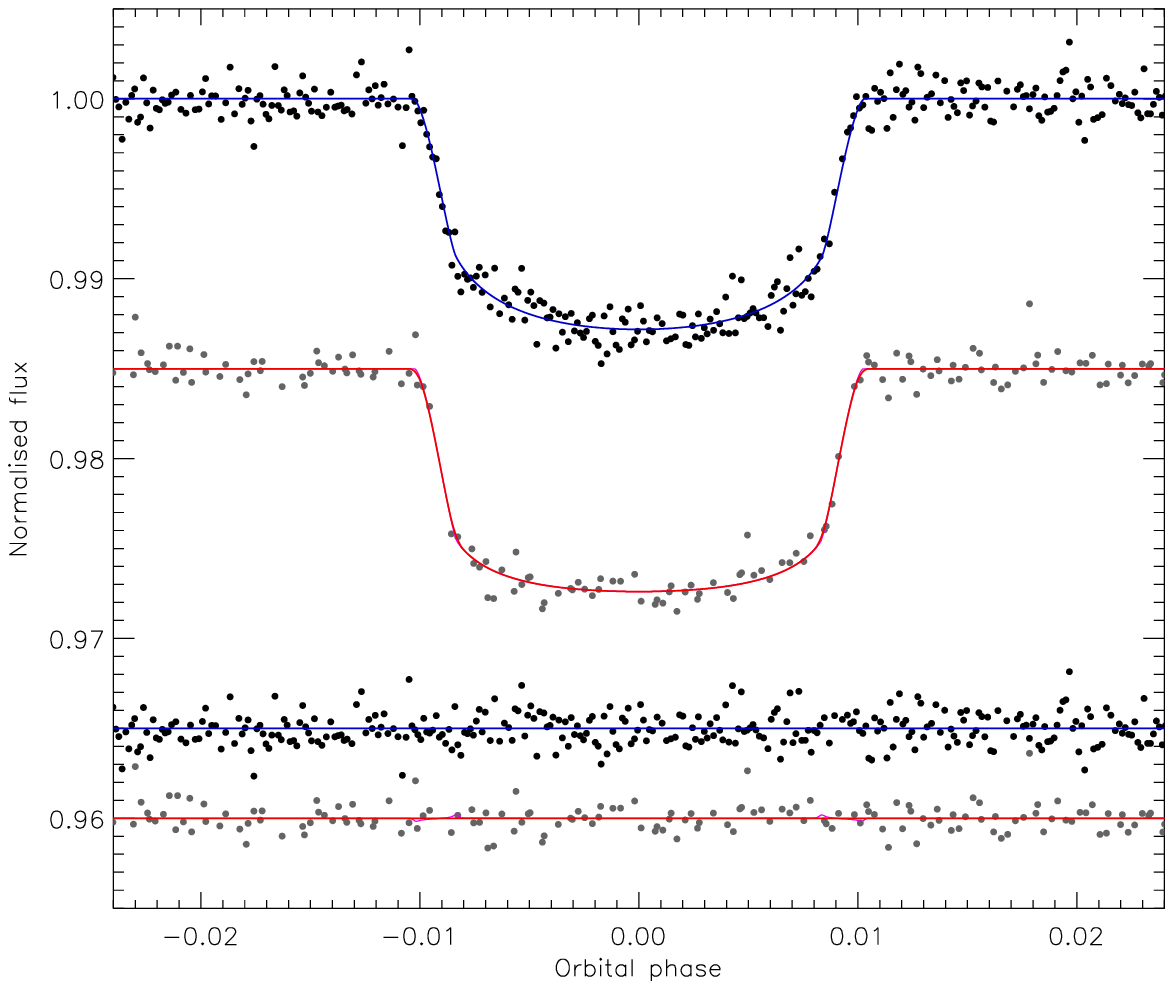}
\caption{\label{fig:corot4:lc} The 32\,s (upper) and 512\,s (lower)
\corot\ light curves of \corot-4. Other comments are the same as
Fig.\,\ref{fig:corot1:lc}.} \end{figure}

\corot-4 was discovered by \citet{Aigrain+08aa} and \citet{Moutou+08aa}. It has a relatively long orbital period of 9.2\,d, and thus one of the lower \Teq s despite having a late-F host star. There are three transits each in the 32\,s and the 512\,s data. The 32\,s data were phase-binned by a factor of 10 after 4$\sigma$ clipping to remove a small number of outliers. The 512\,s data were modelled using $N_{\rm int} = 3$.

The {\sc jktebop} solutions favour a central transit, which causes the photometric parameters to have asymmetric errorbars. Correlated noise is unimportant for the phase-binned 32\,s data but is important for the 512\,s data. In both cases the LD-fit/fix results are the best (Tables A\arabic{appref}\addtocounter{appref}{1} and A\arabic{appref}\addtocounter{appref}{1}). These were combined into final photometric parameters by multiplying their probability density functions (Table\,A\arabic{appref}\addtocounter{appref}{1}). The best fits are shown in Fig.\,\ref{fig:corot4:lc}. The agreement between my results and those of \citet{Aigrain+08aa} is excellent.

The physical properties of the \corot-4 system are given in Table\,A\arabic{appref}\addtocounter{appref}{1} and show good agreement with those of \citet{Moutou+08aa} except for larger errorbars for some properties (notably $M_{\rm A}$ and $R_{\rm A}$). I provide the first published measurements of $\rho_{\rm A}$, $g_{\rm b}$ and \safronov. The long orbital period and short observing run conspire together to allow only six transits to be observed by \corot: a better light curve would be useful, as would a more precise $K_{\rm A}$.


\subsection{\corot-5}                                                                                                         \label{sec:teps:corot5}

\begin{figure} \includegraphics[width=\columnwidth,angle=0]{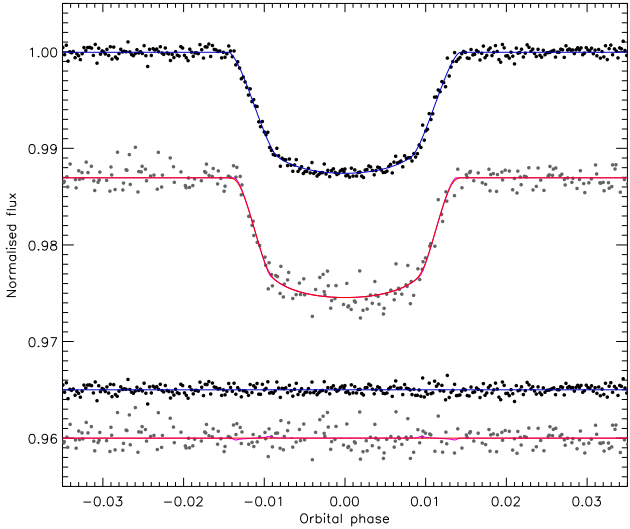}
\caption{\label{fig:corot5:lc} The 32\,s (upper) and 512\,s (lower)
\corot\ light curves of \corot-5. Other comments are the same as
Fig.\,\ref{fig:corot1:lc}.} \end{figure}

Discovered by \citet{Rauer+10aa}, \corot-5 is a fairly normal system containing a low-density TEP. The 32\,s data contain 23 transits, which were cut out of the light curve, 4$\sigma$ clipped and phase-binned as usual. The 512\,s light curve contains six transits and these data were solved using $N_{\rm int} = 3$. \citet{Rauer+10aa} found a preliminary third light of 8.6\% (no uncertainty) so I adopted $L_3 = 0.086 \pm 0.020$ for the {\sc jktebop} analysis. Note that \citet{Rauer+10aa} removed this contaminating light from the light curve prior to modelling it, so neglected the uncertainty in $L_3$. Eccentricity is significant at the 3$\sigma$ level so this was included via the constraints $e\cos\omega = \er{-0.057}{0.048}{0.020}$ and $e\sin\omega = \er{-0.071}{0.147}{0.130}$.

The {\sc jktebop} results are shown in Tables A\arabic{appref}\addtocounter{appref}{1} and A\arabic{appref}\addtocounter{appref}{1}, and the best fits are plotted in Fig.\,\ref{fig:corot5:lc}. In both cases correlated noise was unimportant and the LD-fit/fix solutions were adopted. The two sets of results do not appear to agree very well, but are consistent to within 1$\sigma$. The errorbars are large primarily due to the poorly constrained orbital parameters: $e\cos\omega$ and $e\sin\omega$ are both uncertain and the latter is significantly correlated with $r_{\rm A}$ and $r_{\rm b}$. The combined results (Table\,A\arabic{appref}\addtocounter{appref}{1}) are in poor agreement with those of \citet{Rauer+10aa}, presumably due to differences in the analysis methods.

The {\sc jktabsdim} results are given in Table\,A\arabic{appref}\addtocounter{appref}{1} and show a good agreement between the different model sets (and with the dEB constraint discussed in Sect.\,\ref{sec:absdim:eb}). As expected given the discrepant light curve results, I find physical properties which differ from those of \citet{Rauer+10aa} by up to 2$\sigma$. I find star and planet radii which are notably smaller than those of \citet{Rauer+10aa}, and errorbars which are substantially larger (by a factor of 5 for $M_{\rm A}$). The properties of \corot-5 are more uncertain than previously thought, and additional RV measurements are the best route to fixing this.


\subsection{\corot-6}                                                                                                         \label{sec:teps:corot6}

\begin{figure} \includegraphics[width=\columnwidth,angle=0]{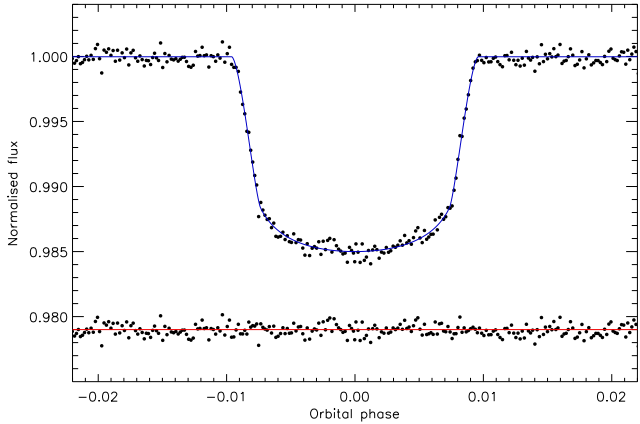}
\caption{\label{fig:corot6:lc} The 32\,s light curve of \corot-6.
See Fig.\,\ref{fig:corot1:lc} for details.} \end{figure}

Comparatively speaking, \corot-6 is a massive (2.96\Mjup) and dense (1.66\pjup) TEP with a long orbital period (8.89\,d) around a moderately active F9\,V star. The discovery light curve from \corot\ \citep{Fridlund+10aa} contains 331\,397 datapoints at short cadence, covering 14 complete transits. A study of the starspot activity has been given by \citet{Lanza+11aa}.

I adopted $e = 0$ and $L_3 = 0.028 \pm 0.007$ \citep{Fridlund+10aa}. 40 out of the 23\,802 near-transit datapoints were rejected by a 4$\sigma$ clip and the remainder were phase-binned by a factor of 50 into 476 normal points. The {\sc jktebop} results are given in Table\,A\arabic{appref}\addtocounter{appref}{1} and are in good agreement with those of \citet{Fridlund+10aa}. Correlated noise is \reff{completely negligible} and the LD-fit/fix solution is the best choice (Table\,A\arabic{appref}\addtocounter{appref}{1}). The best fit is plotted in Fig.\,\ref{fig:corot6:lc}.

The physical properties of \corot-6 are given in Table\,A\arabic{appref}\addtocounter{appref}{1}, where the agreement between theoretical models, the dEB constraint and with \citet{Fridlund+10aa} is excellent. I provide the first measurements of $\rho_{\rm A}$, $g_{\rm b}$ and \safronov. The system could do with more spectroscopy to obtain an improved $K_{\rm A}$ and \FeH.


\subsection{\corot-7}                                                                                                         \label{sec:teps:corot7}

\begin{figure} \includegraphics[width=\columnwidth,angle=0]{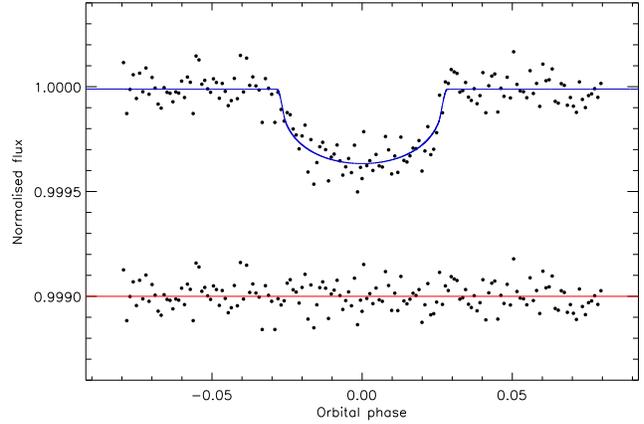}
\caption{\label{fig:corot7:lc} The 32\,s light curve of \corot-7.
See Fig.\,\ref{fig:corot1:lc} for details.} \end{figure}

\corot-7 is one of the most important known planets. At the time of discovery it was both the smallest and least massive TEP discovered so far. The \corot\ light curve shows a transit of depth only 0.034\% recurring on a period of only 0.85\,d. A detailed analysis of these data, plus two high-precision RVs, proved the planetary nature of this object \citep{Leger+09aa}. Extensive RV measurements using the HARPS spectrograph \citep{Queloz+09aa} confirmed the nature of the planet, with a mass of $4.8 \pm 1.8$\Mearth, and allowed the discovery of a second one which is more massive (8.4\Mearth) and on a longer-period orbit (3.70\,d). A study of the atmospheric parameters of the star by \citet{Bruntt+10aa} yielded a smaller $R_{\rm A}$ and therefore $R_{\rm b}$.

\corot-7\,A is a G9 dwarf star with significant chromospheric activity. This activity causes larger RV variations that those of the planets' orbits, making the properties of the system somewhat controversial. \citet{Lanza+10aa} studied the stellar activity and starspots and found a false-alarm probability of $<$10$^{-4}$ that the RV oscillations attributed to \corot-7\,b and \corot-7\,c are spurious effects of noise and activity. \citet{Hatzes+10aa} performed a comprehensive re-analysis of the RVs, finding a larger mass for \corot-7\,b ($6.9 \pm 1.4$\Mearth), confirming the RV signal of the second planet (\corot-7\,c) and tentatively detecting a third RV signal which could be caused by another planet, \corot-7\,d, with a mass of $16.7 \pm 0.4$\Mearth\ and a period of 9.02\,d. However, \citet{Pont++11mn} performed a similar analysis, including a phenomenological model to describe the properties and evolution of many starspots, and found the RV signal of \corot-7\,b to be a lot smaller and only significant at the 1.2$\sigma$ level ($M_{\rm b} = 2.3 \pm 1.5$\Mearth). The huge number of parameters in their starspot model could be expected to lead to more hazy results compared to those of other researchers. \citet{Ferraz+10xxx} showed that previous analyses have tended to remove some of the planetary RV signal when squashing the effects of the stellar activity, and thus underestimate the masses. They found $M_{\rm b} = 8.5 \pm 1.5$\Mearth\ and $M_{\rm c} \sin^3 i_{\rm c} = 13.5 \pm 1.5$\Mearth, and that \corot-7\,d is an artefact rather than an astrophysical signal. \citet{Boisse+11aa} also modelled the starspot-induced RV variations and confirmed planets b and c with false-alarm probabilities of $< 5 \times 10^{-4}$. They found masses of $M_{\rm b} = 5.7 \pm 2.5$\Mearth\ and $M_{\rm c} \sin^3 i_{\rm c} = 13.2 \pm 4.1$\Mearth. Just before the present paper was submitted, \citet{Hatzes+11apj} produced a re-analysis which confirmed their previous results for \corot-7\,b. The difference in $K_{\rm A}$ ($5.15 \pm 0.95$\ms\ versus $5.06 \pm 1.06$\ms) is so small that there was no need  to recalculate my results.

The \corot\ light curve contains 308\,947 datapoints, all at 32\,s cadence. The 47\,702 points near transits were phase-binned by a factor of 300 to give 160 normal points. The transit is so shallow (Fig.\,\ref{fig:corot7:lc}) that there was no point in calculating LD-fitted models. The results for the LD-fit/fix solutions (Table\,A\arabic{appref}\addtocounter{appref}{1}) show a good mutual agreement whereas the LD-fixed solutions do not. I therefore adopted the LD-fit/fix solutions. Compared to \citet{Leger+09aa}, my results are in good agreement but have much larger errorbars (Table\,A\arabic{appref}\addtocounter{appref}{1}). This wraps over into the physical properties (Table\,A\arabic{appref}\addtocounter{appref}{1}). My analysis approach may not be optimal for this system, as I do not apply any external constraints to the light curve model (e.g.\ and spectroscopically derived \logg\ or $\rho_{\rm A}$), but changing this would cause an inhomogeneity with the other TEPs treated in this work. \corot-7 would benefit from improved photometric and RV observations, although these are observationally highly demanding.


\subsection{\corot-8}                                                                                                         \label{sec:teps:corot8}

\begin{figure} \includegraphics[width=\columnwidth,angle=0]{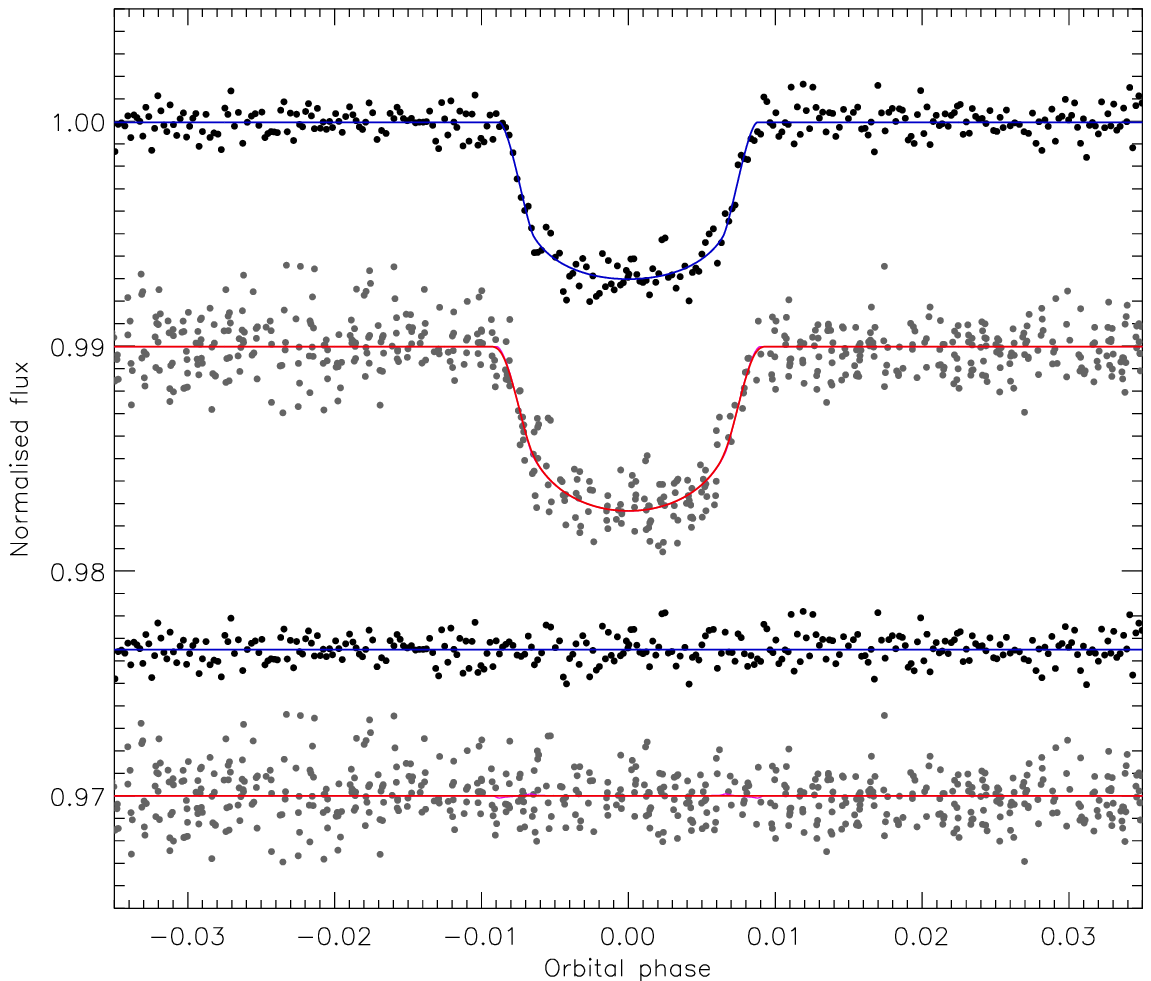}
\caption{\label{fig:corot8:lc} The 32\,s (upper) and 512\,s (lower)
\corot\ light curves of \corot-8. Other comments are the same as
Fig.\,\ref{fig:corot1:lc}.} \end{figure}

\corot-8 was discovered by \citet{Borde+10aa} and is a small and low-mass planet in a relatively long-period orbit (6.21\,d) around a metal-rich K dwarf. \citet{Borde+10aa} quote a third light value of 0.9\% without an errorbar; I adopted $L_3 = 0.009 \pm 0.003$ in my photometric analysis. The 32\,s data cover 12 transits with 19\,413 datapoints. 46 were rejected by a 4$\sigma$ clip and the remainder were phase-binned by a factor of 40 to give 483 normal points. The 512\,s observations harbour 11 transits; nine out of 946 points were rejected by a 4$\sigma$ clip and the remaining data were solved using $N_{\rm int} = 3$.

In the course of extracting the transits from the \corot\ data it became clear that the ephemeris in \citet{Borde+10aa} predicts transits to occur too early. I therefore binned up the 32\,s data by a factor of 16 to match the 512\,s data and fitted both datasets together to get a new orbital ephemeris:
$$ T_0 = {\rm HJD} \,\, 2\,454\,239.03311 (78) \, + \, 6.212381 (57) \times E $$
where $E$ is the number of orbital cycles after the reference epoch and the bracketed quantities denote the uncertainty in the final digit of the preceding number. The errobars come from Monte Carlo and residual permutation simulations, which are in good agreement. Compared to the ephemeris of \citet{Borde+10aa} I obtain larger errorbars, a consistent orbital period, and a $T_0$ which is later by 0.059\,d (67$\sigma$). \reff{P.\ Bord\'e (private communication) has kindly confirmed that this discrepancy has been noted elsewhere, by two amateur astronomers who have found that the transits of \corot-8 occur later than predicted from the ephemeris in the discovery paper.}

The 32\,s and 512\,s datasets were then fitted individually, the latter with $N_{\rm int} = 3$. For both, correlated noise is unimportant and the LD-fit/fix solutions are adopted (Tables A\arabic{appref}\addtocounter{appref}{1} and A\arabic{appref}\addtocounter{appref}{1}). The best fits are shown in Fig.\,\ref{fig:corot8:lc}. The combined solution of the two datasets (Table\,A\arabic{appref}\addtocounter{appref}{1}) does not agree well with the results of \citet{Borde+10aa}, in particular $r_{\rm A}$ (1.6$\sigma$) and $k$ (2.8$\sigma$). The errorbars found by \citet{Borde+10aa} seem to be too small (e.g.\ 0.1$^\circ$ compared to my 0.6$^\circ$ for $i$), especially in light of the disagreement.

The physical properties of \corot-8 were not straightforward to derive. \reff{The long evolutionary timescale of the host star caused the best fits for the {\it Teramo} and {\it VRSS} models to be slightly discrepant with ages formally in excess of a Hubble time}. However, reasonable solutions could be found by restricting the age range to 0--10\,Gyr. My results (Table\,A\arabic{appref}\addtocounter{appref}{1}) again differ from those of \citet{Borde+10aa}, most obviously for the radii of the star ($0.90 \pm 0.09$ versus $0.77 \pm 0.02$) and planet ($0.71 \pm 0.08$ versus $0.57 \pm 0.02$). The masses of the components agree well with my results, and I present the first measurements of $g_{\rm b}$, \Teq\ and \safronov. I find that the density of the planet is less than half the value in the discovery paper. Further light curves and RVs would be useful for this object, in order to improve the system parameters and confirm the orbital ephemeris determined above.


\subsection{\corot-9}                                                                                                         \label{sec:teps:corot9}

\begin{figure} \includegraphics[width=\columnwidth,angle=0]{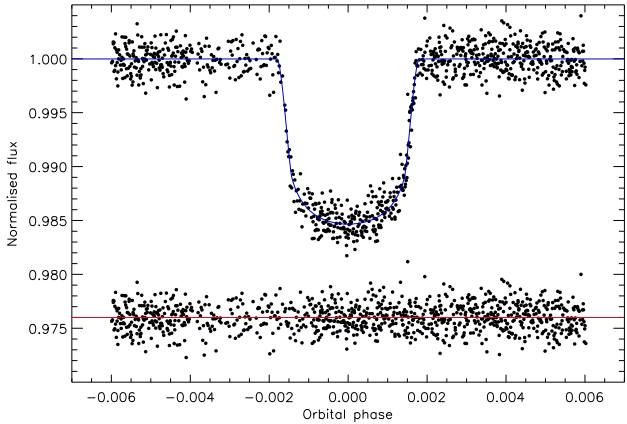}
\caption{\label{fig:corot9:lc} The 32\,s light curve of \corot-9.
See Fig.\,\ref{fig:corot1:lc} for details.} \end{figure}

The discovery of the 93.7-d period \corot-9 system was announced by \citet{Deeg+10natur}: the system has the longest \Porb\ of all TEPs bar HD\,80606 (which was discovered in the course of an RV rather than a photometric survey). The \corot\ data cover only two transits of which one lacks coverage of ingress. The data were cut from the full light curve and phase-binned by a factor of 5 to yield 946 datapoints (Fig.\,\ref{fig:corot9:lc}). The orbital shape was constrained using $e = 0.11 \pm 0.04$ and $\omega = \er{37}{9}{37}$ \citep{Deeg+10natur}. The {\sc jktebop} solutions show that correlated noise is not important and that the LD-fit/fix alternative is best (Table\,A\arabic{appref}\addtocounter{appref}{1}). The agreement with the results of \citet{Deeg+10natur} is excellent, both for the photometric parameters (Table\,A\arabic{appref}\addtocounter{appref}{1}) and the physical properties (Table\,A\arabic{appref}\addtocounter{appref}{1}). \corot-9 would benefit from further light curves and RVs.


\subsection{\corot-10}                                                                                                       \label{sec:teps:corot10}

\begin{figure} \includegraphics[width=\columnwidth,angle=0]{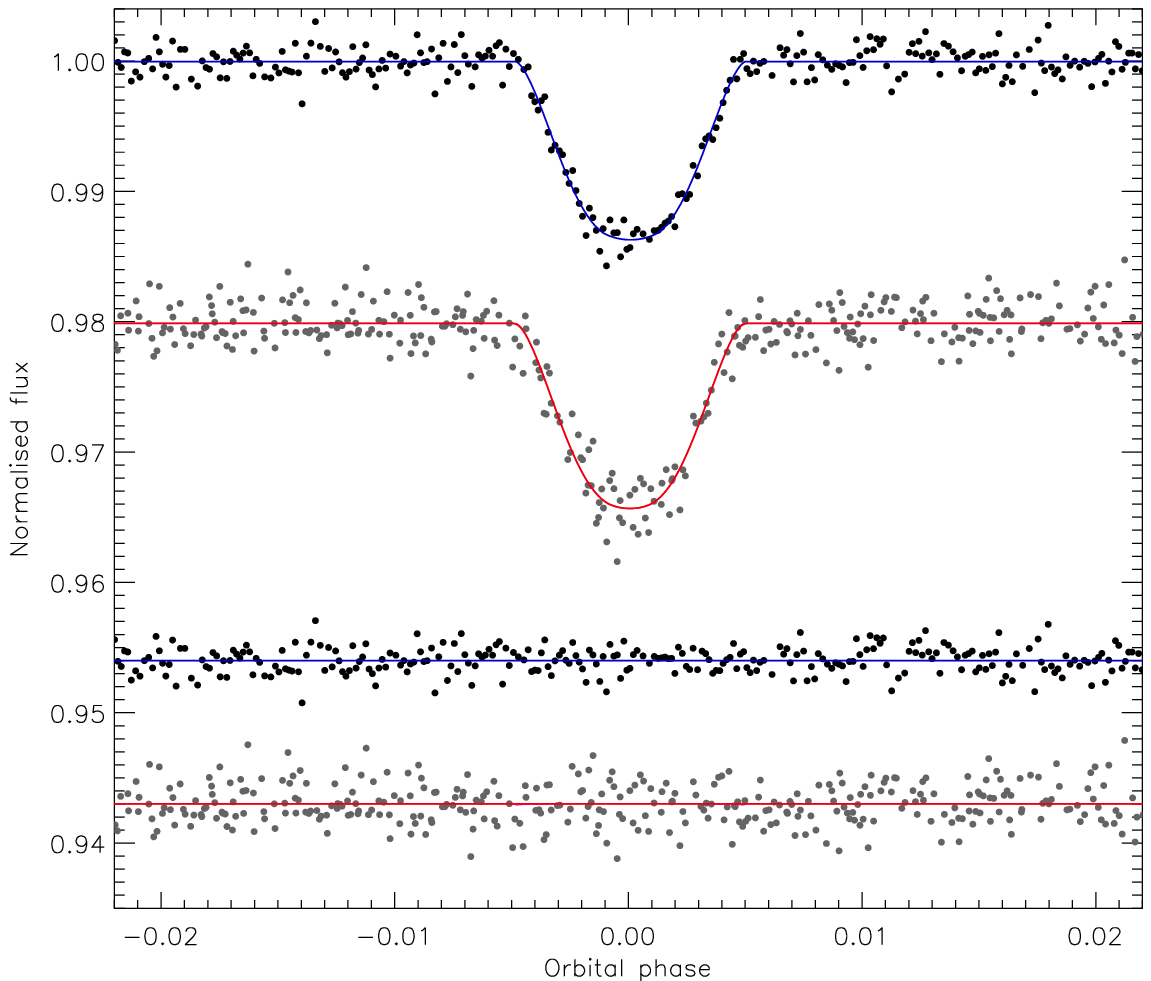}
\caption{\label{fig:corot10:lc} The \corot\ 32s (upper) and 512s (lower) light curves
of \corot-10. Other comments are the same as Fig.\,\ref{fig:corot1:lc}.} \end{figure}

\corot-10 was discovered by \citet{Bonomo+10aa} and comprises a relatively massive TEP (2.78\Mjup) in a long-period (13.24\,d) and highly eccentric ($e=0.53 \pm 0.04$) orbit around a K1 dwarf. Five transits were observed at 512\,s sampling and five more at 32\,s sampling, of which one was ignored because it is affected by a data jump. A third light of $L_3 = 0.055 \pm 0.003$ was found by \citet{Bonomo+10aa}.

The 32\,s data have 7190 points in the vicinity of a transit, of which 16 were rejected by a 4$\sigma$ clip and the rest phase-binned by a factor of 20 to give 369 normal points. No clipping or binning was needed for the 512\,s data, which were solved using $N_{\rm int} = 3$. The best fits are shown in Fig.\,\ref{fig:corot10:lc} and the solutions arranged in Tables A\arabic{appref}\addtocounter{appref}{1} and A\arabic{appref}\addtocounter{appref}{1}. My final photometric parameters (Table\,A\arabic{appref}\addtocounter{appref}{1}) agree well with those of \citet{Bonomo+10aa}.

The physical properties of the system are not precisely defined (Table\,A\arabic{appref}\addtocounter{appref}{1}). It appears to be a young system, resulting in edge effects with the model grids (see \corot-2) and therefore larger systematic errorbars than is typical, especially for $M_{\rm A}$. However, the agreement with the \citet{Bonomo+10aa} parameters is good. A better light curve would be beneficial.


\subsection{\corot-11}                                                                                                       \label{sec:teps:corot11}

\begin{figure} \includegraphics[width=\columnwidth,angle=0]{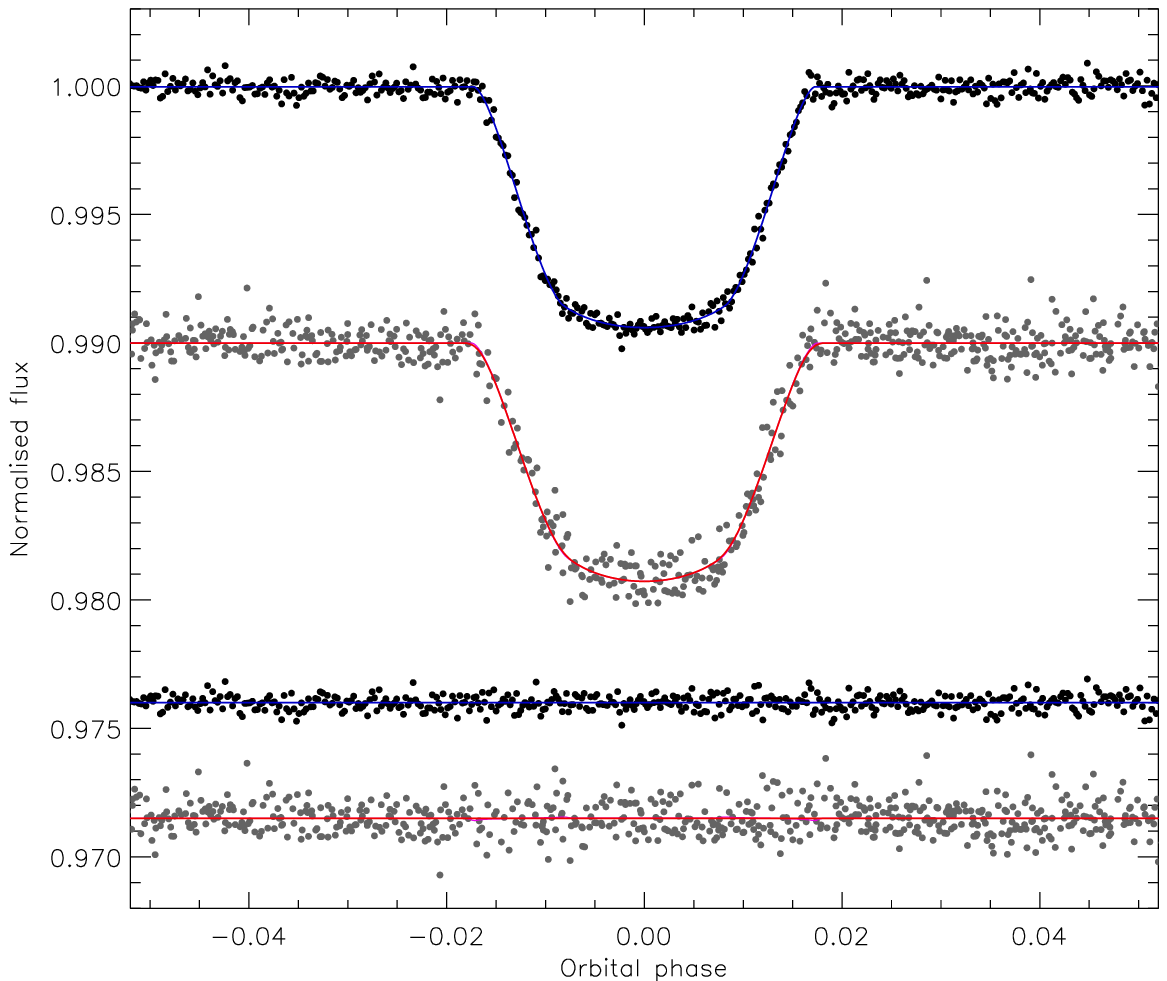}
\caption{\label{fig:corot11:lc} The \corot\ 32\,s (upper) and 512\,s (lower) light curves
of \corot-11. Other comments are the same as Fig.\,\ref{fig:corot1:lc}.} \end{figure}

The star in the \corot-11 system is one of the earliest-type (F6\,V) and most rapidly rotating ($40 \pm 5$\kms) TEP hosts known \citep{Gandolfi+10aa}. The planet itself is comparatively massive (2.34\Mjup) and has a high \Teq\ (1735\,K). \citet{Gandolfi+10aa} found a third light of $L_3 = 0.130 \pm 0.015$ which is included in the {\sc jktebop} analysis.

The \corot\ 32\,s data have 26\,382 points near transits, of which 46 were rejected by a 4$\sigma$ clip and the rest phase-binned by a factor of 50 to get 527 normal points. The 512\,s data contain 707 near transit of which 4 were rejected by a 4$\sigma$ clip and the remainder modelled using $N_{\rm int} = 3$. For both light curves the LD-fit/fix solutions are the best and correlated noise is inconsequential (Tables A\arabic{appref}\addtocounter{appref}{1} and A\arabic{appref}\addtocounter{appref}{1}). The fitted data are exhibited in Fig.\,\ref{fig:corot11:lc}. The results for the two datasets are unusual in that $k$ is the most consistent, whilst the other parameters agree to within  less good 1.1$\sigma$. They were combined with this trifling disagreement accounted for in the errorbars. The final results are very similar to those of \citet{Gandolfi+10aa}, and are among the better results for the known TEPs (TableA\,\arabic{appref}\addtocounter{appref}{1}). This is helped by the relatively low $i$, which means the light curve fits are well-constrained.

\reff{The {\sc jktabsdim} results for different model sets and for the dEB constraint (Table\,A\arabic{appref}\addtocounter{appref}{1}) agree well with each other and with \citet{Gandolfi+10aa}}. Further spectroscopic study of \corot-11 would be profitable.


\subsection{\corot-12}                                                                                                       \label{sec:teps:corot12}

\begin{figure} \includegraphics[width=\columnwidth,angle=0]{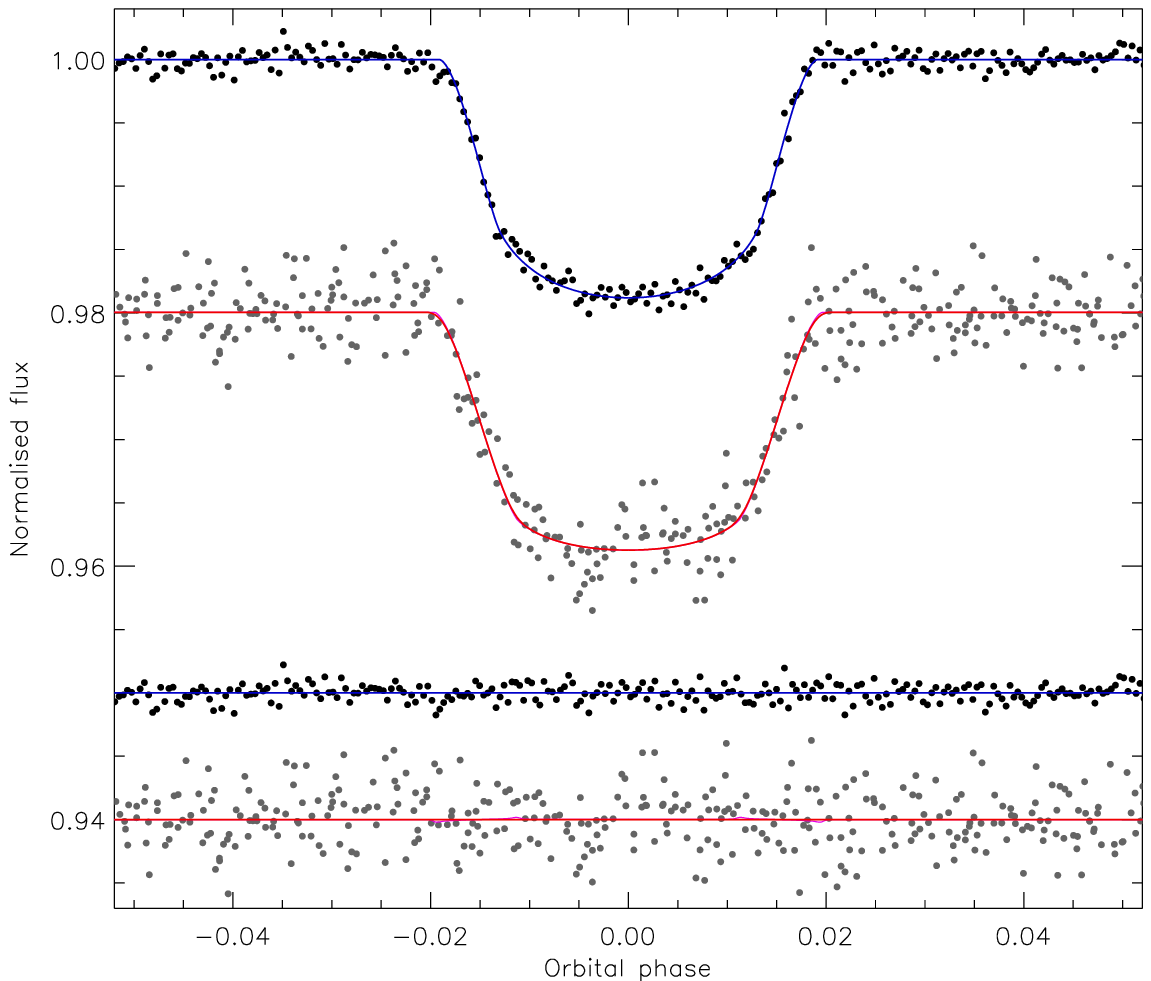}
\caption{\label{fig:corot12:lc} The \corot\ 32\,s (upper) and 512\,s (lower) light
curves of \corot-12. Other comments are the same as Fig.\,\ref{fig:corot1:lc}.} \end{figure}

The discovery paper for the \corot-12 system \citep{Gillon+10aa2} presents a system which is rather typical of the known population. The 32\,s data comprise 242\,558 datapoint of which 29\,114 are adjacent to one of the 36 transits. A 4$\sigma$ clip discards 76 points and the remainder end up in 291 phase-binned normal points. The 512\,s data cover 11 transits of which 413 points remain after rejecting 2780 which are away from eclipse and six which are over 4$\sigma$ away from a preliminary fit.

For the {\sc jktebop} fits I adopted a circular orbit because the $e\cos\omega$ and $e\sin\omega$ values in \citet{Gillon+10aa2} are consistent with zero at the 1$\sigma$ level. A third light of $L_3 = 0.033 \pm 0.005$ was used and $N_{\rm int} = 3$ was employed for the 512\,s data. The best fits are shown in Fig.\,\ref{fig:corot12:lc} and the model solutions in Tables A\arabic{appref}\addtocounter{appref}{1} and A\arabic{appref}\addtocounter{appref}{1}. In both cases correlated noise is unimportant and the LD-fit/fix solutions the most reliable. The results (Table\,A\arabic{appref}\addtocounter{appref}{1}) again show a slight disagreement between the two cadences, with $k$ the worst offender (1.7$\sigma$) and the other parameters divergent by an acceptable 1.2$\sigma$. The combined results agree reasonably well with those of \citet{Gillon+10aa2}. A better light curve would be useful.

The final physical properties (Table\,A\arabic{appref}\addtocounter{appref}{1}) agree very well with \citet{Gillon+10aa2}. Unusually, many of my error estimates are smaller than those found in the literature, which is attributable to the standard assumption of $e=0$ which neglects any uncertainty in the orbital shape. \corot-12 could do with further spectroscopic observations for both RV and spectral synthesis studies.


\subsection{\corot-13}                                                                                                       \label{sec:teps:corot13}

\begin{figure} \includegraphics[width=\columnwidth,angle=0]{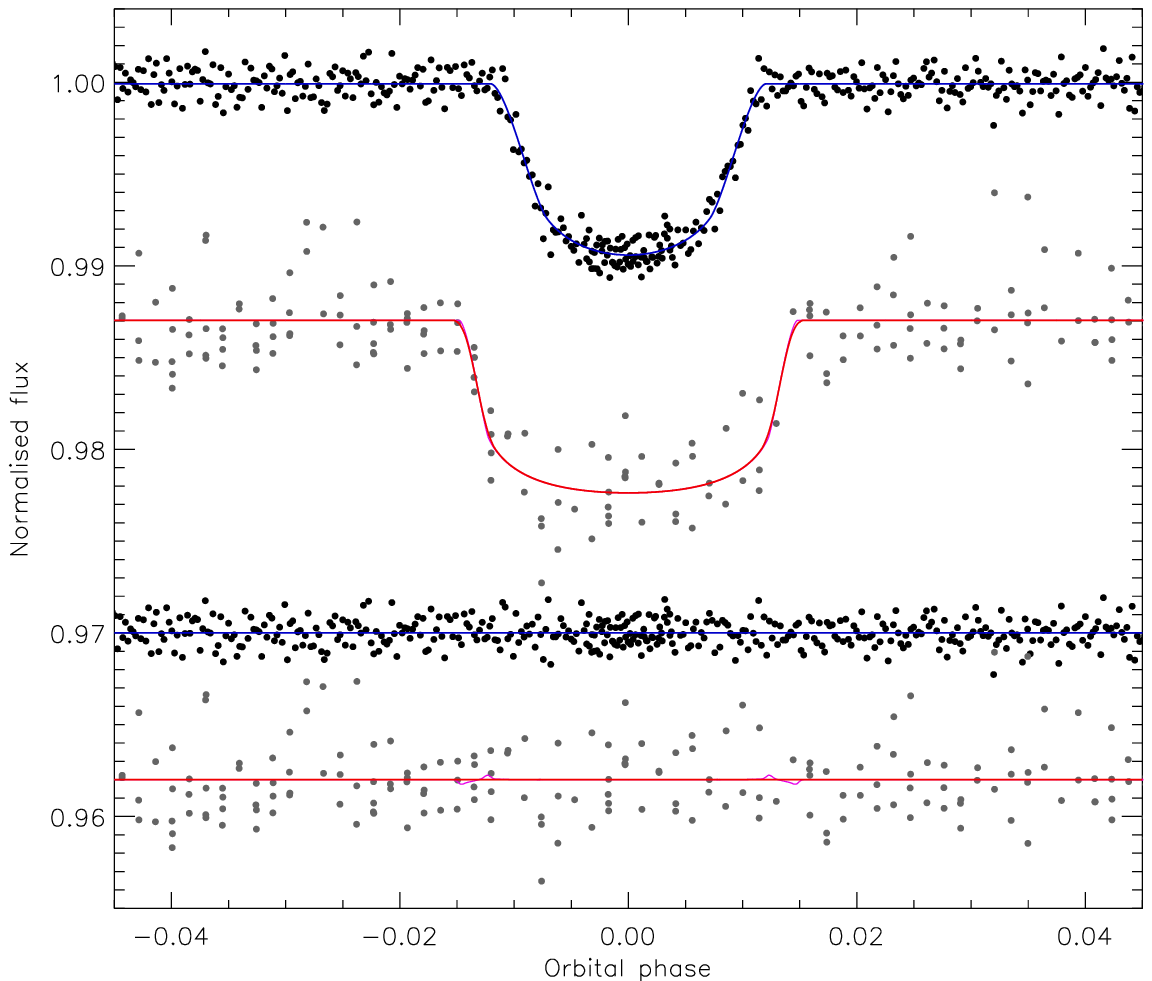}
\caption{\label{fig:corot13:lc} The \corot\ 32\,s (upper) and 512\,s (lower) light
curves of \corot-13. Other comments are the same as Fig.\,\ref{fig:corot1:lc}.} \end{figure}

\begin{figure} \includegraphics[width=\columnwidth,angle=0]{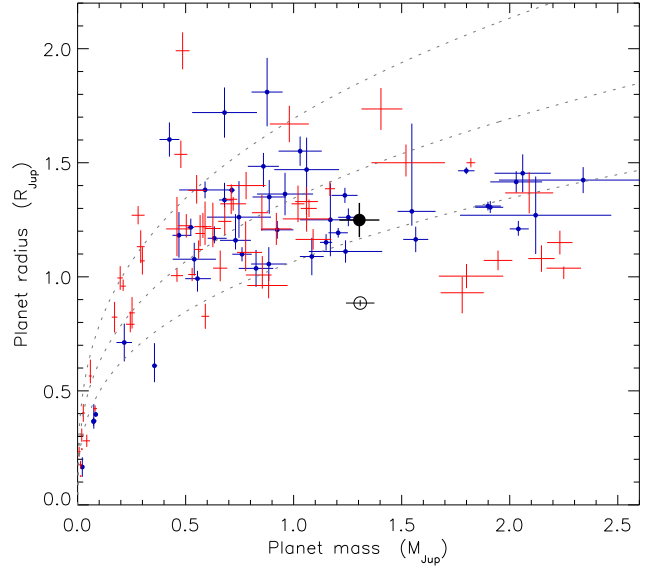}
\caption{\label{fig:corot13:mr} Plot of the masses and radii of the known TEPs
(dark blue crosses for those studied in this series of papers and light red
crosses for other objects). The black open circle shows \corot-13\,b (results
from \citealt{Cabrera+10aa}) and the black filled circle shows \corot-13\,b
(results from this work). The grey dotted lines show the loci where
density is equal to \pjup, $0.5\pjup$ and $0.25\pjup$.} \end{figure}

The discovery paper of this TEP is \citet{Cabrera+10aa}. \corot\ observed 217\,186 datapoints at 32\,s cadence, covering 23 transits. Of the 22\,020 points near transit, 54 were 4$\sigma$ clipped and the rest phase-binned by a factor of 50 to get 441 normal points. Of the 2403 512\,s datapoints, 196 are near transit and were solved using $N_{\rm int} = 3$. The quality of the 512\,s light curve is relatively poor due to the faintness of the star. A third light of $L_3 = 0.11 \pm 0.01$ was adopted from the discovery paper.

The best fits of the 32\,s and the 512\,s data are strikingly different (Fig.\,\ref{fig:corot13:lc}); I have verified that they arrived in the same datafile. The solutions for the 32\,s data are in Table\,A\arabic{appref}\addtocounter{appref}{1} and show that correlated noise is unimportant and that LD-fit/fix is to be preferred. Models with LDCs fixed at theoretical values have stronger LD and return $r_{\rm A}$ and $r_{\rm b}$ smaller by 1$\sigma$.

The 512\,s light curve was difficult to model and suffered from instability of solution. This can be seen in most clearly by inspecting the inclination values in Table\,A\arabic{appref}\addtocounter{appref}{1}, which veer from grazing in the quadratic LD-fit/fix and square-root LD-fit/fix solutions to equatorial in the other 12 solutions. Other parameters are similarly affected. A greater consistency might be obtained by guiding the offending best fits towards the area of parameter space inhabited by the other best fits, but at the expense of mathematical rigour. A better idea is to reject the 512\,s data, which are in any case of much lower weight than the much more extensive and better-sampled 32\,s data. For my final photometric parameters I accordingly adopt the 32\,s LD-fit/fix solution. Correlated noise is not important for this light curve.

A comparison with the results of \citet{Cabrera+10aa} is given in Table\,A\arabic{appref}\addtocounter{appref}{1} and shows a poor agreement, most likely because the solution in the discovery paper rests on the unreliable 512\,s data as well as the reliable 32\,s data. My results are therefore to be preferred, and also have larger and more representative errorbars.

\reff{The final physical properties of \corot-13 (Table\,A\arabic{appref}\addtocounter{appref}{1}) unsurprisingly do not agree well with those of \citet{Cabrera+10aa}.} I find radii which are much larger: 26\% (3.4$\sigma$) for $R_{\rm A}$ and 41\% (4.9$\sigma$) for $R_{\rm b}$. In turn, $g_{\rm b}$ and $\rho_{\rm b}$ are smaller by over a factor of 2. The $\log g_{\rm A}$ also decreases from $4.46 \pm 0.05$ to $4.26 \pm 0.04$, and the spectroscopically derived \logg\ for the host star ($4.30 \pm 0.10$) is in slightly better agreement with my value. A comparison of the different results for \corot-13\,b in the context of other TEPs is given in Fig.\,\ref{fig:corot13:mr} and shows that my results put it in a region of parameter space occupied by many other TEPs whereas the \citet{Cabrera+10aa} results make it unusually dense. An improved light curve for the system is urgently needed to verify the revised parameters that I find for the system.


\subsection{CoRoT-14}                                                                                                        \label{sec:teps:corot14}

\begin{figure} \includegraphics[width=\columnwidth,angle=0]{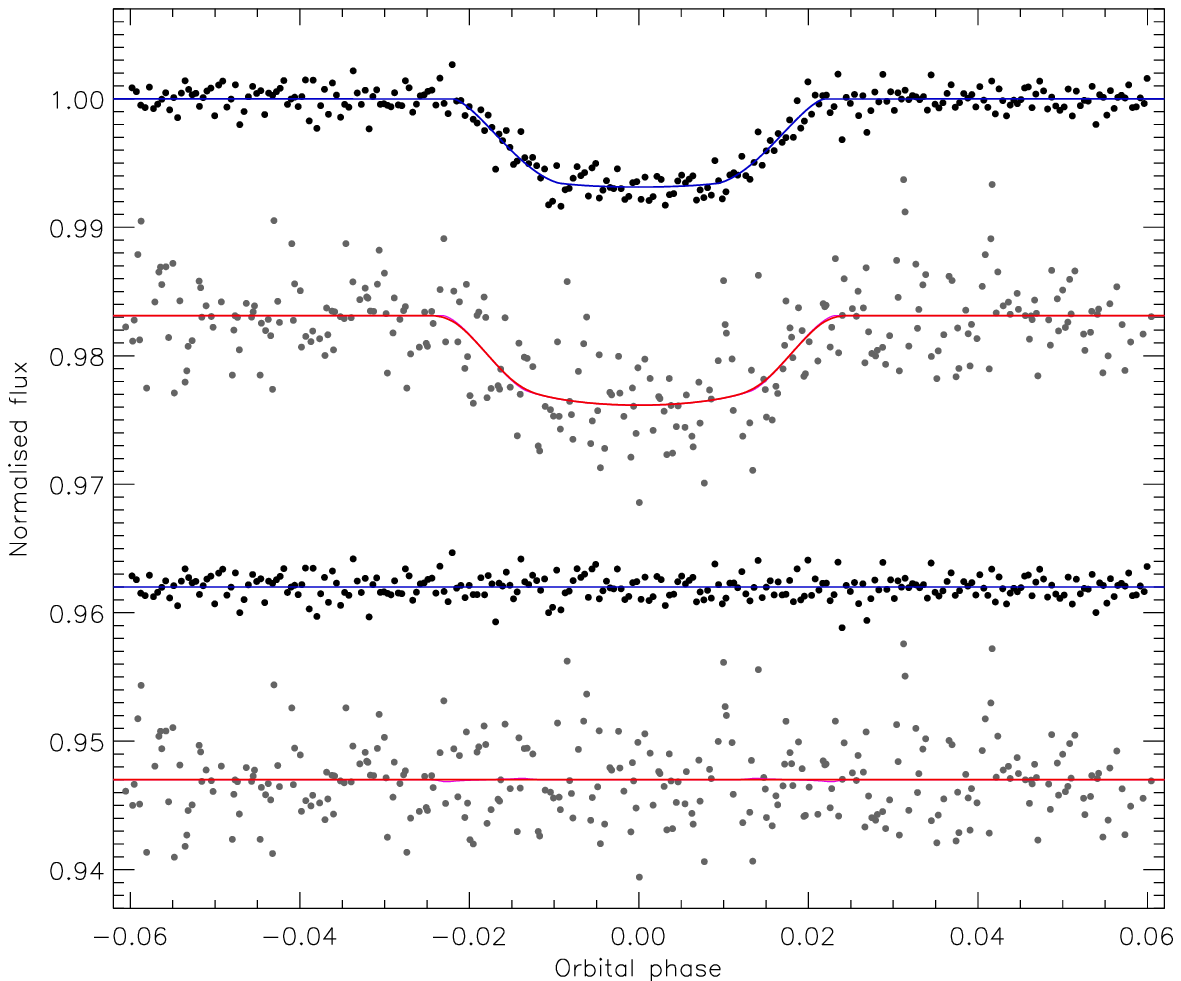}
\caption{\label{fig:corot14:lc} The \corot\ 32\,s (upper) and 512\,s (lower) light
curves of \corot-14. Other comments are the same as Fig.\,\ref{fig:corot1:lc}.} \end{figure}

Discovered by \citet{Tingley+11aa}, \corot-14 contains a very hot Jupiter ($\Teq = 1936$\,K) on a short-period orbit around an F9\,V star. The planet's large mass ($M_{\rm b} = 7.7$\Mjup) is unusual and is second only to WASP-18 \citep{Hellier+09natur,Me+09apj} for objects with $\Porb < 2$\,d. It is rather faint ($V=16.0$) so the light curve is quite scattered, but the RVs are quite sufficient due to its large mass.

The 32\,s data total 217\,262 points and 26\,279 of these cover 60 transits; after 4$\sigma$-clipping 49 points and phase-binning by a factor of 100 I obtained 263 normal points. The 512\,s data comprise 2400 points of which 291 are near transit. One of the 14 transits was rejected due to partial coverage. The remaining 512\,s data were modelling using $N_{\rm int} = 3$. A third light of $L_3 = 0.07 \pm 0.005$ \citep{Tingley+11aa} was taken into account.

The best fits are shown in Fig.\,\ref{fig:corot14:lc} and tabulated in Tables A\arabic{appref}\addtocounter{appref}{1} and A\arabic{appref}\addtocounter{appref}{1}; correlated noise is unimportant. The LD-fit/fix solutions of the 32\,s data are good, but only the LD-fixed solutions are reliable for the 512\,s data. The two light curve solutions agree well with each other (Table\,A\arabic{appref}\addtocounter{appref}{1}) and with \citet{Tingley+11aa}, as do the resulting physical properties (Table\,A\arabic{appref}\addtocounter{appref}{1}). Improved photometry and spectroscopy would be beneficial.


\subsection{CoRoT-15}                                                                                                        \label{sec:teps:corot15}

\begin{figure} \includegraphics[width=\columnwidth,angle=0]{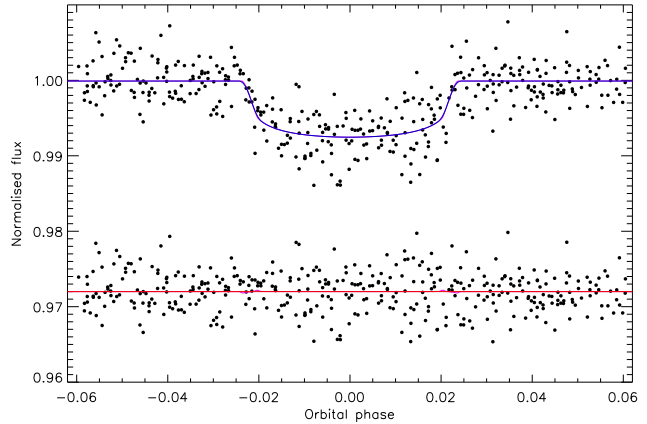}
\caption{\label{fig:corot15:lc} The 512\,s light curve of \corot-15.
See Fig.\,\ref{fig:corot1:lc} for details.} \end{figure}

The \corot-15 system contains a transiting brown dwarf \citep{Bouchy+11aa} with a mass (65\Mjup) very similar to those of WASP-30\,b (61\Mjup; \citealt{Anderson+11apj}) and LHS\,6343\,Ab (63\Mjup; Sect.\,\ref{sec:teps:lhs6343}). \corot-15 is the most difficult to study because of its faintness ($V = 15.5$). The \corot\ light curve lasts only 31.7\,d and covers ten transits at 512\,s cadence. Of the 395 datapoints near transit, four are rejected by a 3$\sigma$ cut and the rest are modelled using $N_{\rm int} = 3$ and $L_3 = 0.019 \pm 0.003$ \citep{Bouchy+11aa}.

The {\sc jtkebop} solutions point to a central transit, and therefore asymmetric errorbars. The LD-fixed and LD-fit/fix solutions agree well (LD-fitted solutions were not tried) so the latter are adopted. \reff{Unusually for the \corot\ data, correlated noise was found to be important}, with the residual-permutation errorbars slightly larger than the Monte-Carlo ones. The best fit is in Fig.\,\ref{fig:corot15:lc} and the model details are in Table\,A\arabic{appref}\addtocounter{appref}{1}. My photometric parameters (Table\,A\arabic{appref}\addtocounter{appref}{1}) are concordant with those of \citet{Bouchy+11aa}. I find a similarly good match for the physical properties of \corot-15 (Table\,A\arabic{appref}\addtocounter{appref}{1}), which could be improved by further observations of all relevant types.


\subsection{HAT-P-4}                                                                                                           \label{sec:teps:hatp4}

\begin{figure} \includegraphics[width=\columnwidth,angle=0]{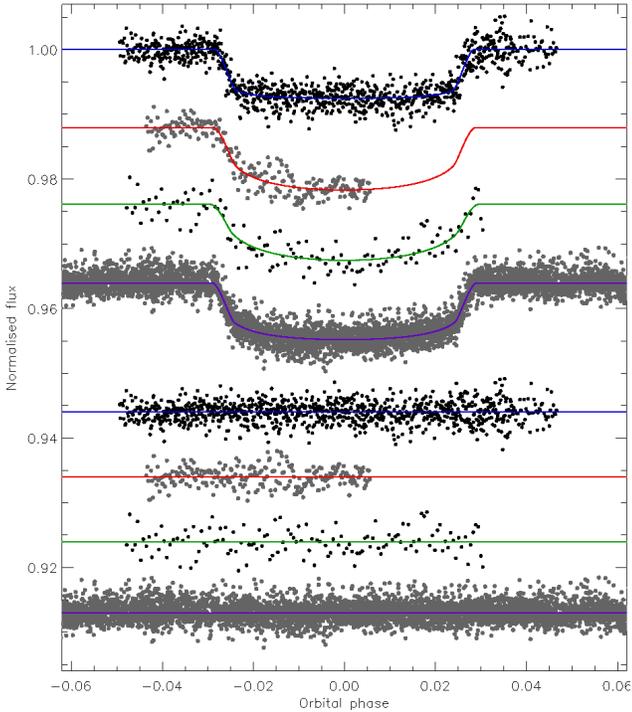}
\caption{\label{fig:hatp4:lc} Phased light curves of HAT-P-4 compared to
the best fits found using {\sc jktebop} and the quadratic LD law. The transit
light curves are, from top to bottom, $z$-band from \citet{Kovacs+07apj},
FLWO $i$-band from \citet{Winn+11aj}, FTN $i$-band from \citet{Winn+11aj},
and the EPOCH dataset from \citet{Christiansen+11apj}. The residuals are
plotted at the base of the figure, offset from zero.} \end{figure}

We now leave the \corot\ objects behind and turn to a TEP system with a light curve from the EPOCH project performed by the NASA {\em Deep Impact} satellite \citep{Christiansen+11apj}. HAT-P-4 was discovered by \citet{Kovacs+07apj} to be a low-density TEP orbiting a metal-rich late-F star. \citeauthor{Kovacs+07apj} obtained follow-up photometry using the FLWO 1.2\,m telescope and KeplerCam, which was subsequently reanalysed by \citet{Torres++08apj}. Further light curves were collected by \citet{Winn+11aj}, who also presented RV observations consistent with alignment of the planetary orbital and stellar rotational axes. \citet{Christiansen+11apj} analysed the EPOCH data, which cover seven consecutive transits followed by three more taken five months later.

The light curve from \citet{Kovacs+07apj} contains 985 datapoints spread over two transits, whose uncertainties I have multiplied by 1.67 to get $\chir \approx 1$. The LD-fit/fix solutions are clearly better than LD-fixed (larger residuals) and LD-fitted (unphysical LDCs). Correlated noise is slightly important. The parameters are given in Table\,A\arabic{appref}\addtocounter{appref}{1} and the best fit is shown in Fig.\,\ref{fig:hatp4:lc}.

\citet{Winn+11aj} observed two transits in the $i$-band, a full one with FTN and a partial one with KeplerCam. I binned the 662 points in the first dataset by a factor of 5 to get 133 normal points. The second dataset was modelled using constraints on the time of transit midpoint \citep{Me++07aa}. In both cases LD-fitted solutions were not attempted (Tables A\arabic{appref}\addtocounter{appref}{1} and A\arabic{appref}\addtocounter{appref}{1}).

I rejected one of the EPOCH transits due to insufficient observational coverage. Of the 6704 points near the other transits, 68 were lost to a 4$\sigma$ clip. The data contain substantial correlated noise due to pointing wander, so I did not phase-bin them. The residual-permutation errorbars are indeed 50\% larger than the Monte-Carlo ones. Table\,A\arabic{appref}\addtocounter{appref}{1} shows the results, of which the LD-fit/fix were adopted.

The four final light curve solutions are in good agreement ($\chir < 0.04$) except for $k$ ($\chir = 1.3$), and were combined by multiplying their probability density functions to get final photometric parameters (Table\,A\arabic{appref}\addtocounter{appref}{1}). These are comparatively imprecise because the transit is central, and accord well with literature results.

The {\sc jktabsdim} results show a significant systematic error from different theoretical model sets, but the final results are reasonable and agree with literature values (Table\,A\arabic{appref}\addtocounter{appref}{1}). HAT-P-4 would benefit from a better light curve.


\subsection{HAT-P-7}                                                                                                           \label{sec:teps:hatp7}

\begin{figure} \includegraphics[width=\columnwidth,angle=0]{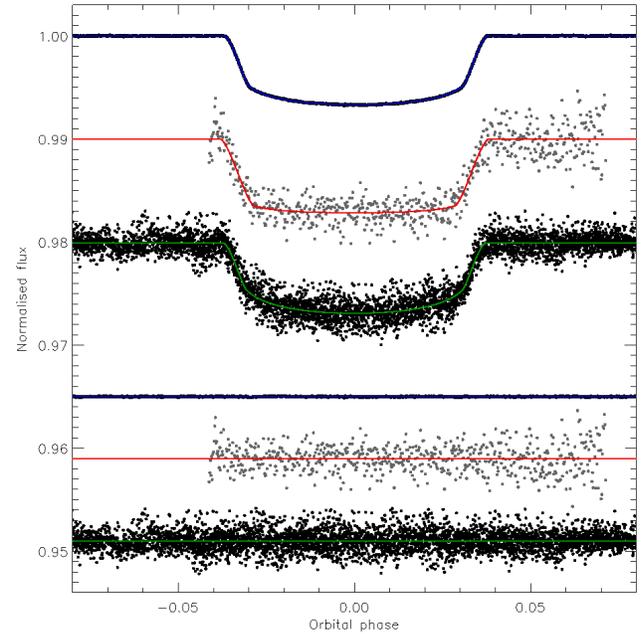}
\caption{\label{fig:hatp7:lc} Phased light curves of HAT-P-7 from the \kepler\
satellite (upper), \citet{Winn+09apj3} (middle) and EPOCH (lower) compared to
the best fits found using {\sc jktebop} and the quadratic LD law. The residuals
are plotted at the base of the figure, offset from zero.} \end{figure}

Discovered by \citet{Pal+08apj}, HAT-P-7 was the second TEP found in the \kepler\ field. It is a relatively massive planet (1.8\Mjup) orbiting a relatively massive star (1.5\Msun), and was the second planet found to have a retrograde orbit from RM observations \citep{Winn+09apj3} after WASP-17 \citep{Anderson+10apj}. \kepler\ has been observing it at short cadence since the start of the mission, and here I have analysed the public data from Q0, Q1 and Q2. These data have already  been used to establish the optical phase curve of the planet \citep{Borucki+09sci} and the ellipsoidal effect of the star \citep{Welsh+10apj}. The \kepler\ data have also be subjected to an asteroseismic investigation by \citet{Christensen+10apj}, who used the oscillation spectrum of the data to determine $\rho_{\rm A}$ to high precision. This constraint was not made use of in the current analysis, in order to retain homogeneity of approach. The follow-up light curve from the discovery paper covers only part of the transit and cannot compete with the \kepler\ data, so was not modelled here.

The \kepler\ observations are stunning (Fig.\,\ref{fig:hatp7:lc}) and cover 59 transits (five in Q0, 15 in Q1, 39 in Q2). 37\,611 of the 186\,786 original datapoints are near a transit, of which 55 were rejected by a 4$\sigma$ clip and the rest were phase-binned by a factor of 50 to get 753 normal points. The LD-fitted solutions are good (Table\,A\arabic{appref}\addtocounter{appref}{1}) and show that both LDCs need to be fitted to account for data with this level of precision, although correlated noise is formally important.

In order to provide a consistency check on the \kepler\ data I have also modelled the $i$-band observations of one transit (Table\,A\arabic{appref}\addtocounter{appref}{1}) by \citet{Winn+09apj3} and the EPOCH space-based light curve (Table\,A\arabic{appref}\addtocounter{appref}{1}) presented by \citet{Christiansen+10apj}. Whilst the latter again have substantial red noise, they agree well with the \kepler\ data and the two solutions are therefore combined to obtain the final photometric parameters. The level of agreement with the results of \citet{Welsh+10apj} is not as good as expected (Table\,A\arabic{appref}\addtocounter{appref}{1}). Despite this, HAT-P-7 is one of the best-measured TEP systems, alongside HD\,209458 (Paper\,I) and TrES-2 (Sect.\,\ref{sec:teps:tres2}).

The resulting physical properties are contained in Table\,A\arabic{appref}\addtocounter{appref}{1} and are highly unusual in that the radii of the star and planet are measured to precisions approaching 1\%. The agreement with literature results is good overall but with one caveat. \citet{Christensen+10apj} obtained a measurement of $\rho_{\rm A} = 0.1926 \pm 0.0023$\psun\ by analysing the oscillation spectrum of HAT-P-7\,A, which is 2.9$\sigma$ adrift from my transit-derived value of $\rho_{\rm A} = 0.2023 \pm 0.0024$\psun. This discrepancy may either gradually sort itself out as additional data appear from \kepler, or alternatively might signal the need to include more sophisticated physics in one or both of the analysis methods.


\subsection{HAT-P-11}                                                                                                         \label{sec:teps:hatp11}

\begin{figure} \includegraphics[width=\columnwidth,angle=0]{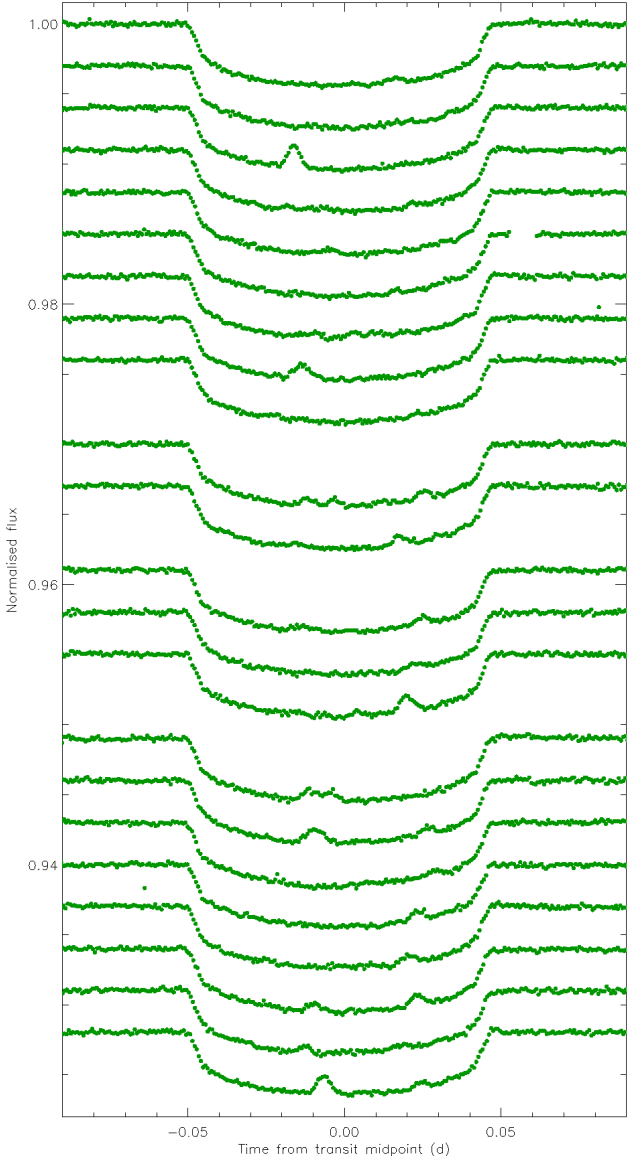}
\caption{\label{fig:hatp11:transits} Individual transits of HAT-P-11 as seen by \kepler.
Successive transits are offset by $-$0.03 flux units for clarity.} \end{figure}

\begin{figure} \includegraphics[width=\columnwidth,angle=0]{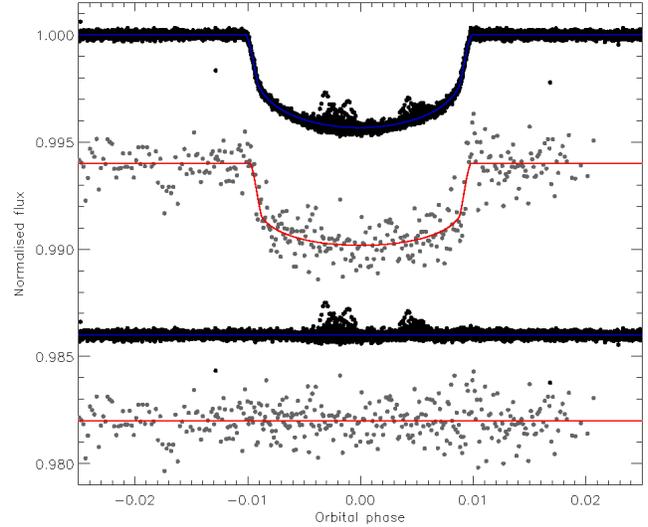}
\caption{\label{fig:hatp11:lc} Phased light curves of HAT-P-11 from the
\kepler\ satellite (upper) and in the $z$-band (lower) from \citet{Bakos+10apj}.
See Fig.\,\ref{fig:hatp7:lc} for further details.} \end{figure}

The third TEP discovered in the \kepler\ field \citep{Bakos+10apj} is a low-mass planet (0.084\Mjup) in an eccentric orbit around a low-mass star (0.81\Msun). It has been found to have a very oblique orbit through RM observations \citep{Winn+10apj,Hirano+11pasj}. \kepler\ has observed it in short cadence from the start of the mission. As with HAT-P-7 above, \citet{Christensen+10apj} analysed the Q0 and Q1 data and measured a $\rho_{\rm A}$ value which has a very a high precision but an undetermined accuracy.

Analysis of the \kepler\ data is not straightforward, because there is clear evidence of spot activity on the star. Of the 25 transits observed (three in Q0, six in Q1 and 16 in Q2), many are marvellous examples of the phenomenon of a planet transiting a starspot or starspot complex and all are affected to some degree (see Fig.\,\ref{fig:hatp11:transits}). The ratio of the planetary orbital to stellar spin periods is close to 6.0, so every sixth transit will cross over nearly the same part of the stellar surface and thus be similarly affected by spots which evolve on a timescale of 29\,d or less, a situation that was predicted by \citet{Winn+10apj}. Fig.\,\ref{fig:hatp11:transits} certainly shows that there are two preferred orbital phases for spot activity in the \kepler\ data. A detailed analysis of the spot characteristics will be very interesting.

In the current work I model the \kepler\ data using my usual approach. The effects of the starspot will therefore be treated as correlated noise. The oblique orbit of \reff{HAT-P-11\,b} means that it transits, at some point, a much greater fraction of the stellar surface than an aligned system would\reff{, and with a wide distribution of latitudes}. The assumption that the transited parts of the stellar surface on average behave the same as the non-transited areas is therefore more justifiable than in cases such as \corot-2. Due to LD, the effects of starspots are much stronger near the centre of the star than its limb \citep[e.g.][]{Sanchis+11apj}. In contrast, starspot deviations have the greatest effect on a photometric model when they occur at the partial phases of the transit. Treating them as correlated noise is therefore more likely to lead to an overestimate rather than an underestimate of the errorbars.

The \kepler\ Q0 to Q2 data comprise 186\,802 datapoints of which 12\,535 are adjacent to a transit. They were not phase-binned, as this might affect the estimation of the starspot-induced correlated residuals of the fit. The \kepler\ data don't agree exactly with previous orbital ephemerides of the system so the $T_0$ from \citet{Bakos+10apj} was included as a constraint following the approach of \citet{Me++07aa}. I also accounted for orbital eccentricity using $e\cos\omega = 0.261 \pm 0.082$ and $e\sin\omega = 0.085 \pm 0.043$ \citep{Winn+10apj}. The LD-fitted solutions are poor and have unphysical LDCs so the LD-fit/fix solutions were preferred (Table\,A\arabic{appref}\addtocounter{appref}{1}). The residual-permutation errors were larger than the Monte Carlo ones for the inclination but not for the other photometric parameters, supporting the approach taken to fit the data. The best fit is in Fig.\,\ref{fig:hatp11:lc}.

Additional light curves are available from \citet{Bakos+10apj}, comprising eight transits observed in the $z$-band using KeplerCam, one in the $r$-band with the same instrument, and three in the $I$-band with the Konkoly Schmidt. The first of these datasets is worth solving as a check of the \kepler\ data: the 4110 datapoints were phase-binned by a factor of ten to obtain 410 normal points. Correlated noise was not found to be important, and the LD-fit/fix solutions were adopted (Table\,A\arabic{appref}\addtocounter{appref}{1}). The agreement with the \kepler\ data is good (Table\,A\arabic{appref}\addtocounter{appref}{1}). Literature results are in moderate agreement, but the current results are to be preferred as they are the first to be based on the \kepler\ observations.

The {\sc jktabsdim} results suggest that the measured \Teff\ of the host star is quite high ($4780 \pm 50$\,K), causing the system to occupy the zero-age edge of the model grids. A hefty decrease of 300\,K in \Teff\ would be needed to assuage this problem, and would shift $M_{\rm A}$ from 0.82 to 0.75\Msun\ with other parameters less strongly affected. I therefore acquired some alternative \Teff\ estimates from a number of sources. The Tycho $B-V$ value \citep{Hog+97aa} supplemented by the calibration of \citet{Sousa+08aa} returns $\Teff = 4852$\,K. $B-V$ is not generally regarded as a good \Teff\ indicator for late-type dwarfs: the same value for this colour index equates to a K5 star with $\Teff = 4410$\,K using the tables of \citet{Zombeck90book}. Using the 2MASS $JHK_s$ \citep{Skrutskie+06aj} magnitudes is a better bet: the $V-K_s$ colour index and the calibration of \citet{Casagrande++08mn} yields $\Teff = 4765$\,K. Dr.\ B.\ Smalley has kindly calculated the luminosity of HAT-P-11\,A from the available broad-band optical and infrared photometry and the {\it Hipparcos} parallax \citep{Perryman+97aa} to be $\log(L/\Lsun) = -0.61$. This agrees with the {\sc jktabsdim} solutions for the measured \Teff\ of the star ($-0.63$) but not with those for the lower \Teff\ ($-0.77$). The \Teff\ measurement from \citet{Bakos+10apj} is supported by the investigations above.

Table\,A\arabic{appref}\addtocounter{appref}{1} shows my calculated physical properties for \reff{HAT-P-11\,b} compared to literature solutions. \citet{Bakos+10apj} find smaller radii for the two components which may be down to differences in analysis. In the HAT methodology the stellar parameters are forced to agree with a theoretical stellar model (usually interpolated from the {\it Y$^2$} isochrones); in my solution process I find the point of closest agreement but do not require this to exactly reproduce a point in a grid of theoretical predictions. I find a very young but poorly constrained age, which is in accordance with the starspot activity seen in both the \kepler\ and the HAT data. \citet{Christensen+10apj} derived $\rho_{\rm A} = 1.7846 \pm 0.0006$\psun\ (errorbar does not account for systematic errors) from the oscillation spectrum of \reff{HAT-P-11\,A} in the \kepler\ light curve. This indicates a much less dense star than I find ($2.415 \pm 0.097$\psun), a similar situation to HAT-P-7 but with a much stronger discrepancy. Further investigation is needed to understand the disagreement; extra RV measurements would be useful to refine the $K_{\rm A}$, $e\cos\omega$ and $e\sin\omega$ values.


\subsection{HD 17156}                                                                                                        \label{sec:teps:hd17156}

\begin{figure} \includegraphics[width=\columnwidth,angle=0]{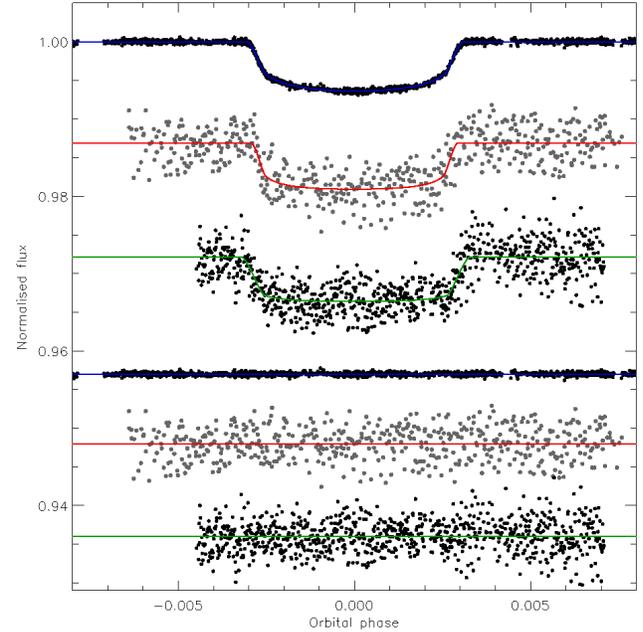}
\caption{\label{fig:17156:lc} The light curves of HD\,17156 compared to the
{\sc jktebop} best fit. Top is HST/FGS, middle and bottom are $(b+y)/2$ and
$z$ from \citet{Winn+09apj4}. See Fig.\,\ref{fig:hatp7:lc} for further
details.} \end{figure}

The planet orbiting HD\,17156 was discovered by the RV method by \citet{Fischer+07apj}, who also searched for transits but did not detect them or rule out their existence. A transit was soon observed using three telescopes by \citet{Barbieri+07aa}, as part of the {\tt transitsearch.org} network \citep{Seagroves+03pasp}. Further transit light curves were obtained by \citet{Gillon+08aa}, \citet{Barbieri+09aa} and \citet{Irwin+08apj}. Two good light curves were presented by \citet{Winn+09apj4}, covering the transit on Christmas Day in 2007. The RM effect has been observed on multiple occasions \citep{Narita+08pasj,Cochran+08apj,Barbieri+09aa,Narita+09pasj} and overall indicates axial alignment.

An extensive analysis of HD\,17156 was performed by \citet{Nutzman+11apj} and \citet{Gilliland+11apj} based on data obtained using the HST/FGS. The asteroseismic study \citep{Gilliland+11apj} yielded a mean density and age for the star. The theoretical uncertainty of the former quantity is probably small, but is significant for the latter quantity. The HST/FGS data cover three transits, which were used by \citet{Nutzman+11apj} to measure the physical properties of the system. Analyses including the asteroseismic $\rho_{\rm A}$ as a constraint are in good agreement with standard analyses, but with smaller errorbars (by a factor of three for $R_{\rm A}$ and $R_{\rm b}$). I did not include the asteroseismic constraint in my analysis, in order to retain homogeneity with the results for other systems.


I have solved the two light curves from \citet{Winn+09apj4}, which are in the $(b+y)/2$ and $z$ passbands, each covering one transit. I also modelled the HST data from \citet{Nutzman+11apj}, which comprises partial coverage of each of three transits, using a light curve which has been corrected for instrumental effects. The FGS data were taken with the F583W filter which covers 440--710\,nm, so I adopted for LDCs appropriate for a combined $g$+$r$ filter. The orbital shape of HD\,17156 was accounted for by adopting $e = 0.6768 \pm 0.0304$ and $\omega = 121.71 \pm 0.43$\,deg \citep{Nutzman+11apj}. The best fits are exhibited in Fig.\,\ref{fig:17156:lc}.

The results for the HST light curve are given in Table\,A\arabic{appref}\addtocounter{appref}{1} and LD-fit/fix is the best alternative. Correlated noise is marginally important. The $(b+y)/2$ data prefer a higher $i$ and thus smaller $r_{\rm A}$ and $r_{\rm b}$ (Table\,A\arabic{appref}\addtocounter{appref}{1}), whereas the $z$ observations yield the opposite situation (Table\,A\arabic{appref}\addtocounter{appref}{1}). The two datasets agree with the HST results overall, so the HST ones are adopted as the final parameters (Table\,A\arabic{appref}\addtocounter{appref}{1}). Their agreement with published values is reasonable.

The physical properties of HD\,17156 are shown in Table\,A\arabic{appref}\addtocounter{appref}{1} and reveal some model-dependent error which stems primarily from the {\em DSEP} models. The agreement with literature values is again good, and I provide the first measurement of \Teq\ and \safronov. HD\,17156 is a well-characterised system which is not in need of further observations.


\subsection{HD 80606}                                                                                                          \label{sec:teps:80606}

\begin{figure} \includegraphics[width=\columnwidth,angle=0]{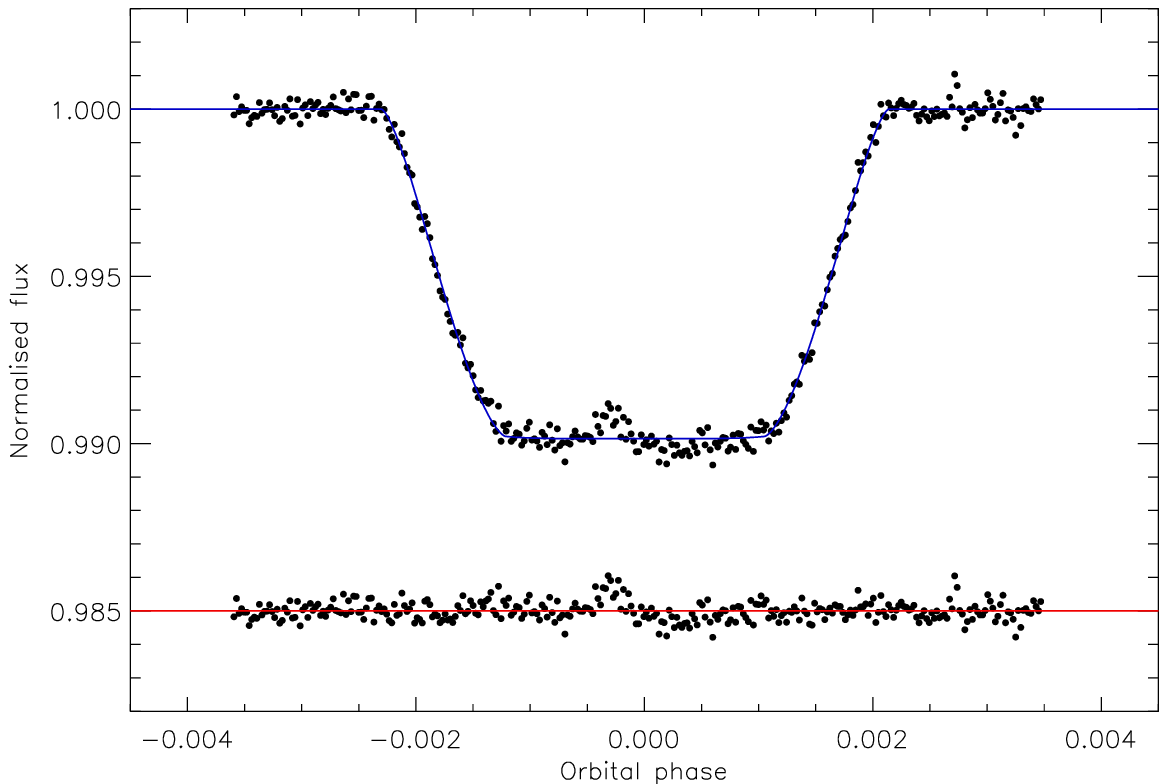}
\caption{\label{fig:80606:lc} The \spitzer\ light curve of HD\,80606
\citep{Hebrard+10aa}. See Fig.\,\ref{fig:hatp7:lc} for further details.}
\end{figure}

HD\,80606\,b was found to be a planetary-mass object with a period of 111.8\,d by \citet{Naef+01aa} from extensive RV observations. At the time of discovery it was the most eccentric exoplanet known ($e=0.927$). \citet{Laughlin+09natur} observed an occultation of the planet whilst using \spitzer\ to probe the heating of the planet around periastron. Based on the parameters then known, \citeauthor{Laughlin+09natur} calculated a probability of 15\% that a transit also occurs, and encouraged follow-up observations to detect it. The system was monitored over the next predicted time of transit and duly found to be a TEP by \citet{Moutou+09aa}, \citet{GarciaMccullough09apj} and \citet{Fossey++09mn}.

RM observations of HD\,80606 have been obtained and studied by \citet{Moutou+09aa}, \citet{Pont+09aa}, \citet{Hebrard+10aa} and \citet{Winn+09apj2}, with a good agreement that the orbit is oblique to high confidence. Coupled with the high eccentricity, this has bearing on its formation and evolution mechanisms \citep{Naef+01aa,MardlingLin02apj,Matsuo+07apj}. The relatively high mass of the planet (4.11\Mjup) agrees with the established pattern that TEPs in eccentric orbits are generally the massive ones \citep{Me+09apj}.

Spectral synthesis analyses of the parent star have been performed by \citet{Naef+01aa}, \citet{Santos++04aa} and \citet{Gonzalez++10mn}: I adopt the \Teff\ from \citet{Santos++04aa} as it is the middle of the three values and its errorbar encompasses the other two; I use the \citet{Gonzalez++10mn} \FeH\ value because all three agree and it has an equitable errorbar. I take the $K_{\rm A}$ value from the most recent detailed RV analysis, \citet{Winn+09apj2}.

Good light curves of HD\,80606 are difficult to obtain due to the long transit duration (nearly 12 hours) and orbital period. The transit discovery light curves \citep{Moutou+09aa,GarciaMccullough09apj,Fossey++09mn} all missed ingress as it occurred during daylight. \citet{Winn+09apj2} combined data from eight observing sites spread through mainland USA and Hawaii to obtain full coverage of the 2009 June transit, but the resulting light curve was still a long way from definitive. \citet{Shporer+10apj} and \citet{Hidas+10mn} performed similar observing campaigns covering several transits, but the data are unfortunately strongly affected by systematics. \citet{Hebrard+10aa} used \spitzer\ to obtain a complete and high-quality light curve of the 2010 January transit in the IRAC 4.5\,$\mu$m passband. This is by far the best light curve of any transit of HD\,80606, and is the only one analysed here.

The \spitzer\ data were binned by a factor of 100 to lower the number of datapoints\footnote{The \spitzer\ data I used had already been binned by a factor of five before being lodged with the CDS.} from 31\,767 to 318. The resulting time resolution is 215\,s, which is reasonable for such a long-duration transit. A disturbance near mid-eclipse is noticable and can be attributed to starspots (Fig.\,\ref{fig:80606:lc}); this was treated as correlated noise in the {\sc jktebop} analysis (see Sect.\,\ref{sec:teps:hatp11}). The flux from the planet in this near-infrared passband must be accounted for, which was done by including a light ratio of $0.001 \pm 0.001$ with the method described by \citet{Me++07aa}. This light ratio gives a secondary eclipse of similar depth to that observed by \citet{Laughlin+09natur}, with a conservative errorbar to \reff{allow for inefficient energy redistribution within the planetary atmosphere}. The orbital shape was accounted for using $e\cos\omega = 0.4774 \pm 0.0018$ and $e\sin\omega = -0.8016 \pm 0.0017$ \citep{Hebrard+10aa}. The LD-fitted solutions yield unphysical LDCs so the LD-fit/fix solution was adopted (Table\,A\arabic{appref}\addtocounter{appref}{1}). Correlated noise is important, due to the treatment of the starspot. The final photometric parameters (Table\,A\arabic{appref}\addtocounter{appref}{1}) are in reasonable agreement with published values.

The {\sc jktabsdim} results show that the systematic error (model disagreement) is relatively high, and dominates the error budget for $M_{\rm A}$, $M_{\rm b}$ and $a$ (Table\,A\arabic{appref}\addtocounter{appref}{1}). The solution with the dEB constraint is also somewhat different to the solutions using theoretical models. The source of these problems is not obvious, but the discrepancies are small enough that HD\,80606 is one of the best-characterised TEP systems.


\subsection{Kepler-4}                                                                                                        \label{sec:teps:kepler4}

\begin{figure} \includegraphics[width=\columnwidth,angle=0]{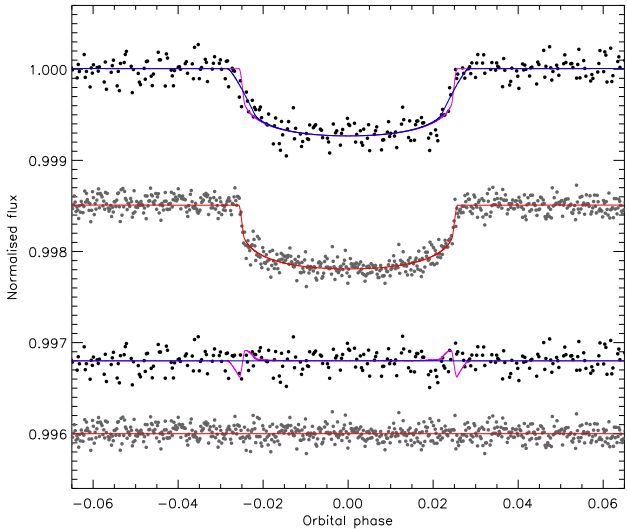}
\caption{\label{fig:kepler4:lc} Phased long-cadence (top) and short-cadence
(second from top) light curves of Kepler-4 compared to the best fit found
using {\sc jktebop} and the quadratic LD law. For the long-cadence data the
fit is shown by a blue line, and an evaluation of the same model but without
numerical integration is shown by a purple line. The residuals are plotted
at the base of the figure, offset from unity. The purple line through the
residuals shows the difference between the fit to the data and an evaluation
of the same model but without numerical integration.} \end{figure}

Kepler-4 was the first TEP system to be announced as a discovery of the \kepler\ satellite \citep{Borucki+10apj} and is a very low-mass object (0.075\Mjup) orbiting a slightly evolved star ($\logg = 4.17$). It has also been studied by \citet{KippingBakos11apj}. Eccentricity is significant at only 2$\sigma$ and additional RV measurements are necessary to investigate this further. The available \kepler\ data cover 13 shallow transits at long cadence (29.4\,min) in Q0 and Q1 and 26 transits at short cadence (58.8\,s) in Q2.

The addition of Q2 data over Q0 and Q1 allows an improvement of the orbital ephemeris. The short-cadence data were binned to the time resolution of the long-cadence data. The long-cadence and binned short-cadence data were then modelled using {\sc jktebop} and $N_{\rm int} = 10$ to find:
$$ T_0 = {\rm HJD} \,\, 2\,454\,956.61132 (92) \, + \, 3.213658 (38) \times E $$
where the bracketed quantities give the uncertainties in the preceding digit, and the uncertainties come from 100 MC simulations.

The long-cadence transit data were cut from the full light curve and normalised as described in Sect.\,\ref{sec:data}, giving 323 datapoints out of the original 2101. The 123\,536 short-cadence observations were treated similarly, then phase-binned by a factor of 25 to yield 708 normal points. I assumed a circular orbit and a third light of $L_3 = 0.02 \pm 0.02$. The long-cadence data do not constrain the fit well; parameter perturbations could not be applied in the Monte Carlo analysis so these errorbars may be optimistic. The long-cadence data were solved using $N_{\rm int} = 10$ (Table\,A\arabic{appref}\addtocounter{appref}{1}) and the short-cadence data were solved without using numerical integration (Table\,A\arabic{appref}\addtocounter{appref}{1}). The best fits are given in Fig.\,\ref{fig:kepler4:lc}. In both cases the LD-fit/fix solutions were best. The residual-permutation errorbars were much larger than the Monte-Carlo ones for the long-cadence light curve, which is likely due to the inability to apply parameter perturbations in this case.

The solutions for the two datasets agree well and were combined by multiplying their probability density functions to give final photometric parameters (Table\,A\arabic{appref}\addtocounter{appref}{1}). The photometic parameters from \citet{KippingBakos11apj} disagree with those from \citet{Borucki+10apj}: the latter find a lower inclination and thus larger $r_{\rm A}$ and $r_{\rm b}$. My results support those of \citet{Borucki+10apj} and I am unable to completely reproduce the \citet{KippingBakos11apj} results by modifying my treatment of eccentricity or numerical integration.

The physical properties of Kepler-4 (Table\,A\arabic{appref}\addtocounter{appref}{1}) show a reasonable agreement with literature results albeit with substantially larger errorbars in some cases. This is despite the availability of much more extensive (and higher-cadence) observations available for my analysis. The \kepler\ satellite continues to observe the system at short cadence so a much improved light curve will gradually accumulate. Additional spectroscopy to improve the \Teff\ and $K_{\rm A}$ measurements should be a high priority.


\subsection{Kepler-5}                                                                                                        \label{sec:teps:kepler5}

\begin{figure} \includegraphics[width=\columnwidth,angle=0]{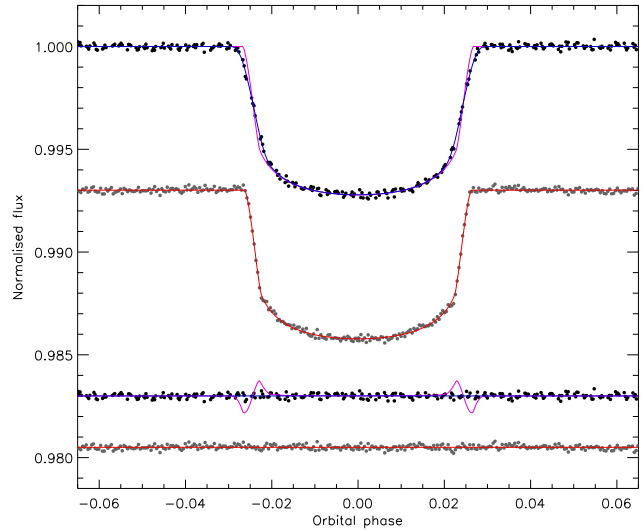}
\caption{\label{fig:kepler5:lc} Phased light curves of Kepler-5.
See Fig.\,\ref{fig:kepler4:lc} for further details.} \end{figure}

The discovery of Kepler-5 was announced by \citet{Koch+10apj2}: the planet is relatively massive (2.04\Mjup) and the star is quite evolved ($\logg = 4.17$). It has also been studied by \citet{KippingBakos11apj}. The available \kepler\ data now comprise 12 transits studied at long cadence during Q0 and Q1, of which 330 of the 2101 datapoints occur near to a transit, and 23 transits at short cadence after rejection of one transit due to systematic noise. The 123\,536 short-cadence datapoints were reduced into 379 normal points. A revised orbital ephemeris was measured from the Q0, Q1 and Q2 data:
$$ T_0 = {\rm HJD} \,\, 2\,454\,955.90059 (36) \, + \, 3.548469 (15) \times E $$
using the same approach as for Kepler-4.

A third light of $L_3 = 0.02 \pm 0.002$ and a circular orbit were assumed \citep{Koch+10apj2} and the long-cadence data were modelled using $N_{\rm int} = 10$. The {\sc jktebop} solutions (Tables A\arabic{appref}\addtocounter{appref}{1} and A\arabic{appref}\addtocounter{appref}{1}) show that correlated noise is unimportant and that the LD-fitted results are viable. Agreement between the two datasets is not good and can be attributed to information loss at the long cadence. I therefore adopt the short-cadence solutions as final (Table\,A\arabic{appref}\addtocounter{appref}{1}). I find a notably smaller $r_{\rm A}$ and $r_{\rm b}$ than \citet{Koch+10apj2} and \citet{KippingBakos11apj}, which reflects the difference between the long-cadence data (which were available for those studies) and the better short-cadence observations (which were not). The uncertainties given by \citet{Koch+10apj2} are too low. The best fits are shown in Fig.\,\ref{fig:kepler5:lc}.

The {\sc jktabsdim} results are contained in Table\,A\arabic{appref}\addtocounter{appref}{1} and expectedly disagree with literature studies based on only the long-cadence data: I find smaller masses and radii for both star and planet. The discovery paper \citep{Koch+10apj2} obtained two solutions corresponding to different evolutionary stages for the host star (with masses 1.21 and 1.38 \Msun), and endorsed the more evolved alternative. I do not find this problem, due to my refined photometric parameters. Additional photometric observations are needed, are currently being obtained by the satellite, and will make Kepler-5 one of the best-charaterised TEP systems.


\subsection{Kepler-6}                                                                                                        \label{sec:teps:kepler6}

\begin{figure} \includegraphics[width=\columnwidth,angle=0]{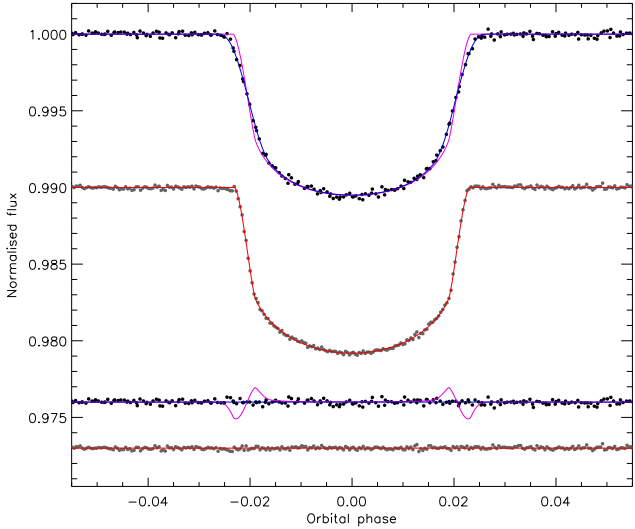}
\caption{\label{fig:kepler6:lc} Phased light curves of Kepler-6.
See Fig.\,\ref{fig:kepler4:lc} for further details.} \end{figure}

Kepler-6 was found to be a TEP by \citet{Dunham+10apj} and is interesting because of the high metallicity of the host star ($\FeH = 0.34 \pm 0.05$).
The same long-cadence data were studied by \citet{KippingBakos11apj}, who found a solution with a higher orbital inclination and 2.6$\sigma$ signficant signal in the TTV peridogram which may indicate stellar activity. The \kepler\ Q0 and Q1 data cover 13 transits at long cadence, whereas the Q2 data include 41 transits at short cadence. A new orbital ephemeris was determined from these data:
$$ T_0 = {\rm HJD} \,\, 2\,454\,954.485805 (64) \, + \, 3.2347020 (33) \times E $$

The long cadence data comprise 2102 points of which 246 are near a transit. These were modelled using $N_{\rm int} = 10$ to yield the results in Table\,A\arabic{appref}\addtocounter{appref}{1}. The solutions are very sensitive to the treatment of LD: LD-fixed gives a poor internal agreement and LD-fitted returns unphysical LDCs, so the LD-fit/fix results were adopted. The 123\,536 short-cadence datapoints were reduced into 34 normal points and the LD-fit/fix solutions were best (Table\,A\arabic{appref}\addtocounter{appref}{1}). In both cases $L_3 = 0.033 \pm 0.004$ \citep{Dunham+10apj} was incorporated and correlated noise was found to be inconsequential. The two datasets agree well so the parameters were combined (Table\,A\arabic{appref}\addtocounter{appref}{1}). Published studies agree well with my results, although their errorbars are questionable. The best fits are shown in Fig.\,\ref{fig:kepler6:lc}.

The physical properties of Kepler-6 are given in Table\,A\arabic{appref}\addtocounter{appref}{1}. The {\it DSEP} model solutions disagree with the others so are not included in the final results. I find a somewhat smaller planet and star compared to \citet{Dunham+10apj}, whereas \citet{KippingBakos11apj} agree with my results within the errors. \kepler\ continues to observe Kepler-6 at short cadence and additional RV measurements would also be useful.


\subsection{Kepler-7}                                                                                                        \label{sec:teps:kepler7}

\begin{figure} \includegraphics[width=\columnwidth,angle=0]{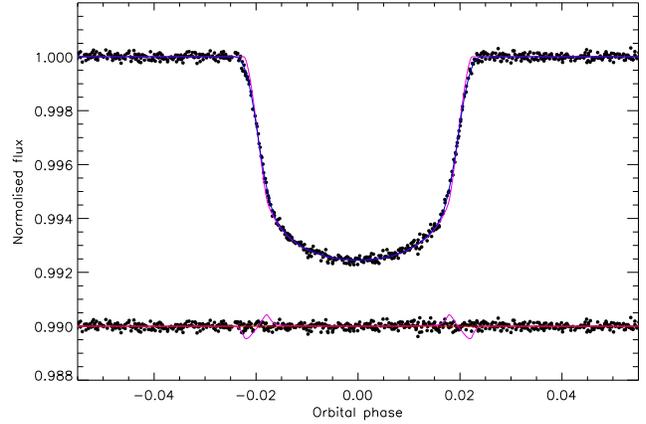}
\caption{\label{fig:kepler7:lc} Phased long-cadence light curve of Kepler-7.
See Fig.\,\ref{fig:kepler4:lc} for further details.} \end{figure}

Discovered by \citet{Latham+10apj}, this a very low-density TEP (0.10\pjup) around a slightly evolved star ($\logg = 3.96$). The same data were also studied by \citet{KippingBakos11apj}, who detected an occultation with a significance level of 3.5$\sigma$. The Q2 data are long-cadence, so no short-cadence data are available for analysis. The Q0, Q1 and Q2 observations cover 26 transits, and 947 of the 6177 datapoints were retained for the {\sc jktebop} analysis. $L_3 = 0.025 \pm 0.005$, a circular orbit and $N_{\rm int} = 10$ were adopted. The updated orbital ephemeris is:
$$ T_0 = {\rm HJD} \,\, 2\,454\,967.27598 (11) \, + \, 4.8854948 (82) \times E $$
The LD-fitted results are reliable and have reduced residuals compared to the LD-fit/fix solutions (Table\,A\arabic{appref}\addtocounter{appref}{1}). Correlated noise is insignificant. I find a lower inclination and therefore a higher $r_{\rm A}$ and $r_{\rm b}$ than previous studies (Table\,A\arabic{appref}\addtocounter{appref}{1}), and investigations reveal that this is primarily due to inclusion of the Q2 data. The best fits are shown in Fig.\,\ref{fig:kepler7:lc}.

The physical properties of Kepler-7 (Table\,A\arabic{appref}\addtocounter{appref}{1}) reveal that the star is one of the most evolved known to host a TEP, and the planet itself is the second-most rarefied known after WASP-17 \citep{Anderson+10apj,Anderson+11xxx}. As expected from the photometric parameters, my physical properties do not agree well with those previously published. Confirmation of this will be possible soon, as Kepler-7 has been observed in short cadence by \kepler\ from Quarter 3 onwards. New RV measurements would also be worthwhile.


\subsection{Kepler-8}                                                                                                        \label{sec:teps:kepler8}

\begin{figure} \includegraphics[width=\columnwidth,angle=0]{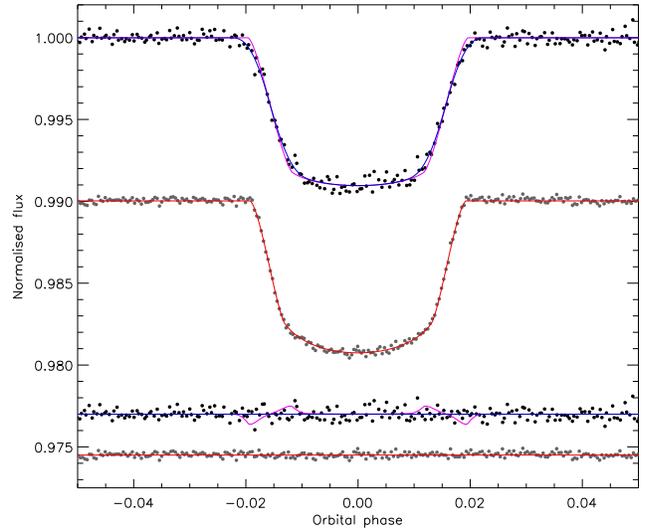}
\caption{\label{fig:kepler8:lc} Phased light curves of Kepler-8.
See Fig.\,\ref{fig:kepler4:lc} for further details.} \end{figure}

Like Kepler-7, Kepler-8 is a low-density TEP (0.21\pjup) orbiting a slightly evolved star ($\logg = 4.18$). \citet{Jenkins+10apj3} and \citet{KippingBakos11apj} have studied the \kepler\ Q0 and Q1 long-cadence data, whereas the Q2 short-cadence observations are now available. A refined orbital ephemeris was established as:
$$ T_0 = {\rm HJD} \,\, 2\,454\,954.11844 (18) \, + \, 3.5225047 (76) \times E $$

The long-cadence observations cover 13 transits, and I retain 265 of the original 2098 datapoints (solved using $N_{\rm int} = 10$). The Q2 data consist of 123\,536 points covering 24 transits, which I phase-binned down to 299 normal points. The third light value given by \citet{Jenkins+10apj3} does not have an errorbar, so I adopt $L_3 = 0.0075 \pm 0.0075$ to be conservative. The best fits can be seen in Fig.\,\ref{fig:kepler8:lc}. Models of the long-cadence data are rather sensitive to the treatment of LD and are also in comparatively poor agreement with the short-cadence models (Tables A\arabic{appref}\addtocounter{appref}{1} and A\arabic{appref}\addtocounter{appref}{1}). I therefore adopt the LD-fit/fix flavour of the latter (Table \,A\arabic{appref}\addtocounter{appref}{1}), which is also in accord with and more precise than literature studies. The errorbars found by \citet{Jenkins+10apj3} are smaller than expected.

\reff{The absolute dimensions of Kepler-8 (Table\,A\arabic{appref}\addtocounter{appref}{1}) agree well with published values and establish it as a well-understood system. \kepler\ continues to observe Kepler-8 photometrically and further RV observations are merited.}


\subsection{KOI-428}                                                                                                          \label{sec:teps:koi428}

\begin{figure} \includegraphics[width=\columnwidth,angle=0]{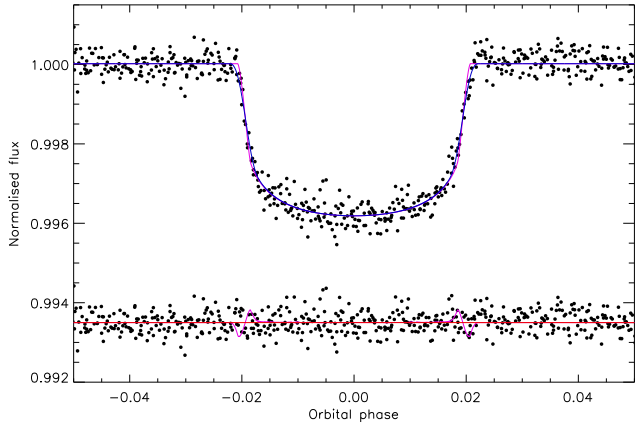}
\caption{\label{fig:koi428:lc} Phased light curve of KOI-428.
See Fig.\,\ref{fig:kepler4:lc} for further details.} \end{figure}

KOI-428 was one of the 306 \kepler\ Objects of Interest that was presented to the astronomical community by \citet{Borucki+11apj}; the 400 brightest ones were retained for in-house analysis by the \kepler\ Team. Follow-up spectroscopic observations by \citet{Santerne+11aa} have subsequently shown this to be a system containing a relatively massive planet (2.1\Mjup) orbiting a comparatively hot star (6510\,K).

The \kepler\ Q1 and Q2 data contain 12 transits observed at long cadence, compared to only four in the Q1 data available to \citet{Santerne+11aa}. 741 of the 5708 datapoints were retained; one of the 12 transits was rejected because it is deformed by instrumental artefacts arising from a pointing jump. These were solved using $N_{\rm int} = 10$, a circular orbit and no $L_3$ (Table\,A\arabic{appref}\addtocounter{appref}{1}). Correlated noise is negligible. I find a lower inclination and thus larger $r_{\rm A}$ and $r_{\rm b}$ compared to \citet{Santerne+11aa}, but the values are within their errorbars (Table\,A\arabic{appref}\addtocounter{appref}{1}). Fig.\,\ref{fig:koi428:lc} shows the best fit to the light curve. The revised orbital ephemeris is:
$$ T_0 = {\rm HJD} \,\, 2\,455\,005.51858 (50) \, + \, 6.873130 (75) \times E $$

{\sc jktabsdim} returns a noticably larger but less massive star compared to \citet{Santerne+11aa}, as expected given the slightly different photometric parameters (Table\,A\arabic{appref}\addtocounter{appref}{1}). The model fits prefer a \Teff\ lower by 80\,K (0.8$\sigma$), so this may indicate that analysis of the \kepler\ data in isolation results in a star which is too large. KOI-428 remains on the \kepler\ target list so additional data will be available soon. The errorbars quoted by \citet{Santerne+11aa} are too small. Further spectroscopic study of this object is warranted.


\subsection{LHS 6343}                                                                                                        \label{sec:teps:lhs6343}

\begin{figure} \includegraphics[width=\columnwidth,angle=0]{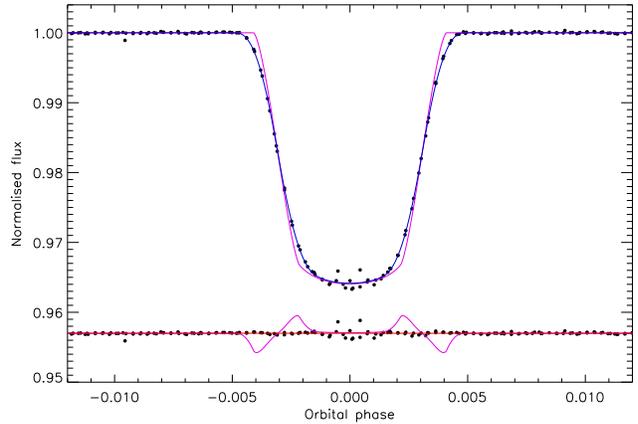}
\caption{\label{fig:lhs6343:lc} Phased light curve of LHS\,6343.
See Fig.\,\ref{fig:kepler4:lc} for further details.} \end{figure}

LHS\,6343 is a nearby M dwarf \citep{Luyten79book} which was found to host a small transiting object by \citet{Johnson+11apj} based on \kepler\ data. Follow-up imaging and spectroscopic observations \citep{Johnson+11apj} revealed that the system contains two M dwarfs (A and B) separated by 0.55\as\, and that the brighter component A hosts a likely brown-dwarf companion with an orbital period of 12.71\,d. The \kepler\ Q0, Q1 and Q2 data contain a total of 11 transits observed at long cadence.

The \kepler\ observations are of the combined flux of the three components, and a light ratio of B versus A has not been directly observed. The adaptive-optics imaging obtained by \citet{Johnson+11apj} give magnitude differences in the $JHK_s$ passbands of $0.49 \pm 0.05$, $0.49 \pm 0.05$ and $0.45 \pm 0.05$, respectively. The characteristics of these passbands are such that an extrapolation to the wide \kepler\ passband is reasonable for stars as similar as A and B. This was performed using the process outlined in Paper\,III \citep[see also][]{Me+10mn}, resulting in $L_3 = 0.29 \pm 0.09$. Measurements of the flux ratio at optical wavelengths would be useful in refining these numbers.

\citet{Johnson+11apj} obtain $\Teff = 3130 \pm 20$\,K for component A from the calculated $V$ and observed $K_s$ magnitudes of the star applied to the photometric calibrations of \citet{Casagrande++08mn}. \reff{The quoted errorbar is clearly a precision rather than a true uncertainty}. I therefore extrapolated the infrared light ratios into the $V$ band to obtain the $V-K_s$ colour index which, with the calibrations from \citet{Casagrande++08mn}, gives $\Teff = 3300\pm 200$\,K for component A where the errorbar is conservative. My \Teff\ value was used in the analyses below.

The 6177 long-cadence datapoints from the \kepler\ Q0, Q1 and Q2 observations were slimmed down to 268 points nearby a transit. These were modelled with {\sc jktebop} using $L_3 = 0.29 \pm 0.09$ (see above), $e = 0.056 \pm 0.032$, $\omega = 337^\circ \pm 56^\circ$ \citep{Johnson+11apj} and $N_{\rm int} = 10$. I included \Porb\ and $T_0$ as fitted parameters constrained by the $T_0$ value found by \citet{Johnson+11apj} from a ground-based observation of one transit. The ensuing orbital ephemeris is:
$$ T_0 = {\rm HJD} \,\, 2\,454\,995.358014 (44) \, + \, 12.7138107 (73) \times E $$
The LD-fixed solutions are poor (Table\,A\arabic{appref}\addtocounter{appref}{1}). The LD-fit/fix alternatives are better, but the quality of fit is hampered by scattered data mischievously placed right at the transit midpoint. \kepler\ continues to observe LHS\,6343 so an improved light curve is in the pipeline. Correlated noise is not important. Compared to \citet{Johnson+11apj} I find a smaller $k$ and larger $r_1$, as well as larger errorbars despite having many more photometric observations (Table\,A\arabic{appref}\addtocounter{appref}{1}). The best fit is in Fig.\,\ref{fig:lhs6343:lc}.

Derivation of the physical properties of the star and its substellar transiting companion is difficult due to the low mass of the former object. \reff{Theoretical stellar models are unreliable in this regime (see fig.\,4 in Paper\,III) so an additional systematic error should be added to those quoted in Table\,A\arabic{appref}\addtocounter{appref}{1}}. Compared to \citet{Johnson+11apj} I find the star to be more massive, which propagates into a correspondingly larger mass for the companion of $M_{\rm b} = 70 \pm 6$\Mjup\ close to the stellar/substellar boundary. Further spectroscopic and spatially resolved optical observations would be useful in pinning down the \Teff\ of the host star and the mass of its companion.


\subsection{TrES-2}                                                                                                            \label{sec:teps:tres2}

\begin{figure} \includegraphics[width=\columnwidth,angle=0]{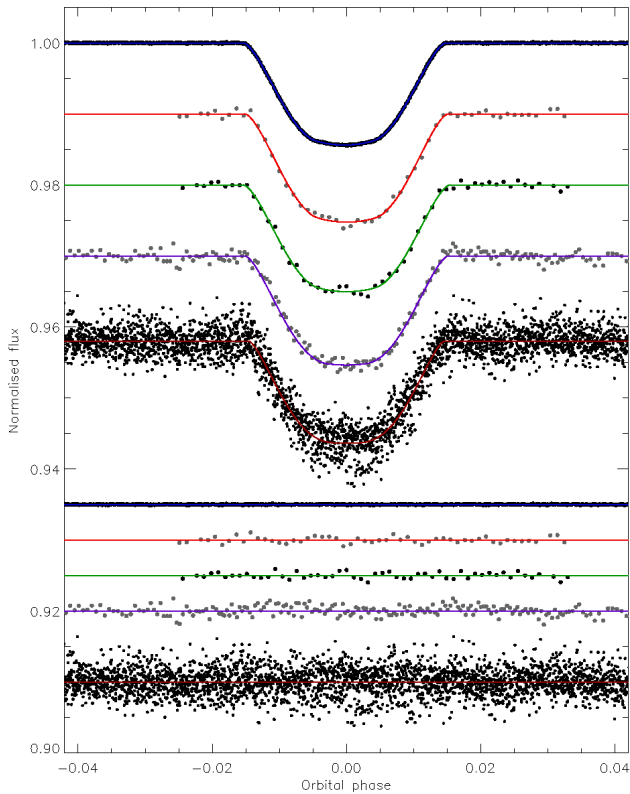}
\caption{\label{fig:tres2:lc} Phased light curves of TrES-2 compared to the best
fits found using {\sc jktebop} and the quadratic LD law. From top to bottom the
data are from \kepler, the 790.2\,nm then 794.4\,nm sets from \citet{Colon+10mn},
$R$-band from \citet{Rabus+09aa}, and EPOCH from \citet{Christiansen+11apj}.
See Fig.\,\ref{fig:80606:lc} for further details. See Paper\,III (sect.\,4.7
and fig.\,11) of Paper\,III for previous results.} \end{figure}

TrES-2 was discovered by the Trans-Atlantic Exoplanet Survey \citep{Odonovan+06apj} and subsequently treated in Paper\,I and Paper\,II. TrES-2 was the first TEP discovered in the \kepler\ field, and a light curve of stunning quality from this satellite is now available \citep{Gilliland+10apj}. An analysis of these data has been given by \citet{KippingBakos11apj2}. In addition, excellent ground-based light curves have been obtained by \citet{Colon+10mn} and it was one of the targets of the EPOCH project using the NASA {\it Deep Impact} spacecraft \citep{Christiansen+11apj}. We also add to this dataset the Johnson $R$-band light curve obtained by \citet{Rabus+09aa} in the course of a TTV study. The system offers one complication: a fainter star at a separation of 1.09\as\ \citep{Daemgen+09aa}. TrES-2 was revisited in Paper\,III to account for this situation.

The \kepler\ data cover four transits in Q1, 14 in Q1 and 34 in Q2 (ignoring one near an instrumental artefact), all at short cadence. The 186\,802 original datapoints were reduced to 18\,310 by rejecting observations far from transit, then to 18\,297 by a 4$\sigma$ clip, and then to 311 normal points by phase-binning by a factor of 30. $L_3 = 0.0258 \pm 0.0008$ was used and the LD-fit/fix solutions adopted (Table\,A\arabic{appref}\addtocounter{appref}{1}). Correlated noise was not important. Although \kepler\ continues to observe TrES-2, it is already the photometrically best-measured TEP known. The light curve fits are shown in Fig.\,\ref{fig:tres2:lc}, which is best viewed in conjunction with Fig.\,11 in Paper\,III.

The two light curves from \citet{Colon+10mn} cover the same transit in two very narrow passbands, at 790.2 and 794.4 nm, obtained using the 10.4\,m GranTeCan and OSIRIS imager equipped with a tunable filter. $L_3 = 0.0355 \pm 0.0005$ was used for both passbands. For the 790.2\,nm data the LD-fixed solutions had to be adopted, and correlated noise was found to be moderately important (Table\,A\arabic{appref}\addtocounter{appref}{1}). The LD-fit/fix solutions could be adopted for the 794.4\,nm data, for which correlated noise was found to be unimportant (Table\,A\arabic{appref}\addtocounter{appref}{1}.

The $R$-band data from \citet{Rabus+09aa} comprise data from five transits which are supplied already phased and binned. I scaled the errorbars by a factor of 0.485 to obtain $\chir \approx 1$ and modelled them using $L_3 = 0.0287 \pm 0.0007$. The LD-fit/fix solutions are reasonable (Table\,A\arabic{appref}\addtocounter{appref}{1}) and correlated noise is incidental (as expected given the phase-binning process).

The EPOCH data \citep{Christiansen+11apj} cover eight transits. 3534 of the original 27\,724 datapoints were retained, of which 17 were subsequently rejected by a 3.5$\sigma$ clip. They were not phase-binned as correlated noise was correctly expected to be important. $L_3 = 0.026 \pm 0.002$ was assumed. I had to adopt the LD-fixed solutions as the LD-fit/fix and LD-fitted alternatives gave anomalous results and negative LDCs (Table\,A\arabic{appref}\addtocounter{appref}{1}).

The overall results for each light curve are given in Table\,A\arabic{appref}\addtocounter{appref}{1} and show an excellent agreement overall. $k$ is as expected the least concordant parameter, but even here the agreement is at the level of $\chir= 0.9$. The final photometric parameters are the weighted means of the ones for each light curve. Their agreement with published values is in general excellent except, perplexingly, for those of \citet{KippingBakos11apj2} which are the only other ones to be derived from the \kepler\ data of TrES-2. Similar concerns have been noted for Kepler-4 to Kepler-8, so there may be a small systematic difference in the results from Kipping \& Bakos compared to other researchers.

TrES-2 is of particular interest because \citet{MislisSchmitt09aa} and \citet{Mislis+10aa} have found evidence for a decrease in the system's orbital inclination. The \kepler\ data rule out an effect of the expected size, and do not provide evidence of changes in any of the photometric parameters (see also \citealt{KippingBakos11apj2}). A natural explanation of the previous detection of a change in inclination would be the presence of subtle systematic errors in transit light curves.

The {\sc jktabsdim} results are given in Table\,A\arabic{appref}\addtocounter{appref}{1} and show that TrES-2 is now very well characterised. The photometric parameters contribute only a small part of the error budget for the measurements of its physical properties. The best way to improve the results further would be to obtain a more precise \FeH\ value.


\subsection{TrES-3}                                                                                                            \label{sec:teps:tres3}

\begin{figure} \includegraphics[width=\columnwidth,angle=0]{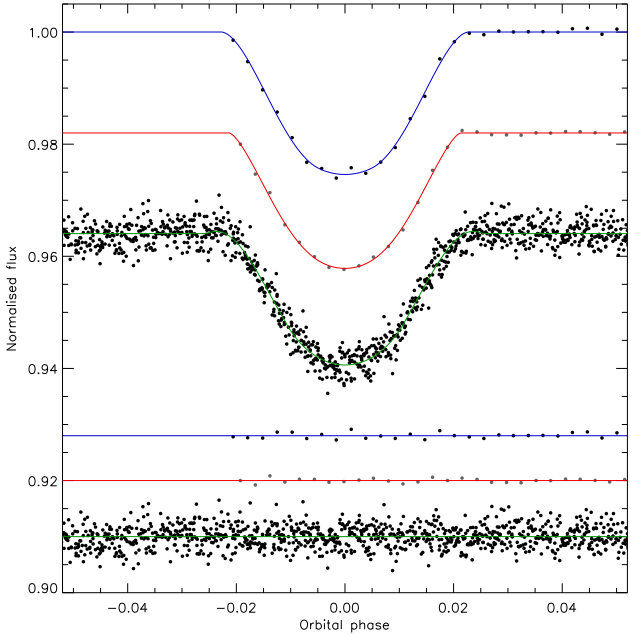}
\caption{\label{fig:tres3:lc} Phased light curves of TrES-3 compared to the best
fits found using {\sc jktebop} and the quadratic LD law. From top to bottom are
the 790.2\,nm then 794.4\,nm datasets from \citet{Colon+10mn}, and the EPOCH data
from \citet{Christiansen+11apj}. See Paper\,III (sect.\,4.8 and fig.\,12) for
previous results. See Fig.\,\ref{fig:80606:lc} for further details.} \end{figure}

TrES-3 was discovered by \citet{Odonovan+07apj} and previously studied in Paper\,III. Since then interleaved light curves of one transit have been published in two narrow passbands, with central wavelengths 790.2\,nm and 794.4\,nm, by \citet{Colon+10mn}, photometry of four transits has been presented by \citet{Lee+11pasj}, and a light curve from the {\it Deep Impact} mission has been obtained in the course of the EPOCH project \citep{Christiansen+11apj}. In the current work I improve upon the results from Paper\,III (see fig.\,12 in that work) by studying the data from \citet{Colon+10mn} and \citet{Christiansen+11apj}.

The 790.2\,nm and 794.4\,nm light curves cover the same transit by alternating between the two passbands using a tunable filter. In both cases (Tables A\arabic{appref}\addtocounter{appref}{1} and A\arabic{appref}\addtocounter{appref}{1}) correlated noise is unimportant and LD-fit/fix provides the best solution.

The EPOCH data cover six transits, of which one has only partial coverage and one has problems with systematic noise. The 1171 datapoints in the region of the remaining four transits were selected and four were rejected by a 3.5$\sigma$ clip. Table\,A\arabic{appref}\addtocounter{appref}{1} shows that the LD-fit/fix solutions return questionable results, so the LD-fixed solutions had to be adopted. Unusually for the EPOCH data, the residual-permutation errorbars were not larger than the Monte-Carlo ones.

The final results for each light curve are given in Table\,A\arabic{appref}\addtocounter{appref}{1}, alongside the results for the seven light curves investigated in Paper\,III. Fig.\,\ref{fig:tres3:lc} shows the best fits. The final photometric parameters were obtained by multiplying the probability density functions for the ten light curves from Paper\,III and the current work. The agreement between light curves and with literature values is excellent. The physical properties of TrES-3 can be found in Table\,A\arabic{appref}\addtocounter{appref}{1} and again show good correspondance with published numbers. TrES-3 is well-characterised, but would benefit from a few more RV measurements.


\subsection{WASP-3}                                                                                                            \label{sec:teps:wasp3}

\begin{figure} \includegraphics[width=\columnwidth,angle=0]{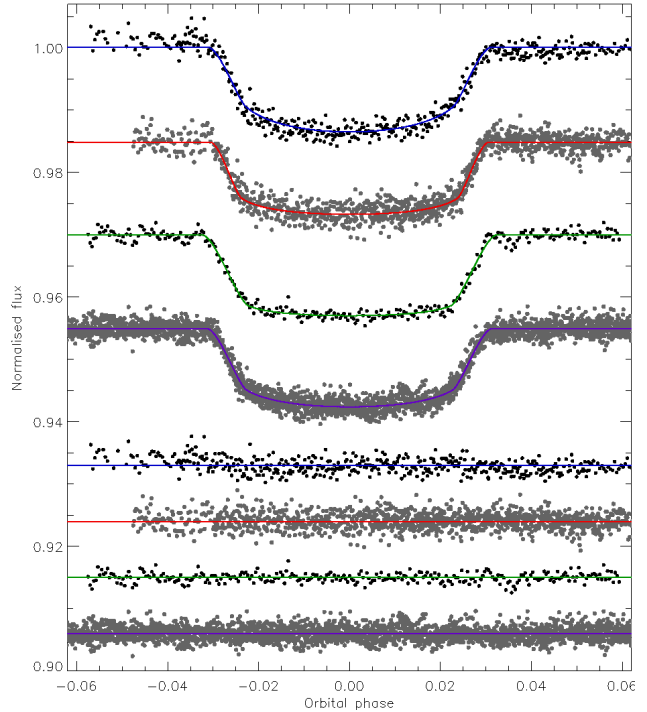}
\caption{\label{fig:wasp3:lc} Phased light curves of WASP-3 compared to the best
fits found using {\sc jktebop} and the quadratic LD law. From top to bottom the
datasets are $g$ then $i$ then $z$ from \citet{Tripathi+10apj}, and the EPOCH
data from \citet{Christiansen+11apj}. See Fig.\,\ref{fig:80606:lc} for further
details. See Paper\,III (fig.\,14 and sect.\,4.10) for previous results.} \end{figure}

WASP-3 was treated in Paper\,III but is revisited here, as it has since been observed from the ground by \citet{Tripathi+10apj} and from space by \citet{Christiansen+11apj}. Six nice transit light curves were obtained by \citet{Tripathi+10apj}, comprising three in $i$ and two in $g$ from the FLWO 1.2\,m and one in $z$ from the University of Hawaii 2.2\,m. The \citet{Christiansen+11apj} data come from EPOCH and cover eight transits with good precision but some systematics due to pointing wander and imperfect flat-fielding. \citet{Maciejewski+10mn} has detected possible TTVs in the WASP-3 system, which are yet to be independently confirmed.

I adopt the same \Teff\ and \FeH\ as in Paper\,I \citep{Pollacco+08mn}. Two RV studies exist for WASP-3, yielding velocity amplitudes of $290.5 \pm 9.5$\ms\ \citep{Tripathi+10apj} and $278.2 \pm 13.6$\ms\ \citep{Miller+10aa}. In Paper\,III I took the former of these two values, but in the current work I adopted instead the weighted mean: $K_{\rm A} = 286.5 \pm 7.8$\ms.

The $g$-band data contain two transits so the \Porb\ was included in the fit to insure against possible bias from TTV effects. The LD-fit/fix solutions are best and the residual-permutation errorbars are about 30\% larger than the Monte-Carlo ones (Table\,A\arabic{appref}\addtocounter{appref}{1}). The $i$-band data cover three transits so \Porb\ was again fitted for. The LD-fit/fix solutions are good and correlated noise is not important (Table\,A\arabic{appref}\addtocounter{appref}{1}). The $z$-band data encompass one transit; LD-fit/fix values were adopted and the residual-permutation errorbars are 25\% larger than Monte-Carlo (Table\,A\arabic{appref}\addtocounter{appref}{1}. The EPOCH data cover eight transits, of which one was rejected due to poor observational coverage, with 18\,622 datapoints. 3397 are near transit, of which 34 fell foul of a 3.5$\sigma$ clip. The LD-fit/fix results were retained and the residual-permutation errorbars are 40\% larger than the Monte Carlo ones (Table\,A\arabic{appref}\addtocounter{appref}{1}).

Fig.\,\ref{fig:wasp3:lc} shows best fits of the four light curves. Table\,A\arabic{appref}\addtocounter{appref}{1} summarises the photometric results from the current work and from Paper\,III. All seven light curves are used to calculate the final photometric parameters, which are mostly in good agreement. The $k$ values are more scattered than they should be ($\chir=5.4$) and this has a knock-on effect on $r_{\rm b}$ which has been accounted for in the errorbars. Other works are in good agreement with my results, albeit with unreasonably small errorbars in some cases. An exception is \citet{Miller+10aa} who find a discrepant solution with high inclination and thus lower $r_{\rm A}$. This situation propagates into the physical properties (Table\,A\arabic{appref}\addtocounter{appref}{1}), which are now substantially improved over those in Paper\,III. Further spectroscopic observations, both for RV and atmospheric parameter measurements, are warranted.


\subsection{Other TEPs}                                                                                                         \label{sec:teps:other}

I have returned to the XO-4 system, which has received a new and substantially improved $K_{\rm A}$ measurement from \citet{Narita+10pasj} since Paper\,III. Table\,A\arabic{appref}\addtocounter{appref}{1} shows the revision in the system parameters this brings.

Finally, I have checked that the modifications to the {\sc jktabsdim} code outlined in Sect.\,\ref{sec:absdim:model} (primarily the much better sampling in age) by rerunning solutions for the WASP-7 system \citep{Me+11aa}. The new results are almost identical to the old ones (Table\,A\arabic{appref}\addtocounter{appref}{1}), except for the correction to $\rho_{\rm b}$ discussed in Sect.\,\ref{sec:absdim:units}. The full set of homogeneous properties for transiting extrasolar planetary systems can now be accessed by considering only the current work and Paper\,III.


\section{Performance of the dEB relation versus constraints from theoretical models}                                               \label{sec:syserr}

\begin{figure*} \includegraphics[width=\textwidth,angle=0]{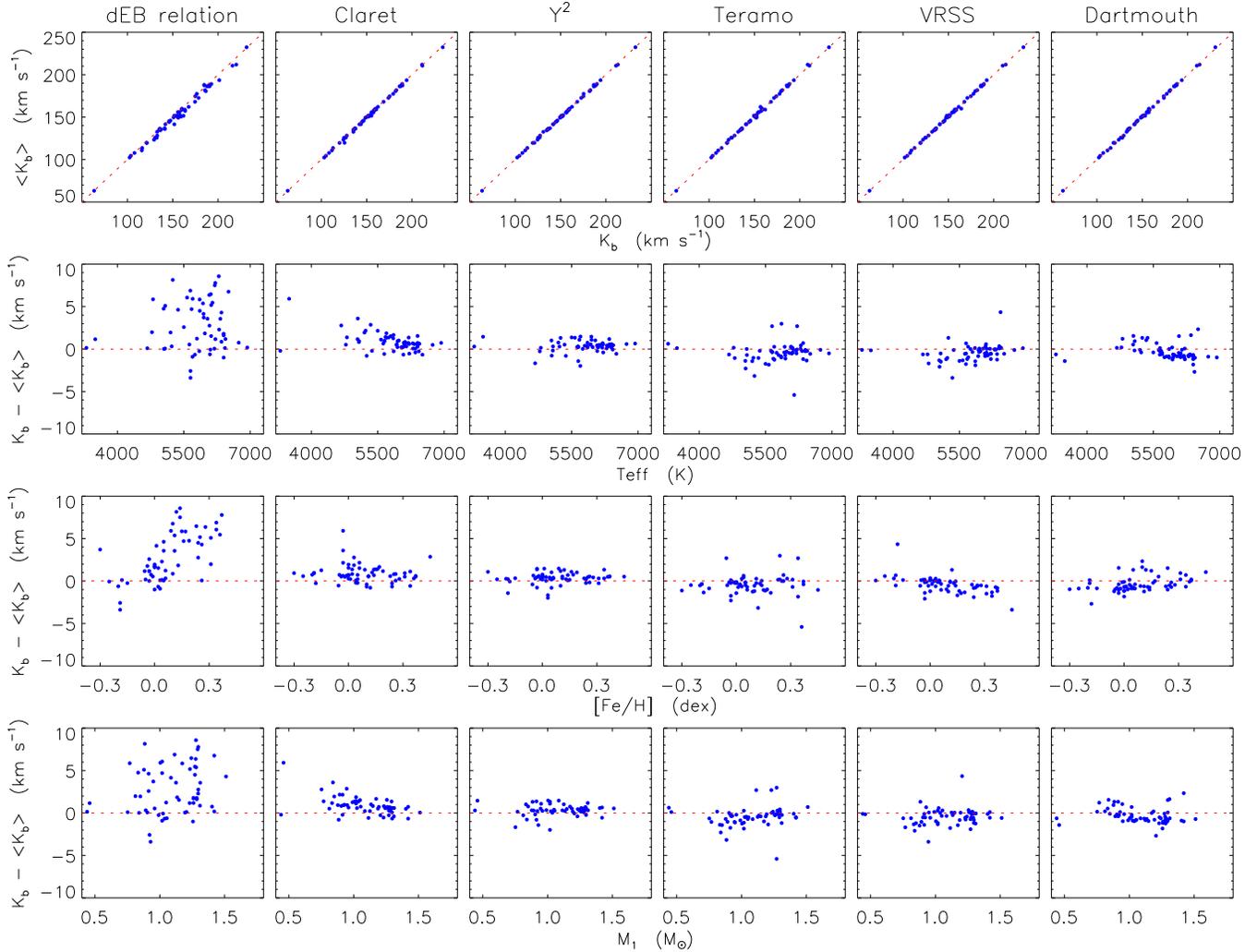}
\caption{\label{fig:syserr} Comparisons between the $K_{\rm b}$ values obtained
using individual sets of stellar evolutionary models and the unweighted mean value,
$\langle$$K_{\rm b}$$\rangle$, for each TEP. From left to right the panels show
results for the dEB constraint and then the five stellar model sets. The top
panels compare $K_{\rm b}$ to $\langle$$K_{\rm b}$$\rangle$ for each model set,
with parity indicated by a dotted line. Lower panels show the difference
($K_{\rm b} - \langle$$K_{\rm b}$$\rangle$) as a function of \Teff, \FeH\
and $M_{\rm A}$.} \end{figure*}

One of the new procedures introduced in the current work is an optional constraint using a $R_{\rm A} = f(M_{\rm A},\Teff,\FeH)$ relation obtained from well-studied dEBs and following the method of \citet{Enoch+10aa}. This replaces the approach used in Papers II and III, which utilised a mass--radius relation from dEBs, which was simpler but did not work very well. The cost is a reliance on \Teff\ and \FeH\ measurements, which incurs a dependence on theoretical model atmospheres. The new approach gives results in much better agreement with those found via theoretical models.

The dEB constraint has been used to calculate physical properties for the 30 TEPs studied in Paper\,III, giving a sample of \reff{58} TEPs with physical properties calculated in several ways: using the dEB constraint, using each of five different theoretical model tabulations, and the nominal results which are an unweighted mean of the ones from the five model sets. The parameter $K_{\rm b}$ is well-suited for comparing the different options, as it is the solution control parameter in {\sc jktabsdim} and wholly encompasses the outside constraints used in calculating the physical properties. A larger $K_{\rm b}$ results in larger numbers for all of the physical properties (see eqs.\ 4 to 12 in Paper\,II) with the exceptions of \safronov\ (which gets smaller) and the three quantities which have no model dependence ($g_{\rm b}$, $\rho_{\rm A}$ and \Teq).

Fig.\,\ref{fig:syserr} presents a detailed visualisation of the $K_{\rm b}$ values obtained from the various solutions for each TEP system. This Figure is a modified version of fig.\,20 in Paper\,III with double the number of systems and with the dEB constraint instead of the mass--radius constraint. Previous assertions can be confirmed: the {\it Claret} models yield a larger $K_{\rm b}$ on average, the VRSS models give a lower $K_{\rm b}$ on average, and the {\it Y$^2$} models show no trends with \Teff, \FeH\ and $M_{\rm A}$ compared to the mean model solution. The dEB constraint is clearly hugely more successful than the mass--radius relation in reproducing the theoretical model results, but tends to return a larger $K_{\rm b}$ particularly at high metallicity. This confirms that it is a useful tool in quick calculation of the properties of transiting planets.


\section{Physical properties of the transiting extrasolar planetary systems}                                                   \label{sec:properties}

\begin{figure} \includegraphics[width=\columnwidth,angle=0]{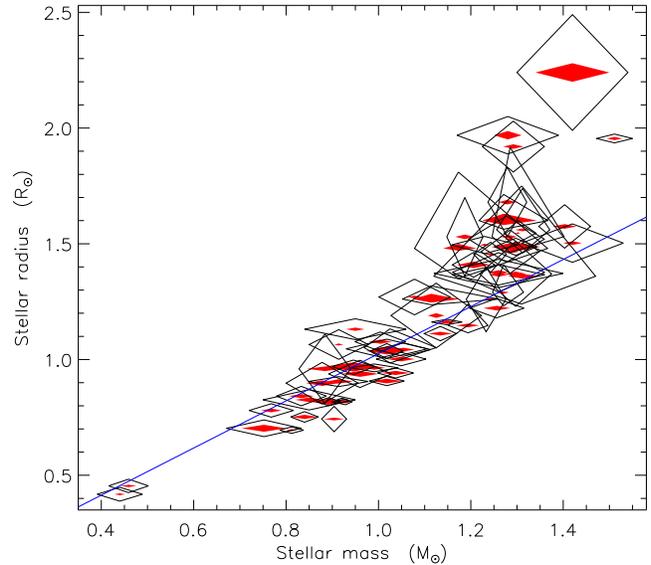}
\caption{\label{fig:absdim:M1R1} Plot of the masses versus the radii of the
stars in the \reff{58} TEPs with homogeneous properties. The statistical uncertainties
are shown by black open diamonds and the systematic uncertainties by red
filled diamonds. The empirical mass--radius relation from Paper\,II is shown
with a blue line.} \end{figure}

\begin{figure} \includegraphics[width=\columnwidth,angle=0]{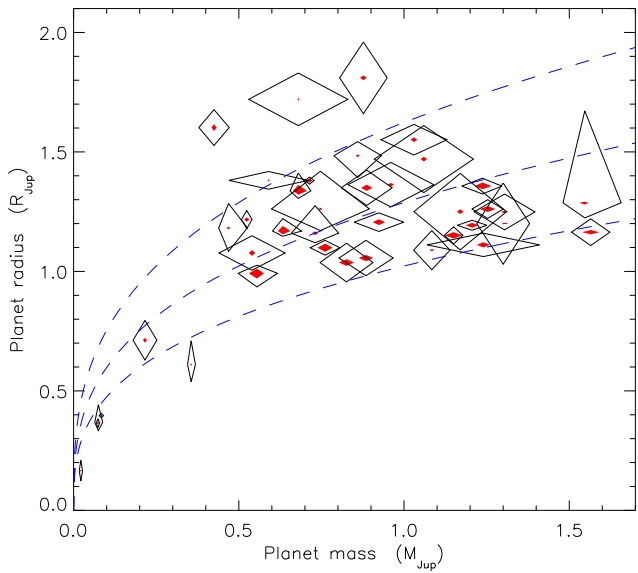}
\caption{\label{fig:absdim:M2R2} Plot of the masses versus the radii of
the planets in the \reff{58} TEPs with homogeneous properties. The statistical
uncertainties are shown by black open diamonds and the systematic
uncertainties by red filled diamonds. Blue dotted lines show where
density is 1.0, 0.5 and 0.25 \pjup.} \end{figure}

\begin{figure*} \includegraphics[width=\textwidth,angle=0]{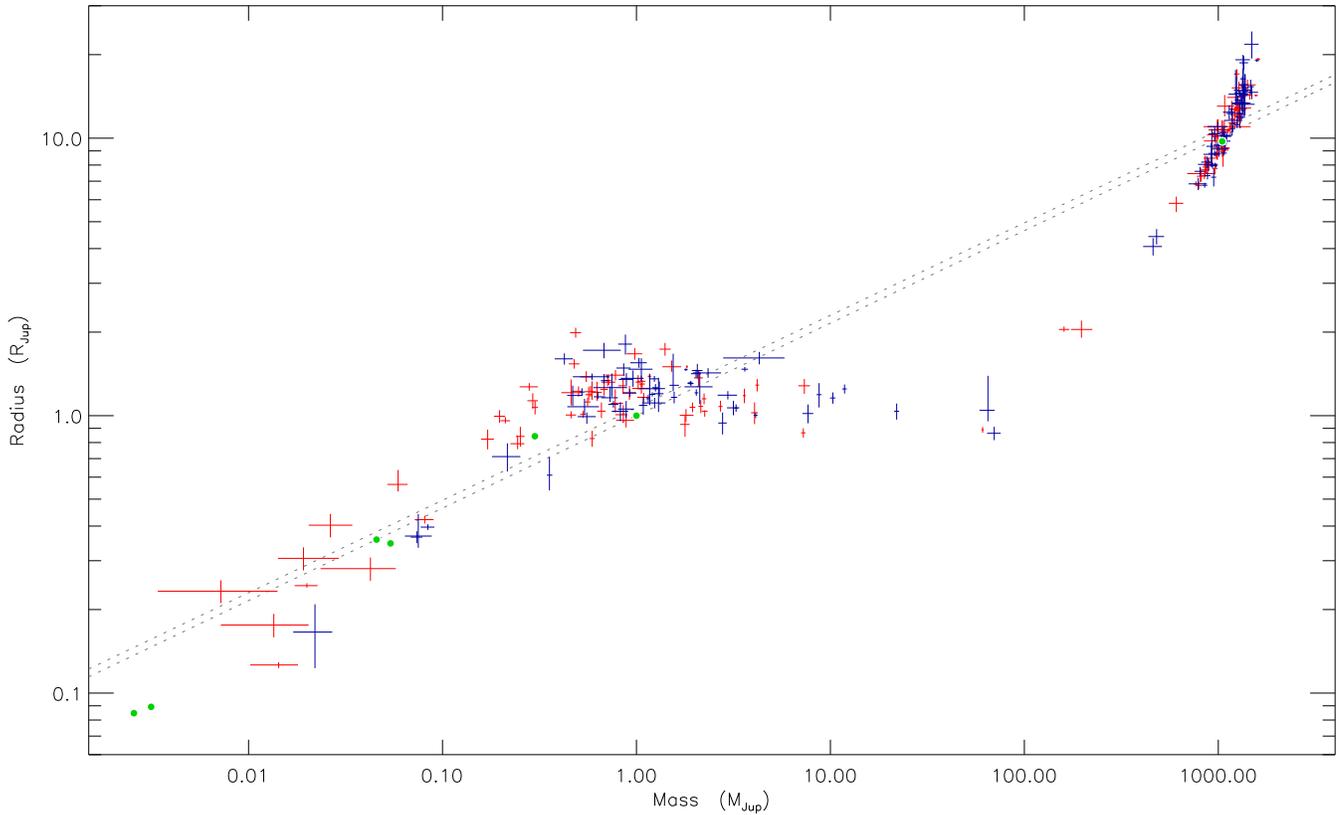}
\caption{\label{fig:absdim:both} Plot of the masses versus the radii of
all published TEPs and their host stars. Blue crosses show the objects
studied in the current series of papers and red crosses show results taken
from the literature. The Solar system bodies are shown by green filled
circles. The grey dotted lines show the loci where density is equal to
\pjup\ (upper line) and \psun\ (lower line).} \end{figure*}

\begin{table*} \caption{\label{tab:absdim:stars} Physical properties of the stellar
components of the TEPs studied in this work. For each quantity the first uncertainty
is derived from a propagation of all observational errors and the second uncertainty
is an estimate of the systematic errors arising from the dependence on stellar theory.}
\setlength{\tabcolsep}{3pt}
\begin{tabular}{l l@{$\pm$}l@{$\pm$}l l@{$\pm$}l@{$\pm$}l l@{$\pm$}l@{$\pm$}l l@{$\pm$}l@{$\pm$}l l@{$\pm$}l@{$\pm$}l l@{$\pm$}l@{$\pm$}l}
\hline \hline
System & \mcc{Semimajor axis (AU)} & \mcc{Mass (\Msun)} & \mcc{Radius (\Rsun)} & \mcc{$\log g_{\rm A}$ [cm/s]} & \mcc{Density (\psun)} & \mcc{Age (Gyr)} \\
\hline
\corot-1    & 0.02536   & 0.00098   & 0.00016     & 0.95      & 0.11      & 0.02        & 1.131     & 0.045     & 0.007       & 4.311     & 0.019     & 0.003       & \mcc{$0.660 \pm 0.019$}             & \ermcc{ 7.8}{ 4.0}{ 3.8}{ 0.7}{ 0.7}    \\
\corot-2    & 0.02854   & 0.00036   & 0.00032     & 1.018     & 0.038     & 0.034       & 0.907     & 0.020     & 0.010       & 4.530     & 0.015     & 0.005       & \mcc{$1.362 \pm 0.064$}            & \ermcc{ 0.6}{ 1.9}{ 2.1}{ 1.5}{ 0.6}    \\
\corot-3    & 0.05783   & 0.00078   & 0.00035     & 1.403     & 0.056     & 0.026       & 1.575     & 0.094     & 0.010       & 4.191     & 0.046     & 0.003       & \mcc{$0.359 \pm 0.058$}            & \ermcc{ 1.5}{ 0.5}{ 0.5}{ 0.3}{ 0.2}    \\
\corot-4    & \ermcc{0.09120}{0.00110}{0.00112}{0.00061}{0.00067}  & \ermcc{1.194}{0.043}{0.043}{0.024}{0.026}            & \ermcc{1.1475}{0.0923}{0.0322}{0.0077}{0.0084}        & \ermcc{4.3959}{0.0237}{0.0678}{0.0029}{0.0032}      & \ercc{0.790}{0.064}{0.161}                           & \ermcc{ 0.8}{ 2.6}{ 1.0}{ 0.6}{ 0.3}     \\[1pt]
\corot-5    & \ermcc{0.05004}{0.00161}{0.00092}{0.00022}{0.00033}  & \ermcc{1.025}{0.100}{0.056}{0.013}{0.020}            & \ermcc{1.0516}{0.0810}{0.0666}{0.0045}{0.0069}        & \ermcc{4.4053}{0.0683}{0.0594}{0.0019}{0.0028}      & \ercc{0.88}{0.21}{0.16}                              & \ermcc{ 3.9}{ 2.6}{ 5.3}{ 0.6}{ 1.0}     \\
\corot-6    & 0.0855    & 0.0016    & 0.0007      & 1.054     & 0.059     & 0.024       & 1.043     & 0.029     & 0.008       & 4.425     & 0.022     & 0.003       & \mcc{$0.929 \pm 0.064$}            & \ermcc{ 2.5}{ 2.1}{ 1.7}{ 0.6}{ 0.7}    \\
\corot-7    & 0.01690   & 0.00036   & 0.00025     & 0.884     & 0.056     & 0.039       & 0.96      & 0.15      & 0.01        & 4.42      & 0.14      & 0.01        & \mcc{$1.00 \pm 0.48$}              & \mcc{unconstrained}                     \\
\corot-8    & 0.0633    & 0.0019    & 0.0008      & 0.878     & 0.078     & 0.035       & 0.898     & 0.090     & 0.012       & 4.475     & 0.077     & 0.006       & \mcc{$1.21 \pm 0.32$}              & \mcc{unconstrained}                     \\
\corot-9    & 0.4027    & 0.0095    & 0.0056      & 0.960     & 0.068     & 0.040       & 0.938     & 0.059     & 0.013       & 4.476     & 0.063     & 0.006       & \mcc{$1.16 \pm 0.24$}              & \mcc{unconstrained}                     \\
\corot-10   & 0.1060    & 0.0011    & 0.0009      & 0.904     & 0.027     & 0.022       & 0.743     & 0.055     & 0.006       & 4.652     & 0.062     & 0.004       & \mcc{$2.20 \pm 0.47$}              & \ermcc{ 0.1}{ 2.0}{ 0.2}{ 0.3}{ 0.1}    \\
\corot-11   & 0.0440    & 0.0016    & 0.0003      & 1.26      & 0.14      & 0.02        & 1.374     & 0.061     & 0.009       & 4.264     & 0.019     & 0.003       & \mcc{$0.488 \pm 0.022$}            & \ermcc{2.0}{0.8}{2.1}{0.4}{0.4}    \\
\corot-12   & 0.0394    & 0.0011    & 0.0004      & 1.018     & 0.088     & 0.029       & 1.046     & 0.042     & 0.010       & 4.407     & 0.029     & 0.004       & \mcc{$0.889 \pm 0.076$}            & \ermcc{ 5.8}{ 3.3}{ 6.7}{ 1.8}{ 1.5}    \\
\corot-13   & 0.0510    & 0.0012    & 0.0005      & 1.086     & 0.077     & 0.035       & 1.274     & 0.077     & 0.014       & 4.264     & 0.040     & 0.005       & \mcc{$0.526 \pm 0.072$}            & \ermcc{5.8}{1.4}{6.2}{0.5}{1.0}    \\
\corot-14   & 0.02687   & 0.00077   & 0.00015     & 1.125     & 0.098     & 0.018       & 1.19      & 0.14      & 0.01        & 4.338     & 0.082     & 0.002       & \mcc{$0.67 \pm 0.19$}              & \ermcc{ 3.7}{ 2.5}{ 5.0}{ 0.7}{ 0.6}    \\
\corot-15   & \ermcc{0.0458}{0.0018}{0.0022}{0.0005}{0.0003}       & \ermcc{1.31}{0.16}{0.19}{0.04}{0.03}                 & \ermcc{1.36}{0.39}{0.12}{0.01}{0.01}                 & \ermcc{4.288}{0.059}{0.191}{0.005}{0.003}            & \ercc{0.52}{0.12}{0.25}                              & \ermcc{ 1.6}{ 4.5}{ 5.9}{ 0.9}{ 1.6}     \\[1pt]
HAT-P-4     & \ermcc{0.04465}{0.00113}{0.00062}{0.00084}{0.00054}  & \ermcc{1.271}{0.096}{0.053}{0.072}{0.046}            & \ermcc{1.600}{0.113}{0.037}{0.030}{0.019}            & \ermcc{4.134}{0.015}{0.038}{0.008}{0.005}            & \ercc{0.310}{0.016}{0.041}                           & \ermcc{ 3.9}{ 0.6}{ 0.9}{ 0.6}{ 1.1}     \\
HAT-P-7     & 0.03805   & 0.00033   & 0.00015     & 1.511     & 0.039     & 0.017       & 1.955     & 0.019     & 0.007       & 4.0354    & 0.0049    & 0.0017      & \mcc{$0.2023 \pm 0.0024$}          & \ermcc{2.0}{0.4}{0.3}{0.3}{0.2}    \\
HAT-P-11    & 0.05259   & 0.00056   & 0.00027     & 0.812     & 0.026     & 0.012       & 0.695     & 0.014     & 0.004       & 4.663     & 0.012     & 0.002       & \mcc{$2.415 \pm 0.097$}            & \mcc{unconstrained}                \\
HD\,17156   & 0.1637    & 0.0019    & 0.0022      & 1.297     & 0.046     & 0.053       & 1.487     & 0.037     & 0.020       & 4.207     & 0.018     & 0.006       & \mcc{$0.395 \pm 0.022$}            & \ermcc{2.8}{1.1}{0.6}{0.4}{0.4}    \\
HD\,80606   & 0.4564    & 0.0054    & 0.0068      & 1.018     & 0.035     & 0.045       & 1.037     & 0.032     & 0.015       & 4.415     & 0.021     & 0.007       & \mcc{$0.913 \pm 0.062$}            & \ermcc{5.9}{1.6}{2.2}{4.1}{2.1}    \\
Kepler-4    & \ermcc{0.0449}{0.0024}{0.0012}{0.0005}{0.0004}       & \ermcc{1.173}{0.193}{0.095}{0.039}{0.033}            & \ermcc{1.48}{0.33}{0.13}{0.02}{0.01}                 & \ermcc{4.168}{0.063}{0.133}{0.005}{0.004}            & \ercc{0.362}{0.081}{0.136}                           & \ermcc{ 5.3}{ 1.5}{ 2.5}{ 0.4}{ 0.4}     \\[1pt]
Kepler-5    & \ermcc{0.04967}{0.00051}{0.00038}{0.00014}{0.00028}  & \ermcc{1.296}{0.040}{0.030}{0.011}{0.022}            & \ermcc{1.544}{0.055}{0.042}{0.004}{0.009}            & \ermcc{4.174}{0.023}{0.024}{0.001}{0.002}            & \ercc{0.352}{0.025}{0.029}                           & \ermcc{ 2.8}{ 0.3}{ 0.5}{ 0.3}{ 0.3}                       \\
Kepler-6    & \ermcc{0.04438}{0.00181}{0.00081}{0.00080}{0.00053}  & \ermcc{1.114}{0.146}{0.062}{0.061}{0.040}            & \ermcc{1.261}{0.057}{0.024}{0.023}{0.015}            & \ermcc{4.284}{0.017}{0.011}{0.008}{0.005}            & \ercc{0.5555}{0.0076}{0.0209}                        & \ermcc{ 5.7}{ 0.8}{ 2.3}{ 0.3}{ 0.5}     \\
Kepler-7    & 0.0613    & 0.0017    & 0.0006      & 1.28      & 0.11      & 0.03        & 1.969     & 0.081     & 0.018       & 3.959     & 0.024     & 0.004       & \mcc{$0.168 \pm 0.012$}            & \ermcc{4.4}{0.4}{1.6}{0.7}{0.4}    \\
Kepler-8    & 0.0485    & 0.0012    & 0.0002      & 1.230     & 0.072     & 0.01        & 1.495     & 0.037     & 0.005       & 4.178     & 0.022     & 0.002       & \mcc{$0.368 \pm 0.014$}            & \ermcc{3.2}{1.8}{3.1}{0.6}{1.1}    \\
KOI-428     & 0.0795    & 0.0022    & 0.0015      & 1.42      & 0.12      & 0.08        & 2.24      & 0.25      & 0.04        & 3.889     & 0.099     & 0.008       & \mcc{$0.126 \pm 0.042$}            & \ermcc{2.1}{0.6}{0.6}{0.1}{0.3}    \\
LHS\,6343   & 0.0850    & 0.0031    & 0.0007      & 0.440     & 0.049     & 0.012       & 0.418     & 0.030     & 0.004       & 4.839     & 0.035     & 0.004       & \mcc{$6.01 \pm 0.76$}               & \ermcc{4.0}{0.7}{0.0}{2.0}{3.0}    \\
TrES-2      & 0.03567   & 0.00061   & 0.00029     & 0.991     & 0.052     & 0.024       & 0.964     & 0.017     & 0.008       & 4.4660    & 0.0081    & 0.0035      & \mcc{$1.105 \pm 0.011$}            & \ermcc{3.4}{2.0}{2.2}{0.5}{0.5}    \\
TrES-3      & 0.02276   & 0.00012   & 0.00011     & 0.921     & 0.014     & 0.014       & 0.8235    & 0.0098    & 0.0040      & 4.5710    & 0.0064    & 0.0021      & \mcc{$1.648 \pm 0.041$}            & \ermcc{1.0}{0.5}{0.0}{0.0}{0.0}    \\
WASP-3      & 0.03185   & 0.00086   & 0.00020     & 1.26      & 0.10      & 0.02        & 1.366     & 0.044     & 0.008       & 4.268     & 0.018     & 0.003       & \mcc{$0.495 \pm 0.024$}            & \ermcc{2.1}{1.2}{1.2}{0.4}{0.4}    \\
WASP-7      & 0.0619    & 0.0010    & 0.0003      & 1.285     & 0.063     & 0.019       & 1.466     & 0.094     & 0.007       & 4.215     & 0.046     & 0.002       & \mcc{$0.408 \pm 0.068$}            & \ermcc{2.5}{0.8}{0.9}{0.2}{0.4}    \\
XO-4        & \ermcc{0.05474}{0.00162}{0.00056}{0.00020}{0.00031}  & \ermcc{1.285}{0.117}{0.039}{0.014}{0.022}            & \ermcc{1.531}{0.386}{0.068}{0.006}{0.009}            & \ermcc{4.177}{0.034}{0.172}{0.002}{0.002}            & \ercc{0.358}{0.046}{0.160}                           & \ermcc{ 2.7}{ 1.1}{ 0.5}{ 0.2}{ 0.3}                \\
\hline \hline \end{tabular} \end{table*}

\begin{table*} \caption{\label{tab:absdim:planets} Physical properties of the planetary
components of the TEPs studied in this work. For each quantity the first uncertainty is
derived from a propagation of all observational errors and the second uncertainty is an
estimate of the systematic errors arising from the dependence on stellar theory.}
\setlength{\tabcolsep}{4pt}
\begin{tabular}{l l@{\,$\pm$\,}l@{\,$\pm$\,}l l@{\,$\pm$\,}l@{\,$\pm$\,}l l@{\,$\pm$\,}l@{\,$\pm$\,}l l@{\,$\pm$\,}l@{\,$\pm$\,}l l@{\,$\pm$\,}l@{\,$\pm$\,}l l@{\,$\pm$\,}l@{\,$\pm$\,}l}
\hline \hline
System & \mcc{Mass (\Mjup)} & \mcc{Radius (\Rjup)} & \mcc{$g_{\rm b}$ (\mss)} & \mcc{Density (\pjup)} & \mcc{\Teq\ (K)} & \mcc{\safronov} \\
\hline
\corot-1    & 1.03      & 0.10      & 0.01        & 1.551     & 0.064     & 0.010       & \mcc{$10.65 \pm  0.69$}            & 0.259     & 0.021     & 0.002       & \mcc{$1915 \pm   49$}              & 0.0354    & 0.0025    & 0.0002       \\
\corot-2    & 3.62      & 0.14      & 0.08        & 1.470     & 0.028     & 0.016       & \mcc{$41.5 \pm  1.7$}              & 1.066     & 0.057     & 0.012       & \mcc{$1548 \pm   22$}              & 0.1381    & 0.0049    & 0.0016       \\
\corot-3    & 21.96     &  0.65     &  0.27       & 1.037     & 0.069     & 0.006       & \mcc{$506 \pm  67$}                & 18.4      &  3.7      &  0.1        & \mcc{$1695 \pm   57$}              & 1.74      & 0.12      & 0.01         \\
\corot-4    & \ermcc{0.731}{0.072}{0.073}{0.0010}{0.011}           & \ermcc{1.160}{0.116}{0.041}{0.008}{0.009}            & \ercc{13.5}{ 1.6}{ 2.6}                                   & \ermcc{0.438}{0.063}{0.117}{0.003}{0.003}            & \ercc{1058}{  42}{  17}                                   & \ermcc{0.0962}{0.0099}{0.0127}{0.0007 }{0.0006}        \\[1pt]
\corot-5    & \ermcc{0.470}{0.058}{0.031}{0.004}{0.006}            & \ermcc{1.182}{0.102}{0.098}{0.005}{0.008}            & \ercc{8.3}{1.8}{1.3}                                      & \ermcc{0.266}{0.082}{0.058}{0.002}{0.001}            & \ercc{1348}{  50}{  51}                                   & \ermcc{0.0388}{0.0054}{0.0038}{0.0003}{0.0002}        \\
\corot-6    & 2.96      & 0.34      & 0.05        & 1.185     & 0.041     & 0.009       & \mcc{$52.3 \pm  6.4$}              & 1.66      & 0.23      & 0.01        & \mcc{$1025 \pm   16$}              & 0.405     & 0.046     & 0.003        \\
\corot-7    & 0.0220    & 0.0050    & 0.0007      & 0.166     & 0.043     & 0.002       & \mcc{$19 \pm 12$}                  & 4.5       & 4.5       & 0.1         & \mcc{$1910 \pm  140$}              & 0.0051    & 0.0018    & 0.0001       \\
\corot-8    & 0.216     & 0.036     & 0.006       & 0.712     & 0.083     & 0.010       & \mcc{$10.6 \pm  2.9$}              & 0.56      & 0.21      & 0.01        & \mcc{$922 \pm  41$}                & 0.0437    & 0.0084    & 0.0006       \\
\corot-9    & 0.826     & 0.080     & 0.023       & 1.037     & 0.081     & 0.014       & \mcc{$19.1 \pm  3.2$}              & 0.69      & 0.17      & 0.01        & \mcc{$413 \pm  14$}                & 0.668     & 0.076     & 0.009        \\
\corot-10   & 2.78      & 0.14      & 0.05        & 0.941     & 0.085     & 0.008       & \mcc{$78 \pm 14$}                  & 3.13      & 0.88      & 0.03        & \mcc{$647 \pm  24$}                & 0.693     & 0.070     & 0.006        \\
\corot-11   & 2.34      & 0.39      & 0.03        & 1.426     & 0.057     & 0.009       & \mcc{$28.5 \pm  4.2$}              & 0.76      & 0.12      & 0.00        & \mcc{$1735 \pm   34$}              & 0.114     & 0.017     & 0.001        \\
\corot-12   & 0.887     & 0.077     & 0.017       & 1.350     & 0.074     & 0.013       & \mcc{$12.1 \pm  1.3$}              & 0.337     & 0.052     & 0.003       & \mcc{$1410 \pm   28$}              & 0.0508    & 0.0042    & 0.0005       \\
\corot-13   & 1.312     & 0.092     & 0.028       & 1.252     & 0.075     & 0.013       & \mcc{$20.7 \pm  2.5$}              & 0.62      & 0.11      & 0.01        & \mcc{$1432 \pm   39$}              & 0.0983    & 0.0080    & 0.0010       \\
\corot-14   & 7.67      & 0.49      & 0.08        & 1.018     & 0.079     & 0.005       & \mcc{$183 \pm  27$}                & 6.8       & 1.5       & 0.0         & \mcc{$1936 \pm   95$}              & 0.360     & 0.030     & 0.002        \\
\corot-15   & \ermcc{64.9}{ 5.3}{ 6.2}{ 1.3}{ 1.0}                 & \ermcc{1.045}{0.347}{0.091}{0.011}{0.008}            & \ercc{1470}{ 240}{ 620}                              & \ermcc{53}{13}{29}{ 0}{ 0}                           & \ercc{1670}{ 200}{  80}                              & \ermcc{4.34}{0.41}{1.07}{0.03}{0.05}               \\[1pt]
HAT-P-4     & \ermcc{0.680}{0.038}{0.025}{0.026}{0.016}            & \ermcc{1.337}{0.075}{0.032}{0.025}{0.016}            & \ercc{9.42}{0.44}{0.91}                              & \ermcc{0.266}{0.018}{0.038}{0.003}{0.005}            & \ercc{1691}{  46}{  26}                              & \ermcc{0.0357}{0.0012}{0.0021}{0.0004}{0.0007}        \\
HAT-P-7     & 1.799     & 0.038     & 0.014       & 1.465     & 0.015     & 0.006       & \mcc{$20.77 \pm  0.33$}            & 0.535     & 0.011     & 0.002       & \mcc{$2194 \pm   27$}              & 0.0618    & 0.0010    & 0.0002       \\
HAT-P-11    & 0.084     & 0.007     & 0.001       & 0.397     & 0.009     & 0.002       & \mcc{$13.2 \pm  1.1$}              & 1.26      & 0.12      & 0.01        & \mcc{$838 \pm  10$}                & 0.0274    & 0.0022    & 0.0001       \\
HD\,17156   & 3.262     & 0.072     & 0.088       & 1.065     & 0.033     & 0.014       & \mcc{$71.2 \pm  3.7$}              & 2.52      & 0.20      & 0.03        & \mcc{$883 \pm  11$}                & 0.772     & 0.026     & 0.010        \\
HD\,80606   & 4.114     & 0.096     & 0.122       & 1.003     & 0.023     & 0.015       & \mcc{$101.4 \pm   3.9$}            & 3.82      & 0.23      & 0.06        & \mcc{$405.0 \pm   7.0$}            & 3.677     & 0.093     & 0.055        \\
Kepler-4    & \ermcc{0.075}{0.013}{0.011}{0.002}{0.001}            & \ermcc{0.368}{0.074}{0.034}{0.004}{0.003}            & \ercc{13.8}{ 3.3}{ 4.5}                              & \ermcc{1.41}{0.48}{0.62}{0.01}{0.02}                 & \ercc{1620}{ 140}{  60}                              & \ermcc{0.0156}{0.0025}{0.0034}{0.0001}{0.0002}     \\[1pt]
Kepler-5    & \ermcc{2.040}{0.048}{0.040}{0.006}{0.007}            & \ermcc{1.210}{0.035}{0.030}{0.002}{0.002}            & \ercc{34.5}{ 1.7}{ 1.9}                              & \ermcc{1.076}{0.080}{0.086}{0.002}{0.001}            & \ercc{1692}{  29}{  25}                              & \ermcc{0.1286}{0.0036}{0.0040}{0.0002}{0.0002}     \\[1pt]
Kepler-6    & \ermcc{0.633}{0.057}{0.031}{0.023}{0.015}            & \ermcc{1.169}{0.052}{0.022}{0.021}{0.014}            & \ercc{11.48}{ 0.39}{ 0.54}                           & \ermcc{0.370}{0.015}{0.026}{0.004}{0.007}            & \ercc{1451}{  15}{  13}                              & \ermcc{0.0431}{0.0016}{0.0022}{0.0005}{0.0008}        \\
Kepler-7    & 0.425     & 0.046     & 0.008       & 1.602     & 0.075     & 0.014       & \mcc{$4.10 \pm 0.43$}              & 0.097     & 0.013     & 0.001       & \mcc{$1621 \pm   23$}              & 0.0253    & 0.0024    & 0.0002       \\
Kepler-8    & 0.59      & 0.12      & 0.00        & 1.381     & 0.037     & 0.005       & \mcc{$7.7 \pm 1.4$}                & 0.210     & 0.040     & 0.001       & \mcc{$1662 \pm   41$}              & 0.0337    & 0.0063    & 0.0001       \\
KOI-428     & 2.12      & 0.35      & 0.08        & 1.27      & 0.17      & 0.02        & \mcc{$32 \pm 10$}                  & 0.98      & 0.44      & 0.02        & \mcc{$1666 \pm   92$}              & 0.188     & 0.038     & 0.003        \\
LHS\,6343   & 69.9      &  5.6      &  1.2        & 0.864     & 0.048     & 0.007       & \mcc{$2320 \pm  210$}              & 101       &  13       &   0         & \mcc{$352 \pm  22$}                & 31.2      &  2.2      &  0.3         \\
TrES-2      & 1.206     & 0.045     & 0.020       & 1.193     & 0.021     & 0.010       & \mcc{$21.02 \pm  0.31$}            & 0.665     & 0.015     & 0.005       & \mcc{$1466 \pm   12$}              & 0.0727    & 0.0017    & 0.0006       \\
TrES-3      & 1.899     & 0.060     & 0.019       & 1.310     & 0.019     & 0.006       & \mcc{$27.4 \pm  1.1$}              & 0.790     & 0.040     & 0.004       & \mcc{$1638 \pm   22$}              & 0.0716    & 0.0024    & 0.0004       \\
WASP-3      & 2.03      & 0.12      & 0.03        & 1.416     & 0.047     & 0.009       & \mcc{$25.1 \pm  1.2$}              & 0.669     & 0.047     & 0.004       & \mcc{$2020 \pm   35$}              & 0.0724    & 0.0031    & 0.0004       \\
WASP-7      & 0.96      & 0.13      & 0.01        & 1.363     & 0.093     & 0.007       & \mcc{$12.9 \pm  2.4$}              & 0.356     & 0.087     & 0.002       & \mcc{$1502 \pm   47$}              & 0.068     & 0.010     & 0.000        \\
XO-4        & \ermcc{1.547}{0.110}{0.066}{0.011}{0.017}            & \ermcc{1.287}{0.385}{0.063}{0.005}{0.007}            & \ercc{23.1}{ 2.5}{ 9.4}                              & \ermcc{0.68}{0.11}{0.37}{0.00}{0.00}                 & \ercc{1630}{ 170}{  40}                              & \ermcc{0.1023}{0.0065}{0.0240}{0.0006}{0.0004}        \\
\hline \hline \end{tabular} \end{table*}

The major results of this work are the physical properties of \reff{32} transiting extrasolar planetary systems obtained using homogeneous methods and by combining all available photometric data (Table\,\ref{tab:teps:lcpar}) with measured spectroscopic parameters of the host stars (Table\,\ref{tab:teps:spec}). The stellar properties are given in Table\,\ref{tab:absdim:stars} and the planetary ones in Table\,\ref{tab:absdim:planets}. These quantities supplement (and in few cases supersede) the properties for 30 objects found in Paper\,III, giving a total sample of \reff{58} systems. The homogeneous nature of these results means they are well suited for comparing different TEPs, for planning follow-up observations, and for performing detailed statistical studies.

Figs. \ref{fig:absdim:M1R1} and \ref{fig:absdim:M2R2} show the masses and radii of the stars and planets with their random (black open diamonds) and systematic (red filled diamonds) errorbars. It is clear that the property of these four which is most affected by systematic error is $M_{\rm A}$, whereas the masses and radii of the planets are not strongly affected by this model dependence, as previously found in Paper\,III.

Fig.\,\ref{fig:absdim:both} shows the masses and radii of the TEPs and their host stars on the same plot. The plot includes all 118 TEP systems known as of 2011/04/19 and includes the \reff{58} systems studied in this series of papers plus results taken from the literature for the other \reff{60} systems. The Sun, Jupiter, Saturn, Uranus, Neptune, Earth and Venus are also plotted for context, as are lines denoting the points where density is equal to \pjup\ and \psun. Fig.\,\ref{fig:absdim:both} clearly highlights the wide range of parameter space covered by these systems, as well as the fact that the properties of the planets are much more scattered than those of the parent stars.


\section{Follow-up observations}                                                                                                 \label{sec:followup}

\begin{table} \centering \caption{\label{tab:absdim:obs} Summary
of which types of additional observations would be useful for the
thirty TEPs studied in this work. $\star$ denotes where additional
data would be useful, and $\star\star$ indicates where it would be
useful but difficult to either obtain or interpret.}
\setlength{\tabcolsep}{10pt}
\begin{tabular}{lccc}
\hline \hline
System      & Photometric  & Radial       & Spectral     \\
            & observations & velocities   & synthesis    \\
\hline
\corot-1    &              & $\star$      & $\star$      \\
\corot-2    &              & $\star$      & $\star$      \\
\corot-3    & $\star\star$ &              & $\star$      \\
\corot-4    & $\star\star$ & $\star$      &              \\
\corot-5    &              & $\star$      & $\star$      \\
\corot-6    &              & $\star$      & $\star$      \\
\corot-7    & $\star\star$ & $\star\star$ &              \\
\corot-8    & $\star$      & $\star\star$ &              \\
\corot-9    &              &              &              \\
\corot-10   & $\star\star$ &              &              \\
\corot-11   &              & $\star\star$ & $\star$      \\
\corot-12   & $\star\star$ &              & $\star$      \\
\corot-13   &              & $\star$      &              \\
\corot-14   & $\star\star$ &              & $\star\star$ \\
\corot-15   &              & $\star\star$ & $\star\star$ \\
HAT-P-4     & $\star$      &              &              \\
HAT-P-7     &              &              &              \\
HAT-P-11    &              &              &              \\
HD\,17156   &              &              &              \\
HD\,80606   &              &              &              \\
Kepler-4    & $\star$      & $\star$      & $\star$      \\
Kepler-5    &              &              &              \\
Kepler-6    &              & $\star$      &              \\
Kepler-7    & $\star$      &              &              \\
Kepler-8    &              & $\star$      & $\star$      \\
KOI-428     & $\star$      & $\star$      & $\star$      \\
LHS\,6343   & $\star$      & $\star$      & $\star$      \\
TrES-2      &              &              & $\star$      \\
TrES-3      & $\star$      &              &              \\
WASP-3      &              & $\star$      & $\star$      \\
WASP-7      & $\star$      & $\star$      & $\star$      \\
XO-4        & $\star$      &              &              \\
\hline \hline \end{tabular} \end{table}

Most of the TEPs in the current work would benefit from further observations of some sort. In many cases the dominant uncertainty stems from the quality of the light curve. This remains true for many of the \corot\ systems, despite their space-based light curves. It must be remembered that \corot\ has only a 27\,cm diameter telescope and studies relatively faint stars, so is subject to significant photon noise. Also, several of the \corot\ TEPs have few observed transits because they were studied in short runs (\corot-4 and \corot-15) or because they have long orbital periods (\corot-9 and \corot-10). Almost all of the \reff{58} TEP systems in this series of papers which do not need better light curves have been observed from space (as well as from the ground in many cases).

Additional RV measurements are useful too. In many circumstances, particularly for the fainter objects, the RVs are good enough to unambiguously confirm the planetary nature of a system but are the dominant source of uncertainty in the planetary masses. Now over 100 TEPs are known it seems appropriate to concentrate follow-up resources on measuring the physical properties of a golden subset of these to high precision. An additional requirement of RVs is definition of the orbital shape ($e$ and $\omega$), and imprecise measurements of these quantities compromise measurements of the photometric parameters, in particular $r_{\rm A}$.

The physical properties of quite a few of the TEPs are also limited by the precision of the \Teff\ and \FeH\ measurements available. In many cases this can be improved, but in some cases this is not an option because the errorbars are already close to the limit set by our understanding of low-mass stars (taken to be 50\,K in \Teff\ and 0.05\,dex in \FeH). There is no immediate prospect of lowering these thresholds; in fact there is evidence that they are already slightly optimistic \citep{Bruntt+10aa}.

Quite a few of the \corot\ TEPs have ephemerides which will become uncertain over the timescale of a few years, so need to be followed up soon to quash possible ephemeris drift. To investigate this I compiled a catalogue of ephemerides of all known TEPs and identified the first predicted times of transit which were uncertain by one hour, and by half of one transit duration. A list of objects for which one of these dates is earlier than the year 2020 is given in Table\,\ref{tab:absdim:eph}. The \kepler\ planets are separated because they continue to be observed and will have significantly improved ephemerides even with existing (unreleased) data, and in some cases have strong TTVs. The list of the other planets is clearly dominated by \corot\ objects, and it is notable that the ephemerides for \corot-4 and \corot-14 are {\rm already} uncertain by more than one hour. Further photometric observations of these are advocated before the ephemerides deteriorate much further.

\begin{table} \centering \caption{\label{tab:absdim:eph} Limits of
the current ephemerides of the known TEPs. The two dates for each
TEPs indicate the first transits whose midpoints are uncertain by
1\,hour and by half the transit duration. RHJD $=$ HJD $-$ 24000000.
The \kepler\ planets are separated because they have all continued to
be observed by the satellite so are not at risk from ephemeris drift.}
\setlength{\tabcolsep}{4pt}
\begin{tabular}{lcccc}
\hline \hline
TEP   & \mc{1\,hour uncertainty} & \mc{Half-transit uncertainty} \\
      &     RHJD & UT date       &       RHJD & UT date          \\
\hline
    \corot-4 & 55181.1958 & 2009\,12\,15 & 56432.6746 & 2013\,05\,20 \\
   \corot-14 & 55268.5299 & 2010\,03\,13 & 55186.8744 & 2009\,12\,21 \\
    \corot-7 & 55879.0469 & 2011\,11\,13 & 55231.1759 & 2010\,02\,03 \\
   \corot-10 & 57040.6290 & 2015\,01\,18 & 58377.9296 & 2018\,09\,16 \\
    \corot-9 & 57461.5587 & 2016\,03\,14 & 66131.4745 & 2039\,12\,08 \\
     WASP-22 & 58443.6491 & 2018\,11\,21 & 60821.1495 & 2025\,05\,25 \\
 OGLE-TR-211 & 58524.9886 & 2019\,02\,10 & 64688.0429 & 2035\,12\,26 \\
    \corot-8 & 58780.2836 & 2019\,10\,13 & 60451.4141 & 2024\,05\,20 \\
     Qatar-1 & 59217.5962 & 2021\,01\,03 & 58499.0595 & 2019\,01\,15 \\
\hline
   Kepler-9d & 56488.4815 & 2013\,07\,14 & 56466.1816 & 2013\,06\,22 \\
   Kepler-9c & 57381.9738 & 2015\,12\,25 & 59677.5818 & 2022\,04\,08 \\
    Kepler-7 & 57453.9928 & 2016\,03\,06 & 61416.1291 & 2027\,01\,10 \\
  Kepler-11f & 57579.2193 & 2016\,07\,09 & 63602.0693 & 2033\,01\,04 \\
  Kepler-11b & 57609.2652 & 2016\,08\,08 & 60360.3665 & 2024\,02\,19 \\
  Kepler-10c & 57644.0723 & 2016\,09\,12 & 64121.2358 & 2034\,06\,07 \\
    Kepler-4 & 58481.9941 & 2018\,12\,29 & 61933.4628 & 2028\,06\,10 \\
  Kepler-11d & 59473.5186 & 2021\,09\,16 & 67572.8455 & 2043\,11\,19 \\
  Kepler-11g & 59618.6442 & 2022\,02\,08 & 76309.9056 & 2067\,10\,21 \\
  Kepler-11e & 59754.5481 & 2022\,06\,24 & 65289.8388 & 2037\,08\,19 \\
\hline \hline \end{tabular} \end{table}


\section{Summary}                                                                                                            \label{sec:teps:summary}

The physical properties of 32 transiting extrasolar planetary systems have been derived from public light curves and published spectroscopic parameters of the host stars. These include 15 systems observed by the \corot\ satellite, ten by \kepler, and five by the EPOCH project on the {\it Deep Impact} spacecraft. Combined with the 30 objects examined in Paper\,III, a sample of \reff{58} TEPs with homogeneously measured properties is obtained.

All available transit light curves of each TEP were obtained and modelled using the {\sc jktebop} code, with careful attention paid to the treatment of limb darkening, contaminating light, orbital eccentricity, Poisson and correlated noise, and long effective exposure times. The results for each light curve were then amalgamated to yield combined photometric parameters for the system, which were compared with literature results.

The physical properties of the TEPs were calculated from measured quantities by applying constraints from theoretical models, guided by the atmospheric parameters of the host stars. Five different sets of theoretical model tabulations were used, and the final results for each TEP are the unweighted mean of the individual results for each output parameter. Systematic errors were estimated by the interagreement between the individual model results, and statistical errors were propagated using a perturbation analysis. The constants and units needed in this process were tabulated for reference, and an error in the unit used for planetary density was fixed.

I also calculated the physical properties of each TEP system using a constraint obtained from eclipsing binary star systems (see also \citealt{Enoch+10aa}). The constraint was applied in the form of $\lten R = f(\lten\Teff, \lten\rho, \MoH)$, where the precise equation and calibration coefficients were determined using the the measured properties of 90 well-studied detached eclipsing binaries. This gives results in generally good agreement with those from using theoretical stellar models as a constraint, although a trend towards poorer agreement is seen at higher metallicities. It is not obvious whether this trend arises from an imperfection in the calibration equation or source data, or from the physical effects included in the theoretical models.

The resulting physical properties of the 32 TEP systems are typically in good agreement with published results, but exceptions exist. My results for \corot-5 disagree with those of the discovery paper, and this may be related to the treatment of orbital eccentricity. The public light curve of \corot-8 does not match the published orbital ephemeris: I measure a revised ephemeris and somewhat different properties compared to the discovery paper. The two \corot\ light curves (short and long cadence) of \corot-13 are discrepant. After rejection of the much less reliable 512\,s cadence data I find physical properties of the system which are very different to previously thought. The resulting density of the planet is almost a factor of three smaller, moving it from outlier status to more representative of the general population of transiting Hot Jupiters. Many of the error estimates in the literature are far smaller than I find, and are not supported by the intrinsic quality of the data.

My analysis of the TEPs observed by \kepler\ uses data from Quarters 0, 1 and 2 for most of the objects. It is therefore the first analysis of most of the TEPs to include short-cadence data. This allows me to provide updated ephemerides and more reliable physical properties. My results for Kepler-5 are somewhat different to those previously published, due primarily to the inclusion of the Quarter 2 short-cadence data.

Asteroseismic studies are available for the three previously-known TEPs in the \kepler\ field, based on the \kepler\ short-cadence data, and for HD\,17156 based on HST data. These studies use theoretical stellar models to interpret the oscillation spectrum of the star, and measure the stellar density to very high precision. The corresponding values I find from the light curve analysis are in good agreement for HD\,17156 (0.8$\sigma$) and TrES-2 (1.1$\sigma$) but not for HAT-P-7 (2.9$\sigma$) or HAT-P-11 (6.5$\sigma$). This indicates a problem with at least one of the approaches, which might be related to underestimation of the true uncertainties, starspot activity or the measured orbital eccentricity of the HAT-P-11 system.

Finally, the complete error budgets generated for each TEP system allow identification of the observations which would lead to the greatest improvement in our measurement of their physical properties. Many objects would benefit from further photometric observations -- which continue to be obtained for the the TEPs in the \kepler\ field of view -- as well as from spectroscopic radial velocity measurements and spectral synthesis analyses. A list of nine TEP systems is given whose orbital ephemerides will become uncertain by more than one hour within this decade; the transits of \corot-4 and \corot-14 are already not predictable to within one hour.

The homogeneous physical properties obtained in this work will be useful for detailed statistical studies of the extrasolar planet population as well as for planning many types of follow-up observations of these objects. \reff{The primary results from the current work and from previous papers in the series, along with a range of other useful information, have been concatenated and placed in an online catalogue. TEPCat is available at {\tt http://www.astro.keele.ac.uk/$\sim$jkt/tepcat/} in a range of convenient formats for readers to download for reference and further study.}


\section*{Acknowledgments}

I am grateful to Antonio Claret for calculating new theoretical models for me, to Barry Smalley and Pierre Maxted for extensive discussions about transiting planetary systems, and to the anonymous referee for a helpful report. Useful discussions and data were also provided by P.\ Bord\'e, J.\ Christiansen, D.\ Deming, A.\ Dotter and J.\ Winn. I acknowledge financial support from STFC in the form of an Advanced Fellowship. I thank the CDS, MAST, IAS and NSTeD websites for archiving the many datasets now available for transiting planets. The following internet-based resources were used in research for this paper: the ESO Digitized Sky Survey; the NASA Astrophysics Data System; the SIMBAD database operated at CDS, Strasbourg, France; and the ar$\chi$iv scientific paper preprint service operated by Cornell University.


\bibliographystyle{mn_new}


\end{document}